# Screening nature of the van der Waals density functional method: A review and analysis of the many-body physics foundation

Per Hyldgaard, Yang Jiao, and Vivekanand Shukla
*Department of Microtechnology and Nanoscience (MC2), Chalmers University of Technology, SE-41296 Gothenburg, Sweden*

We review the screening nature and many-body physics foundation of the van der Waals density functional (vdW-DF) method [ROPP 78, 066501 (2015)], a systematic approach to construct truly nonlocal exchange-correlation energy density functionals. To that end we define and focus on a class of consistent vdW-DF versions that adhere to the Lindhard screening logic of the full method formulation. The consistent-exchange vdW-DF-cx version [PRB 89, 035412 (2014)] and its spin extension [PRL 115, 136402 (2015)] represent the first examples of this class; In general, consistent vdW-DFs reflect a concerted expansion of a formal recast of the adiabatic-connection formula [PRB 90, 075148 (2014)], an exponential summation of contributions to the local-field response, and the Dyson equation. We argue that the screening emphasis is essential because the exchange-correlation energy reflects an effective electrodynamics set by a long-range interaction. Two consequences are that 1) there are, in principle, no wiggle room in how one balances exchange and correlation, for example, in vdW-DF-cx, and that 2) consistent vdW-DFs have a formal structure that allows them to incorporate vertex-correction effects, at least in the case of levels that experience recoil-less interactions (for example, near the Fermi surface). We explore the extent to which the strictly nonempirical vdW-DF-cx formulation can serve as a systematic extension of the constraint-based semilocal functionals. For validation, we provide a complete survey of vdW-DF-cx performance for broad molecular processes, for the full set of 55 benchmarks in GMTKN55 [PCCP 19, 32181 (2017)] and comparing to the quantum-chemistry calculations that are summarized in that paper. We also provide new vdW-DF-cx results for metal surface energies and work functions that we compare to experiment. Finally, we use the screening insight to separate the vdW-DF nonlocal-correlation term into pure-vdW-interaction and local-field-susceptibility effects and present tools to compute and map the binding signatures.

## I. INTRODUCTION

The van der Waals (vdW) density functional (vdW-DF) method [1–16] is a systematic approach to construct robust, truly nonlocal approximations for the exchange-correlation (XC) energy functional $E_{\rm xc}[n]$ in density functional theory (DFT). It provides a computationally efficient account [17, 18] of dispersion forces while keeping a seamless integration with the local (spin) density approximation [4, 10, 19–24] (LDA) in the homogeneous electron-gas (HEG) limit. The aim is to extend the success of the constraint-based semilocal generalized-gradient approximations [25–27] (GGAs), such as PBE [28] and PBEsol [29], relying exclusively on the nonlocal variation in the electron density $n(\mathbf{r})$. Traditional (GGA-based) DFT works well for systems characterized by dense electron concentrations, e.g., individual (small) molecules and hard (bulk) materials [30, 31]. However, more is needed to accurately describe systems that are sparse [12, 32], that is, have important low-density regions across which vdW forces and other truly nonlocal-correlation effects contribute significantly to the cohesion, structure, and function. The vast classes of molecular crystals and complexes, layered materials, polymers and biochemical systems are important examples of such sparse matter.

The versions and variants [3–5, 8, 9, 11, 16, 40–44] of the vdW-DF method enable strictly parameter-free, nonempirical DFT calculations for general (sparse or dense) matter [12, 30, 32, 45].

The vdW-DF method is based on an exact reformulation of the adiabatic connection formula (ACF) [22, 25, 46], as discussed in Refs. 10, 12, and 47 and as further detailed within. We enforce current conservation on the electron-gas response description [10]. The ACF and current-conservation foundation has a number of advantages. The vdW-DF versions and variants can, for example, be subjected to a formal coupling-constant analysis [14, 15, 48–51] to isolate kinetic-energy, kinetic-correlation-energy, and electron-electron interaction binding contributions. This analysis provides a refined mapping [14, 52] of where, for example, the vdW interactions between molecules and at surfaces originate in space [10, 53–59]. The coupling-constant scaling analysis can also be used to discuss the choice of Fock-exchange mixing in vdW-DF-based hybrids [13, 15].

At the heart of all constraint- and ACF-based XC functional designs lies an analysis of the electron-gas response to an external potential, $\delta\Phi^{\omega}_{\rm ext}$, oscillating at a specific frequency $\omega$, and causing density changes $\delta n^{\omega}_{\lambda}$. The response function $\chi_\lambda(\omega) = \delta n^{\omega}_{\lambda}/\delta\Phi^{\omega}_{\rm ext}$ is investigated as a function of an assumed reduction, $V_\lambda \equiv \lambda V$, of the electron-electron interaction $V$. The ACF then provides a formal determination of the XC energy as an average over the coupling constant $\lambda$,

$$E_{\rm xc} = -\int_0^1 d\lambda \int_0^\infty \frac{du}{2\pi}\,{\rm Tr}\{V\,\chi_\lambda(iu)\} - E_{\rm self}\,. \qquad (1)$$

The trace 'Tr' is often taken in a spatial representation of the response function, $\langle\mathbf{r}|\chi_\lambda(\omega)|\mathbf{r}'\rangle \equiv \chi_\lambda(\mathbf{r},\mathbf{r}';\omega)$,

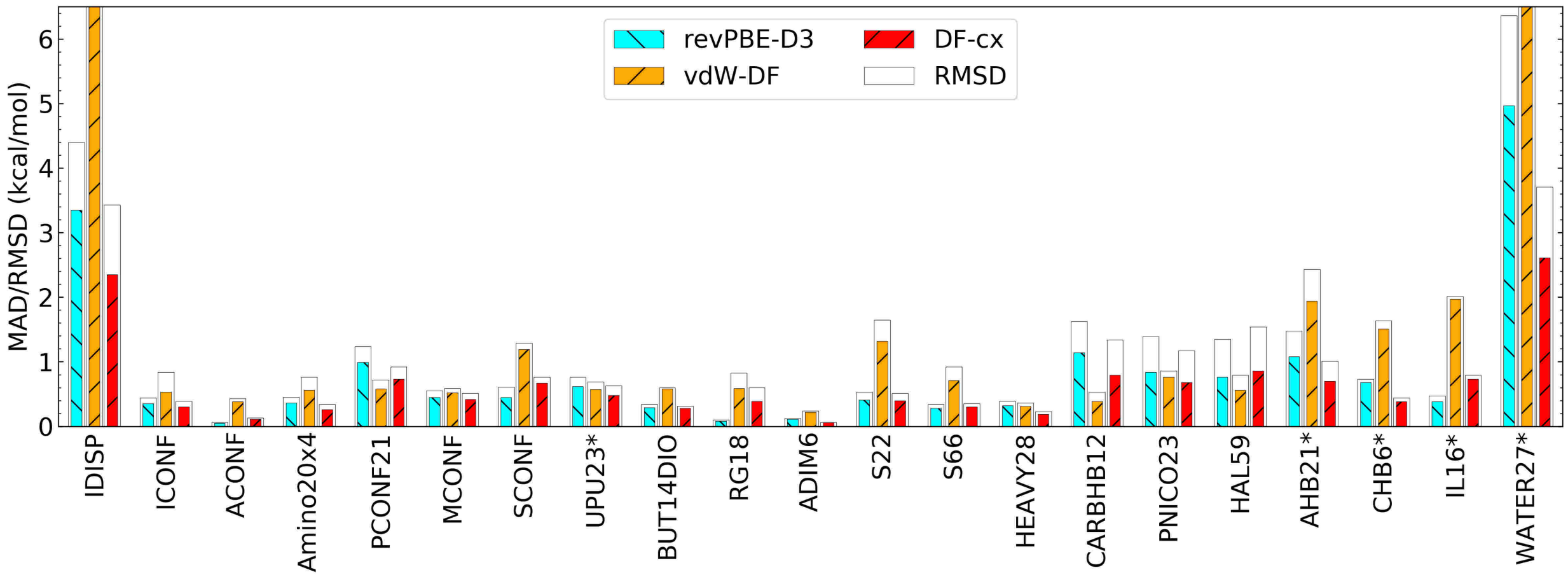

FIG. 1. Comparison of rev-PBE-D3 (the DFT-D3 correction [33] of revPBE [34], light blue), of vdW-DF [4] (the original general-geometry version of the vdW-DF method, orange), and of vdW-DF-cx [9] (the 2014 consistent-exchange method update, red) performance for all of the (intra- plus inter-molecular) noncovalent-interaction benchmark sets defined in the GMTKN55 molecular-testing suite [35]. The bars contrast mean-absolute energy deviation (MAD) and root-mean-square deviation (RMSD) values in kcal/mol, although capped for the IDISP (actual vdW-DF MAD=7.61) and WATER27 (actual vdW-DF MAD=9.53) benchmark sets. We used the spin extension of the vdW-DF method [11], where relevant, and the Quantum Espresso plane-wave code package [36], with a set of previously tested [13, 15, 37] norm-conserving pseudopotentials from the ABINIT package [38]. The Makov-Payne correction scheme [39] were used for charged-system calculations, affecting results for benchmark sets marked by an asterisk.

using the electron-electron interaction matrix element $V(\mathbf{r}-\mathbf{r}') = |\mathbf{r}-\mathbf{r}'|^{-1}$. For an illustration of how the ACF and how individual vdW-DF contributions appear when all spatial arguments are written out, we refer the readers to Ref. 60. Here we stick with a compact notation, noting that traces are invariant under basis-set changes and that our discussion of the role of Lindhard screening [61–63] would otherwise become overly cumbersome. The last term of Eq. (1) is the electron self-energy, given by $E_{\rm self} = {\rm Tr}\{\hat{n}\hat{V}\}/2$, where $\hat{n}(\mathbf{r})$ denotes the operator for the electron density at $\mathbf{r}$.

The vdW-DF method rests on the observation that the vdW interactions are an inherent part of the XC energy functional [1, 64, 65] and should be treated on the same footing as all other electron-electron interaction energy effects. Like the subsequent and related VV10 [66], rVV10 [67], and rVV10-corrected XC formulations [68–72], the vdW-DF versions and variants describe all forces directly on the foundation of the full electron-density variation [9, 10, 14, 45, 47]. We do not rely on an atom-centered vdW account, in contrast to the Grimme dispersion-correction schemes [33, 73–75], the Becke-Johnsson exchange-hole scheme [76–78], the Wannier-based vdW formulation [79], the Tkatchenko-Scheffler formulation [80], and the associated self-consistent screening extension [81]. We also avoid the need for an auxiliary determination of the atomic dipolar (or multipolar) susceptibilities and the choice of a damping function or a Coulomb-range-separation parameter that must otherwise be used to avoid double counting. The vdW-DF method simply helps us design explicit, truly nonlocal XC density functionals for standard, yet vdW-inclusive, ground-state DFT calculations.

For practical approximations, i.e., vdW-DF versions, we start with a nonlocal, double-pole model for the plasmon propagator, termed $S_{\rm xc}$, Refs. 4, 9, 12, and 47. This propagator adapts the HEG plasmon description [19, 20, 82–86], used in early LDA [21, 22, 83], to the logic of the gradient expansion [4, 47, 87, 88] of formal many-body perturbation theory (MBPT). The propagator reflects the local-energy variation of an internal (sometimes called inner) GGA-type semilocal functional [4, 8–10, 12, 89]. It therefore effectively reflects the local-field response $\tilde{\chi}(\omega) = \chi(\omega)/[1+V\chi(\omega)]$ and screening in a weakly perturbed electron gas.

For explicit vdW-DF designs, we typically proceed to extract a nonlocal-correlation energy contribution [4, 8, 9, 12, 45],

$$E_{\rm c}^{\rm nl} \approx \int \frac{du}{4\pi}\,{\rm Tr}\{S_{\rm xc}^2 - (\nabla S_{\rm xc}\cdot\nabla V/4\pi)^2\}\,, \qquad (2)$$

by expanding to second order in terms of $S_{\rm xc}$. The derivation of this truncated expression for $E_{\rm c}^{\rm nl}$ has been documented several places, for example, in Refs. 4, 7, 12, 17, and 60; It will also emerge naturally in our discussion of the full vdW-DF method and of consistent vdW-DF versions, below. For layered geometries we can also retain the full nonlocal-correlation description subject to a slightly different approximation for the local-field response [2, 3, 6].

The vdW-DF approximations for the full XC energy,

that is, vdW-DF versions and variants,

$$E_{\rm xc}^{\rm vdW-DF} = E_{\rm xc}^{\rm in} + \delta E_{\rm x}^{0} + E_{\rm c}^{\rm nl} \,, \qquad (3)$$

contain a cross-over exchange term [9] $\delta E_{\rm x}^0$, reflecting the choice of gradient-corrected exchange. The vdW-DF formulation Eq. (3) captures local and general nonlocal correlation effects in $E_{\rm xc}^{\rm in}$ and $E_{\rm c}^{\rm nl}$, respectively. The exchange effects are captured in the semilocal functional $E_{\rm xc}^0 \equiv E_{\rm xc}^{\rm in} + \delta E_{\rm x}^0$.

This review paper highlights the nature and discusses potential advantages of focusing the vdW-DF framework on the class of consistent designs, i.e., designs that seek to implement the full screening logic of the ACF [9, 10, 45]. The key idea, developed within, is that approximations to exchange are independent of the coupling constant [14, 48, 90] and must completely reside in the approximations to $\tilde{\chi}(\omega)$. This implies that we can use the Dyson equation, reflecting Lindhard screening [61–63], to balance exchange and correlations in the vdW-DF designs. In practice, for consistent vdW-DF designs, there should be no binding-energy consequences of retaining a cross-over exchange component $\delta E_{\rm x}^0$ in Eq. (3). This is effectively true in the consistent-exchange vdW-DF-cx version [9], and in the consistent spin-extension, svdW-DF-cx [11].

Figure 1 demonstrates that there are advantages in terms of accuracy and robustness in moving from the original vdW-DF version [4] and to consistent (spin) vdW-DF-cx. The figure compares the mean absolute deviation (MAD) and root-mean square deviation (RMSD) values obtained for the set of all (inter- and intra-) molecular noncovalent-interaction benchmark sets that are part of (and defined in) the GMTKN55 benchmark suite [35]. The computational details will be given in Sec. V. The performance on individual interaction problems can vary, but the vdW-DF-cx performance is significantly better than that of vdW-DF. The vdW-DF-cx has just a GGA-level exchange and should, in the GMTKN55 survey [35] be compared with (one of) the best-performing dispersion-corrected GGA, namely revPBE-D3 [33, 34] (also shown). Again, the vdW-DF-cx is seen to fare better, statistically speaking. In addition, vdW-DF-cx is seen to perform as well as revPBE-D3 when we broaden the comparison to the full GMTKN55 benchmark suite in Sec. VI.

This revPBE-D3/vdW-DF/vdW-DF-cx performance comparison highlights the importance of balancing exchange and correlation in XC functionals [9, 12, 22, 25, 45]. The first two functionals differ only in the choice of nonlocal correlation, whereas the last two differ only in the exchange, and hence in the exchange-correlation balance. The regions of smooth density variations (regions with low-to-medium values of the scaled density gradient, $s = |\nabla n|/\pi^{1/3}/(2n^{4/3})$) generally dominate in the description of molecular interactions [9, 10]. We have documented that this also holds for noncovalent interactions [9, 10, 14, 53, 56]. This implies that the Langreth-Vosko (LV) MBPT analysis [91] underpins the dominant part of the vdW-DF-cx exchange description [9] – yet LV exchange has not previously impressed anybody for molecules [92]. Meanwhile, the revPBE exchange choice [34] largely follows the hole-based design logic [93] of PBE [28] but avoids one constraint that strengthens PBE and PBEsol relevance for bulk matter that has a dense electron distribution [29].

The revPBE exchange [34] is a good choice for molecules [35], but that is clearly not enough: Fig. 1 shows that vdW-DF performs significantly worse than revPBE-D3. Meanwhile, picking a good nonlocal correlation term (that helps vdW-DF-cx perform slightly better than revPBE-D3 for noncovalent interactions, Fig. 1) is also not enough, for the same reason. Rather, it is the foundation in the Lindhard screening logic [10, 12, 61–63] that puts vdW-DF-cx ahead of both revPBE-D3 and vdW-DF, even for molecular problems.

So far the vdW-DF-cx version has proven itself useful and accurate for descriptions of a range of dense and sparse-matter problems. Successful applications include the study of traditional bulk-matter cohesion, structure, thermo-physical properties [9, 45, 94, 95], and vacancy formation [96], as well as of binding, structure, vibrations, polarization, elastic properties, phase transitions, defects, and dynamics in molecular, polymer, and layered materials [45, 97–112]. This is, for example, relevant for development of ionic liquids, batteries, thermo-electrics, and organic solar-cell materials [98, 113–118]. The vdW-DF-cx version tends to overestimate the binding energy in some layered systems [12, 119, 120], but remains accurate on structure [68, 121]. It works for both intra- and for inter-molecular bonds [9, 11, 13–15, 44, 122–124], for hybrid organic-inorganic perovskites [125], and for analysis and design of porous-materials gas storage [11, 45, 126–128] and of surface pacification [129]. It also works for studies of the role of polycyclic aromatic hydrocarbons in interstellar $H_2$ formation [130], of electric-field driven reactions [131], and of mechanochemical cocrystal formation [132]. Use of vdW-DF-cx is also helping the understanding of the crossover from physisorption to weak chemisorption in molecular adhesion [11, 133–145], of catalysis [146–148], of hydrogen tunneling [149], of magnetic coupling between adsorbates [150], and of charge transfer and electron transport in and through molecular adsorbates [151–153]. It is therefore worth exploring why (spin) vdW-DF-cx works well.

However, the use of screening to balance exchange and correlation (as in vdW-DF-cx) may find use also in future nonlocal XC functional designs. The purpose of this review paper is therefore more than presenting a rederivation of vdW-DF-cx itself. We focus on developing the concepts, and we explore and map the nature of binding contributions in both traditional bulk and in weaker vdW-interaction problems. The paper supplements Ref. 10 that began the work to interpret the vdW-DF method.

The paper has 8 major objectives. We emphasize that there exist 1) a formally exact vdW-DF framework that is closely tied to the screening nature of an effective electrodynamics, 2) powerful guiding principles for making consistent approximations, and 3) a way to simultaneously account for and balance vdW forces [1, 12, 64, 65, 154–158] and the screening [61–63, 159] & vertex-correction effects [86, 160–164] that all affect the electron-gas response and zero-point energy (ZPE) dynamics in concert. Our approach is to 4) discuss what we need for crafting a robust nonlocal-correlation functional and 5) show that those needs can be accommodated, in an Occam approach, leading to so-called consistent vdW-DF versions, like vdW-DF-cx [9]. We find that current conservation [10] and compliance with the Lindhard-Dyson logic of screening [61–63] plays a critical role in balancing exchange and correlation effects in such consistent vdW-DF designs. For consistent vdW-DFs, the current conservation [10] implies an automatic charge conservation for the vdW-DF XC hole, i.e., the essential DFT criterion [10, 45] that the quasi-particles remain neutral [22, 25]. There has only been few prior discussions of the vdW-DF method foundation in a screening approximation [6, 7, 10, 45, 53, 165].

We furthermore 6) identify systems and problems that test the Occam solution strategy behind vdW-DF-cx and 7) summarize the results of such validation checks, whether previously reported or provided here. As such the review and analysis paper also contains new results in the form of a) broad molecular benchmarking, b) tests of performance for metal surface energies and work functions, and c) mappings of the spatial variation in binding in bulk Si, Na, and W as well as in a $C_{60}$ dimer and a graphene bilayer. In addition, the paper contains d) demonstrations of the non-additivity of binding among carbyne wires, carbon nanotubes, and $C_{60}$ fullerenes.

There has only been comparatively few prior numerical explorations of the spatial distribution of the vdW-DF binding contributions [10, 14, 53–59]; Here we are therefore adding 8) a practical approach to separately track the variation in binding contributions arising from pure vdW interactions and from nonlocal vertex corrections (and related screening effects defined by an implicit cumulant expansion [12, 86]). This is done both for the carbyne-wire, the graphene-bilayer, and the Si, Na, & W bulk problems.

We note that to systematically extend the GGA success, we must deliver accuracy in concurrent descriptions of both sparse and dense matter [9, 32, 42, 45]. We hope to simultaneously succeed in predicting diverse properties with just one robust, transferable, and strictly-parameter-free functional. Consistent vdW-DF versions, like vdW-DF-cx [9] (together with its spin extension svdW-DF-cx [11],) aspire to work as such a general-purpose tool for all types of materials and their combination [45].

The paper is organized as follows. The next section II discusses the nature of nonlocal-correlation effects, including vdW interaction effects, but also noting the central role of vertex corrections. Section III provides a formal presentation of the design logic of constraint-based XC functionals, contrasting a direct path (leading to semilocal GGAs) and the indirect approach (leading to the vdW-DF method).

Section IV discusses the design logic of the recent consistent-exchange vdW-DF-cx version. Section IV also interprets the nature of XC functional contributions in such consistent vdW-DF versions, noting that vdW-DF-cx can be sorted into a local-field and a Dyson-correction component; The former reflects vertex corrections (plus related screening effects [86, 163, 164]) and the latter reflects pure vdW interactions. Section IV furthermore highlights the importance of key vdW-DF-cx features: a) seamless integration [4, 166], b) an underlying MBPT analysis of response in the electron gas [4, 7, 11, 91], c) current conservation [10], and, in particular, d) compliance with the Dyson/Lindhard screening logic [61–63].

Section V contains a summary of computational details while Sec. VI contains results and discussions that validate trust in the class of consistent vdW-DF designs. The section also contains an exploration of the role that nonlocal vertex corrections and related screening effects [85, 86, 163, 164] play in material binding.

Finally, Sec. VII contains a summary and conclusions and there are four appendices detailing: A) the relation of an exponential re-summation and screening, B) the vdW-DF double-pole plasmon-propagator model, the electrodynamics-response nature of terms arising in a full (non-truncated) implementation of the vdW-DF method, C) a discussion of how to include higher-order expansion terms in $E_{\mathrm{c}}^{\mathrm{nl}}$, and D) universal-kernel evaluations of both a pure-vdW contribution and of a cumulant contribution that describes nonlocal vertex-type screening effects in consistent vdW-DF versions.

## II. NATURE OF NONLOCAL CORRELATIONS

Figure 2 summarizes the fundamental idea of DFT and the need for developing truly nonlocal, vdW-inclusive functionals. The figure simultaneously highlights that there is something extremely right with the charge-conservation criterion that underpins the design of traditional semilocal XC functionals (i.e., GGAs) and that there are important conceptual reasons for correcting and extending the GGA logic [10, 64, 65, 91, 158].

### A. Ubiquitous vdW interactions challenge DFT

The top left panel of Fig. 2 shows a schematics of a physical systems in which electrons experience full electron-electron interactions $V$; Every electron interacts with each and every other electron, as illustrated by arrows. This interacting problem defines the behavior of molecules and materials but, except in small systems, there is no realistic approach to obtain exact solutions under general conditions. The top right panel shows that DFT nevertheless succeeds with an, in principle, correct yet computationally tractable determination of the ground state density $n(\mathbf{r})$ and energy $E_{\text{tot}}$. It does so by considering an equivalent system of noninteracting quasi-particles (shown as hexagons) moving in some effective single-particle potential (represented by the change in background color). The effective potential is a functional derivative of the XC energy functional $E_{\text{xc}}$.

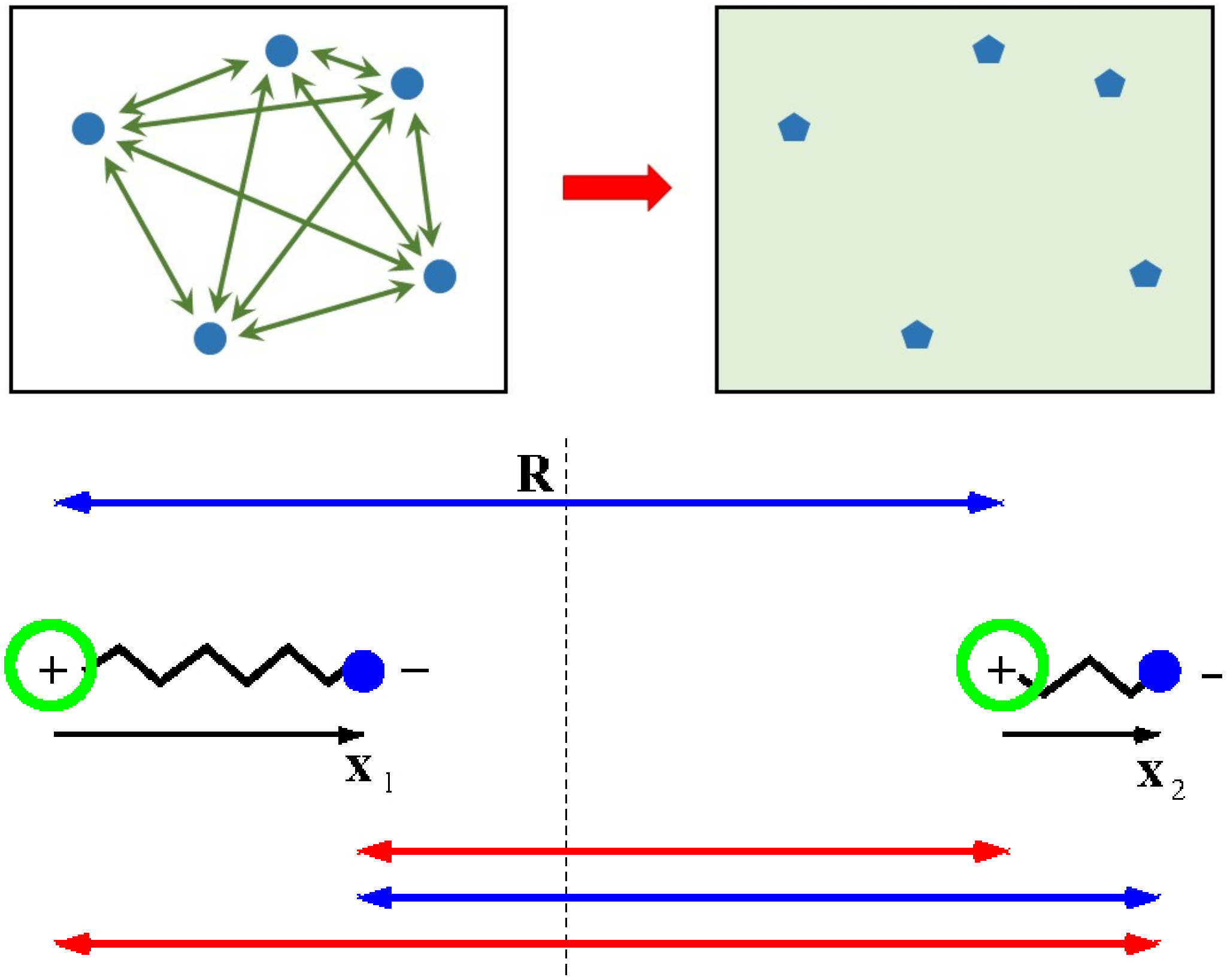

FIG. 2. Nature of the DFT formulation (top right panel), of the electron-electron interaction problem (top left panel) and the DFT need for systematic inclusion of vdW interactions, for example, among disjunct material fragments (bottom panel). The top panels show schematics of the DFT mapping of the interacting system onto a quasi-particle problem with independent-particle dynamics defined by an effective potential (green background). Each electron is wrapped in a screening cloud forming neutral electron-XC-hole composites (shown as pentagons), which in local (LDA) or semilocal (GGA) DFT are fairly compact and cannot reflect the density beyond a low-density region between, for example, molecules [6, 32]. The bottom panel illustrates the inherent limitations of LDA/GGA. The panel shows a schematic of a vdW interaction problem: Two density fragments, each with electrons (exemplified by blue dots) and associated XC holes (green shells), are separated by a characteristic length $R$ (reflecting a typical intermolecular separation). We can ignore the electrostatic and other types of binding across the intermediate density void (represented by a vertical dashed line). However, the neutral electron-XC-hole composites do have an internal zero-point energy (ZPE) dynamics. They act as antennas, and cause mutual electrodynamics couplings that produce a vdW force between the fragments [1, 3, 4, 10, 12, 64, 65, 91, 158].

The top right panel of Fig. 2 also summarizes the core problem for practical use of DFT: deciding exactly how we should place the nut (electron) in its nutshell (a surrounding so-called XC hole). Together, the electron-XC-hole pairs form neutral composites. However, such quasi-particles have an internal structure with charged components (the electron and the associated GGA-type XC hole [10]) and thus an inherent electrodynamics response.

The XC hole describes the tendency of an electron at point $\mathbf{r}$ to inhibit presence of other electrons at point $\mathbf{r}'$. The combination of each electron and of its associated XC hole forms neutral complexes that serve as noninteracting quasi-particles in practical DFT calculations. Screening and exchange effects strongly influence the electron-gas behavior and will wrap an electron (at position $\mathbf{r}$) in a matching XC hole $n_{\text{xc}}(\mathbf{r}; \mathbf{u} = \mathbf{r}' - \mathbf{r})$. An important guideline is that this XC hole must be of exactly opposite charge,

$$\int d\mathbf{u}\, n_{\text{xc}}(\mathbf{r}; \mathbf{u}) = -1 \,. \tag{4}$$

The details in how we approximate this hole is important: The XC energy is given by the Coulomb coupling inside the composite electron-XC-hole quasi-particles

$$E_{\text{xc}} \equiv \frac{1}{2} \int_{\mathbf{r}} \int_{\mathbf{r}'} \frac{n(\mathbf{r})\, n_{\text{xc}}(\mathbf{r}; \mathbf{r}' - \mathbf{r})}{|\mathbf{r} - \mathbf{r}'|} \,. \tag{5}$$

While Eq. (5) may appear as simple electrostatics, it also reflects the ZPE dynamics of the electrons relative to their associated XC holes [10, 64, 65]. The collective excitations, i.e., plasmons identified by some general eigenvalue index $\eta$ and frequency $\omega_\eta$, typically dominate in the specification of the electron gas response. Accordingly, the XC energy approximations must contain a leading term reflecting the plasmon ZPE [10, 25, 158]:

$$E_{\rm xc} \approx E_{\rm pl}^{\rm ZPE} = \sum_\eta \frac{\omega_\eta}{2}\,. \tag{6}$$

Since $\omega_\eta$ depends on the electron-density variation, the leading-term approximation, Eq. (6), allows an elegant representation of gradient effects on the XC energy Eq. (5) at surfaces [25]. Of course, most formulations of the LDA [19–23, 84] and the design of the popular constraint-based GGAs, such as PBE [28] and PBEsol [29], proceed by modeling fairly compact XC holes [93] $n_{\rm xc}(\mathbf{r};\mathbf{u})$; The details are set subject to the charge-conservation criterion Eq. (4). However, the plasmon ZPE picture, Eq. (6), and the XC hole picture, Eq. (5), can be reconciled by noting that we can link the energy density of a semi-local XC functional to an assumed plasmon dispersion, Refs. 2–4, 7, 12, and 25.

The bottom panel of Fig. 2 illustrates the conceptual problems for semilocal density functional approximations. The problems (for the design of vdW-inclusive functionals) are that the XC hole is typically seen as compact, having a modified Gaussian shape in LDA and GGA [19, 20, 84, 93, 167], and as static, without a ZPE dynamics [1, 10, 65, 77, 78]. The first assumption directly affects our description of the electrodynamics, as the XC hole also defines an approximate dielectric function, Eq. (20) below. Maggs, Rapcewicz, and Ashcroft addressed the second assumption, pointing out that a static view of the XC hole is incomplete. LDA and GGAs hide and sometimes ignore the electrodynamics coupling between the electron-XC hole systems [10, 64, 65]. The argument was originally cast in terms of diagrams [64, 65, 91], that is, insight from MBPT that goes beyond the input for specifying the details of the GGA-type descriptions [26, 64, 91, 168]. However, the argument can be summarized by evaluating changes in $E_{\rm pl}^{\rm ZPE}$ induced by the electrodynamics coupling among electron-XC-hole complexes [10, 65].

The bottom panel of Fig. 2 shows a simple one-dimensional model with two disjunct density fragments (assumed to reside on either side of the dashed line). It focuses the discussion on the virtual dynamics of electrons (negative balls on either side) and highlights the central GGA problem: The electron-XC-hole pairs are seen as compact and, for example, entirely confined inside a given electron-density fragment.

For a simple illustration of GGA limitations, we assume that the GGA-type descriptions (for either fragment) can be summarized by just one characteristic plasmon frequency $\omega_{\rm pl,0}$. The frequency $\omega_{\rm pl,0}$ describes the electron-XC-hole dynamics given by relative displacement coordinates $x_{i=1,2}$; For simplicity, we also here assume that the characteristic plasmon dynamics can be represented as a quantum harmonic oscillator with an effective spring-constants $k$ (illustrated in the lower panel) and some effective mass $m_{\rm eff}$, chosen such that $\omega_{\rm pl,0} = \sqrt{k/m_{\rm eff}}$. Adapting Ref. 25, we can represent the GGA-type XC energy for the combined but disjunct double-fragment system using

$$H_{\rm dis} = \sum_{i=1,2} \frac{m_{\rm eff}}{2}\left(\frac{dx_i}{dt}\right)^2 + \frac{1}{2}kx_i^2\,, \tag{7}$$

as we also assume no density overlap of the fragments. However, the model, Eq. (7), is fundamentally flawed.

The point is that each GGA-type quasi-particle, i.e., the electron-XC-hole complex, is a rattler that is formed by charged components (although it is neutral overall). As such, they are antennas transmitting and receiving due to the ZPE dynamics [10, 64, 65, 158]. We cannot ignore the near-field electrodynamics coupling among GGA-type XC-hole/electron systems, even when considering disjunct fragments [10, 65]. Specifically, at electron displacements $x_1$ and $x_2$ and at mean (GGA-type quasi-particle) separation $R_{12}$, there will be a near-field Hamiltonian contribution

$$H_{\rm near-field} = -\frac{2x_1\,x_2}{R_{12}^3}\,. \tag{8}$$

The systems with mutual electrodynamics coupling, Eqs. (7) and (8), can be solved by a simple canonical transformation yielding new coupled-harmonic-oscillator frequencies [56]:

$$\omega_\pm = \sqrt{\omega_{\rm pl,0}^2 \pm \frac{2}{m_{\rm eff}R_{12}^3}}\,. \tag{9}$$

In this simple model, the net ZPE energy gain by the electrodynamics coupling is

$$\Delta E_{\rm xc} \approx \Delta E_{\rm pl}^{\rm ZPE} \sim \delta E_{\rm vdW}\,, \tag{10}$$

$$\delta E_{\rm vdW} = \frac{1}{2}(\omega_+ + \omega_- - 2\omega_{\rm pl,0}) \propto -R_{12}^{-6}\,, \tag{11}$$

and thus represents the asymptotic vdW attraction. The electrodynamics-coupling argument [1, 10, 56, 64, 65, 158], and insight from the underlying MBPT characterizations, cannot easily be cast directly as an explicit refinement of a semilocal XC hole, nor in a semilocal functional [91].

### B. Inclusion of vdW forces in formal DFT

For an introduction on how the vdW-DF method includes such nonlocal-correlation effects, we need formal definitions of the electron-gas problem and for the MBPT solution approach. At any given coupling constant $\lambda$ we separate the Hamiltonian $H_\lambda = H_0 + V_\lambda$ into a single-particle and the electron-electron interaction part. We use $\hat{\psi}(\mathbf{r})$ to denote the operator for removing an electron at $\mathbf{r}$. The operators for the local density and for the kinetic energy are $\hat{n}(\mathbf{r}) = \hat{\psi}^{+}(\mathbf{r})\hat{\psi}(\mathbf{r})$ and $\hat{T} = -\frac{1}{2}\int_{\mathbf{r}} \hat{\psi}^{+}(\mathbf{r})\, \nabla^2_{\mathbf{r}}\hat{\psi}(\mathbf{r})$, respectively. The noninteracting Hamiltonian part $H_0$ also contains a (single-electron) potential $V_{\rm ext}$ representing, for example, the electron-ion interacting. For deriving the ACF and the results behind our coupling-constant analysis, it is convenient to let $H_0$ include an extra single-particle potential term that keeps the solution density independent of $\lambda$ [22, 25, 46]. The temporal arguments of operators, like $\delta\hat{n}(\mathbf{r}, t) = \hat{n}(\mathbf{r}, t) - \langle\hat{n}(\mathbf{r})\rangle$, reflect the time-propagation under $H_\lambda$.

Also, we define time-ordered Green functions and density-density correlation functions

$$g_\lambda(\mathbf{r}, \mathbf{r}'; t - t') = -i\langle T\{\hat{\psi}(\mathbf{r}, t), \hat{\psi}^{+}(\mathbf{r}', t')\}\rangle_\lambda\,, \quad (12)$$

$$\chi_\lambda(\mathbf{r}, \mathbf{r}'; t - t') = -i\langle T\{\delta\hat{n}(\mathbf{r}, t), \delta\hat{n}(\mathbf{r}', t')\}\rangle_\lambda\,. \quad (13)$$

Here '$T$' denotes the operator of time ordering [86, 167, 169] and the subscripts '$\lambda$' identify the coupling constant for which we evaluate ground-state expectation values. The quasi-particle dynamics Eq. (12) is evaluated for a given spin and we assume that $g$ can treated as diagonal in the spin indices, for simplicity of our discussion. As noted in the introduction, we use a compressed operator-type notation for their temporal Fourier transforms, for example, using $g_\lambda(\omega)$ as a shorthand that represents the full coordinate variation

$$g_\lambda(\mathbf{r}, \mathbf{r}'; \omega) = \langle\mathbf{r}|g_\lambda(\omega)|\mathbf{r}'\rangle\,. \quad (14)$$

The Green function $g_0(\omega)$, describing the dynamics of in the absence of the electron-electron interaction $V_\lambda$, is available from an explicit construction from the solutions of $H_0$. The quasi-particle dynamics can then, in principle, be determined by a Dyson equation

$$g_\lambda(\omega) = g^0_\lambda(\omega) + g^0_\lambda(\omega)\sigma(\omega)g_\lambda(\omega)\,, \quad (15)$$

where $\sigma(\omega)$ denotes the self-energy representing the effects of the electron-election interaction $V_\lambda$ on the quasi-particle dynamics [167, 169].

An external-potential fluctuation, $\delta\Phi_{\rm ext}(\mathbf{r}', t) = \delta\Phi^\omega_{\rm ext}(\mathbf{r}')e^{-i\omega t}$ will, at coupling constant $\lambda$, produce a density response

$$\delta n^\omega_\lambda(\mathbf{r}) = \int_{\mathbf{r}'} \langle\mathbf{r}|\chi_\lambda(\omega)|\mathbf{r}'\rangle\delta\Phi^\omega_{\rm ext}(\mathbf{r}')\,, \quad (16)$$

where $\omega > 0$ is assumed, without loss of generality.

The vdW-DF method succeeds in tracking the ZPE-coupling effects of the electron-XC-hole systems by relying on a new (not semi-local) framework for the design of XC energy functionals [2–4, 9, 10, 12, 45, 47]. Formally, the XC energy is given by the electron-electron correlation function $\chi(\omega)$ through the ACF, Eq. (1) or by the XC-hole formulation, Eq. (5), using the formal specification [10, 170]

$$n_{\rm xc}(\mathbf{r}; \mathbf{r}' - \mathbf{r}) = -\delta(\mathbf{r} - \mathbf{r}') - \frac{2}{n(\mathbf{r})} \int_0^\infty \frac{du}{2\pi} \int_0^1 d\lambda\, \langle\mathbf{r}|\chi_\lambda(iu)|\mathbf{r}'\rangle\,. \quad (17)$$

Both the XC hole and the XC energy functional is therefore critically dependent on a correct inclusion of the ZPE dynamics of the collective excitations of the electron gas, i.e., plasmons, defined as poles of $\chi_\lambda(iu)$ or as nodes of the corresponding Lindhard dielectric function $\kappa_\lambda(iu) = (1 + V_\lambda\chi_\lambda)^{-1}$. The tracking of energy shifts by the coupling of the electron and plasmons ZPE dynamics is, for example, the defining step in the early LDA specifications [21, 22], based on a Green function calculation of electron quasi-particle energy shifts [83].

Effectively, in the vdW-DF method we employ a exact recast of the ACF and hence of the true XC energy

$$E_{\rm xc} = \int_0^\infty \frac{du}{2\pi}\, {\rm Tr}\{\ln(\kappa_{\rm ACF}(iu))\} - E_{\rm self}\,. \quad (18)$$

The recast is defined by an effective longitudinal dielectric function [10, 47]:

$$\kappa_{\rm ACF}(\omega) \equiv \exp(-\int_0^1 d\lambda\, V\chi_\lambda(\omega))\,. \quad (19)$$

The derivation fits in this sentence, inserting Eq. (19) into Eq. (18) leads immediately back to Eq. (1). Equations (18) and (19) provide a formulation of the exact XC energy as an electrodynamics problem. This is convenient for discussing dispersive interactions acting in concert with all other interaction effects in the electron gas.

We note in passing that the utility focus is the right way to think of the MBPT progress that we have made in the Chalmers-Rutgers vdW-DF team [1–4, 6–12, 45, 47, 56, 171]. In the vdW-DF method we seek to obtain implicit designs of nonlocal XC functionals, starting from a GGA-level description to model the plasmon dispersion [3, 4, 6, 8, 9, 11]. The vdW-DF team first saw Eq. (18) as an approximative mean-value evaluation of the coupling-constant integration in the ACF [2–4, 8]. We soon realized that discussing the derivation is mute: An explicit derivation from the traditional ACF, Eq. (1), would only be relevant if we actually intended to start from some explicit approximations for $\chi_\lambda(\omega)$. We do not.

The vdW-DF method is crafted in the electron-gas tradition [45]. The complexity of a direct MBPT evaluation has always prompted XC developers to instead look at the formal structure of diagrams and MBPT as a guide [9, 10, 21, 22, 25, 26, 46, 83, 89, 91, 172–174]. The vdW-DF method does not magically vanish the challenges of brute-force MBPT. It does, however, turn the table and make it easier to leverage insight from electrodynamics [1–3, 10, 47, 53, 59, 64, 65, 158, 165, 171, 175–182], conservation laws [4, 10, 45, 83, 183], and the theory of screening [6, 10, 61–63, 157–159, 179, 184–187].

We are free to take Eqs. (18) and (19) as the vdW-DF framework [10] and focus on finding good physically motivated approximations for $\kappa_{\rm ACF}(iu)$. Our framework is, after all, an exact rewrite of the formally exact ACF [10, 47]. The success of the overall vdW-DF program [1–4, 6, 8–10, 12, 32] shows that this is indeed possible. Emphasizing a current-conservation criterion [10] and the Lindhard screening logic [62, 63] leads to better accuracy in the consistent spin vdW-DF-cx version [9, 11, 45]. This is exciting, for it points to the soundness of the vdW-DF design strategy. We do not think it is a total coincidence.

In fact, taking Eqs. (18) and (19) as the starting point for approximations is a direct reflection of some well-known MPBT successes. First, the dielectric function formulation Eq. (19) is inspired by a cluster expansion of electron-gas screening [86, 163]. Second, using Eq. (18) to describe internal-energy contributions, Eq. (18), permits a succinct MBPT discussion of the nature and magnetude of vdW and Casimir forces [158]. Both of these points are detailed further in this review.

The ACF recast as an effective electrodynamics, Eqs (18) and (19), also reflects an explicit construction of an associated effective ($\lambda$-averaged) density-density correlation function $\chi_{\rm ACF}(\omega)$, developed and discussed in Refs. 10, 12, 45, and 47. We first observe that working with an effective ACF electrodynamics function to represent the XC energy effects in DFT has a direct physical meaning to most DFT practitioners: The form of $\kappa_{\rm ACF}(\omega)$ reflects the density correlations in the exact XC hole,

$$n_{\rm xc}(\mathbf{r};\mathbf{r}'-\mathbf{r}) = -\delta(\mathbf{r}-\mathbf{r}') - \frac{1}{2\pi n(\mathbf{r})}\int_0^\infty \frac{du}{2\pi}\nabla^2_{\mathbf{r}}\langle \mathbf{r}|\ln(\kappa_{\rm ACF}(iu))|\mathbf{r}'\rangle \, . \quad (20)$$

To see this, we write out the off-diagonal matrix elements of the ACF dielectric function

$$\langle \mathbf{r}|\ln(\kappa_{\rm ACF}(iu))|\mathbf{r}'\rangle = -\int_0^1 d\lambda \int_{\mathbf{r}''} V(\mathbf{r}-\mathbf{r}'')\langle \mathbf{r}''|\chi_\lambda(iu)|\mathbf{r}'\rangle \, , \quad (21)$$

and use the Gauss-condition on the Coulomb interaction matrix element, $\nabla^2_{\mathbf{r}} V(\mathbf{r}-\mathbf{r}') = -4\pi\delta(\mathbf{r}-\mathbf{r}')$. We find

$$\int_0^1 d\lambda \, \langle \mathbf{r}|\chi_\lambda(iu)|\mathbf{r}'\rangle = \frac{1}{4\pi}\nabla^2_{\mathbf{r}}\langle \mathbf{r}|\ln(\kappa_{\rm ACF}(iu))|\mathbf{r}'\rangle \, , \quad (22)$$

which, in turn, proves Eq. (20).

We also observe that the traditional ACF contains a $\lambda$-averaging of weighted $\chi_\lambda(\omega)$ contributions while we seek a representation of the effective response $\chi_{\rm ACF}(\omega)$ to leverage the electrodynamics insight that underpins vdW interactions [1, 64, 65, 158]. The effective response $\chi_{\rm ACF}(\omega)$ is given in terms of a set of $\lambda$-averaged poles of $\chi_\lambda(\omega)$, i.e., plasmons of the effective ACF electrodynamics. We are not aware of any explicit formulation of the coupling-constant variation of the collective excitations themselves (nor could we be sure that such an approach is tractable). In the vdW-DF method we overcome this challenge by instead setting the effective response form implicitly [10],

$$-V\,\chi_{\rm ACF}(\omega) \equiv 1 - \exp(-\int_0^1 d\lambda\, V\,\chi_\lambda(\omega)) \, , \quad (23)$$

so that it reflects a Lindhard-type relation [10, 62]

$$\kappa_{\rm ACF}(\omega) = (1 + V\chi_{\rm ACF}(\omega))^{-1} \, . \quad (24)$$

This is motivated because there is physical meaning in the effective ACF electrodynamics and deviating from the screening logic is an uncontrolled approximation [63].

In fact, since the ACF recast, given by Eqs. (20) and (18), should be seen as a proper electrodynamics problem, it must reflect an effective long-range interaction, denoted $V_{\rm ACF} \sim V_{\lambda_{\rm eff}}$. Working with an effective long-range interaction always makes it essential to pursue a consistent handling of screening effects [10, 61–63]. As indicated, the ACF electrodynamics ($V_{\lambda_{\rm eff}}$ and $\kappa_{\rm ACF}(\omega)$) is defined by an effective coupling constant value $0 < \lambda_{\rm eff} < 1$, as discussed in Sec. III. For a given external potential change $\delta\Phi^\omega_{\rm ext}$ (assumed to oscillate at frequency $\omega$), the ACF recast leaves no wiggle room in how we balance the external field change $\delta\mathbf{E}^\omega_{\rm ext}(\mathbf{r}) = -\nabla\delta\Phi^\omega_{\rm ext}$ and the resulting induced polarization field $\delta\mathbf{P}^\omega(\mathbf{r})$ (unless we make an uncontrolled approximation in the Dyson equation for the electron-gas response). The ACF dielectric function $\kappa_{\rm ACF}$ also allows us to define a local potential $\delta\Phi^\omega_{\rm loc} \equiv \kappa^{-1}_{\rm ACF}(\omega)\delta\Phi^\omega_{\rm ext}$ and an associated local electric field $\delta\mathbf{E}^\omega_{\rm loc}(\mathbf{r}) = -\nabla\delta\Phi^\omega_{\rm loc}$. Mutually consistent, local-field and external-field susceptibilities must be defined

$$\tilde{\boldsymbol{\alpha}}(\mathbf{r},\mathbf{r}';\omega) = \delta\mathbf{P}^\omega(\mathbf{r})/\delta\mathbf{E}^\omega_{\rm loc}(\mathbf{r}') \, , \quad (25)$$

$$\boldsymbol{\alpha}_{\rm ext}(\mathbf{r},\mathbf{r}';\omega) = \delta\mathbf{P}^\omega(\mathbf{r})/\delta\mathbf{E}^\omega_{\rm ext}(\mathbf{r}') \, . \quad (26)$$

The local-field susceptibility $\tilde{\boldsymbol{\alpha}}(\omega)$ corresponds to an effective polarization insertion [169] $\tilde{\chi}_{\rm ACF}(\omega)$ that must also adhere to the Lindhard screening logic [10]:

$$\kappa_{\rm ACF}(\omega) = 1 - V\tilde{\chi}_{\rm ACF}(\omega) \, . \quad (27)$$

Working with the effective susceptibilities, Eqs. (25) and (26), also provides an intuitive picture. This picture can be used to simplify our discussion of nonlocal-correlation effects in the vdW-DF method and thus help guide approximations to Eq. (18).

We can compute (within an approximation to the ACF electrodynamics description) both the resulting density change $\delta n^\omega$ and the polarization field $\delta P^\omega$ produced by the induced current. However, current-conservation and the continuity equation stipulates that these results must be related [10, 47],

$$\delta n^\omega = \chi_{\rm ACF}(\omega)\delta\Phi^\omega_{\rm ext} = \nabla\cdot\boldsymbol{\alpha}_{\rm ext}(\omega)\cdot\nabla\delta\Phi^\omega_{\rm ext} \quad (28)$$

$$= \tilde{\chi}_{\rm ACF}(\omega)\delta\Phi^\omega_{\rm loc} \, . \quad (29)$$

The second line follows because Eqs. (24) and (27) imply an effective Dyson-type equation [10], which can also be expressed

$$\chi_{\rm ACF}V = \tilde{\chi}_{\rm ACF}V + \tilde{\chi}_{\rm ACF}V\tilde{\chi}_{\rm ACF}V + \cdots \, . \quad (30)$$

Our approximation for the dielectric function must therefore contain explicit longitudinal projections [10]:

$$\chi_{\rm ACF}(\omega)V = \nabla\cdot\boldsymbol{\alpha}_{\rm ext}(\omega)\cdot\nabla V\,, \tag{31}$$

$$\tilde{\chi}_{\rm ACF}(\omega)V = \nabla\cdot\tilde{\boldsymbol{\alpha}}(\omega)\cdot\nabla V\,. \tag{32}$$

The longitudinal projection is, in turn, essential for making robust approximations for the associated dielectric function, $\kappa_{\rm ACF}(\omega)$. Assume that $\boldsymbol{\epsilon}(\omega)=1+4\pi\tilde{\boldsymbol{\alpha}}(\omega)$ denotes an approximation for a dielectric tensor function that aims to reflect the effective electrodynamics. To ensure current conservation [10], we must use the longitudinal projection

$$\kappa_{\rm long}(\omega)\equiv\nabla\cdot\boldsymbol{\epsilon}(\omega)\cdot\nabla G\,, \tag{33}$$

where $G=-V/4\pi$ denotes the Coulomb Green function.

The need for a longitudinal projection has direct implications for the vdW-DF method, as is also further discussed in Appendix B. For consistent vdW-DF approximations, like vdW-DF-cx [9], we focus on making ACF dielectric-function approximations that have this explicit longitudinally projection. That is, we focus on formulations where we effective use

$$\kappa_{\rm ACF}(\omega)\longrightarrow\kappa_{\rm long}(\omega)=\nabla\cdot\epsilon(\omega)\cdot\nabla G\,, \tag{34}$$

inside formal vdW-DF functional specification, Eq. (18).

We observe that using the projected form Eq. (33) is consistent with the vdW-DF framework, Eq. (18). To show this we note that the exact XC functional specification, Eq. (18), can equally well be written

$$E_{\rm xc}=\int_0^\infty\frac{du}{2\pi}\,{\rm Tr}\{\ln(1-\tilde{\chi}_{\rm ACF}(iu)V)\}-E_{\rm self}\,. \tag{35}$$

This follows simply by writing out a Taylor expansion of the logarithm and using the invariance of the trace over spatial coordinates. Given Eq. (35), it is clear that we are always in a position to explicitly enforce the logitudinal projection and to ensure current conservation via Eq. (33). The projection is systematically used in the evaluation of the nonlocal-correlation term, for example, in its truncated form Eq. (2), Ref. 12.

Mahan used the formal '$\ln(\kappa)$' structure of Eq. (18) to compute the interaction of disjunct fragments by an electrodynamics coupling [158]. Mahan's analysis was cast in model susceptibilities. In the vdW-DF method, Eq. (1), we inherit the Mahan '$\ln(\kappa)$' analysis [158], leading us not only to the above-summarized electrodynamics interpretation of the ACF, but also directly to a formal XC-energy evaluation [10, 158]:

$$\int_0^\infty\frac{du}{2\pi}\,{\rm Tr}\{\ln(\kappa_{\rm ACF}(iu))\}=E_{\rm pl}^{\rm ZPE}+E_{\tilde{\boldsymbol{\alpha}}}^{\rm res}\,. \tag{36}$$

The terms are given by the residuals in the complex-frequency contour integral [22, 25, 63, 158], that is implied in the ACF, Eq. (1). The leading component is just the ZPE sum, Eq. (6), of plasmons, defined as the zeros of $\kappa_{\rm ACF}(\omega)$, Ref. 10. The term $E_{\tilde{\boldsymbol{\alpha}}}$ is the sum of residuals produced at the poles of $\tilde{\boldsymbol{\alpha}}(\omega)$. As such, this term is set by the poles of $\kappa_{\rm ACF}(\omega)$, Refs. 10 and 158.

The vdW-DF framework and method are naturally set up to capture the Ashcroft ZPE-coupling mechanism for the vdW interactions because of the electrodynamics recasting. This follows because the ACF contour integral [22, 25] reflects the collective excitations, and the nature of such plasmons depends on the system-specific electrodynamics coupling among the set of electron-XC-hole composites [10, 158]. There is no need to make an assumption of a harmonic behavior for plasmons (as we did for illustration purposes above).

For example, for two disjunct fragments, the binding energy contribution from the XC energy binding reduces to the expected asymptotic form of the vdW interaction [10],

$$\Delta E_{\rm xc}\to\Delta E_{\rm vdW}^{\mathcal{AB}}=-\int\frac{du}{4\pi}{\rm Tr}_{\mathcal{A}}\{\boldsymbol{\alpha}_{\rm ext}^{\mathcal{A}}\cdot T_{\mathcal{AB}}^{\rm dip}\cdot\boldsymbol{\alpha}_{\rm ext}^{\mathcal{B}}\cdot T_{\mathcal{AB}}^{\rm dip}\}\,, \tag{37}$$

where $T^{\rm dip}(\mathbf{r},\mathbf{r}')\equiv-\nabla_{\mathbf{r}}\nabla_{\mathbf{r}'}V(\mathbf{r}-\mathbf{r}')$. In Eq. (37), we have used subscripts on the dipole-dipole coupling, $T_{\mathcal{AB}}^{\rm dip}$, to emphasize that we consider components connecting points in the spatially disjunct regions. The result Eq. (37) follows simply by expanding the logarithm in the formal recast, Eq. (18) to second order in $\chi_{\rm ACF}(iu)$ using Eq. (34) and the nature of an assumed electrodynamics approximation, $\boldsymbol{\epsilon}(\omega)=1+4\pi\tilde{\boldsymbol{\alpha}}(\omega)$.

The result, Eq. (37), corresponds exactly to the starting point of the Zaremba-Kohn description of physisorption [184, 185] and holds as long as we can ignore interaction-induced changes in the density, Ref. 10. At the same time, Eq. (18) and vdW-DF are set up to track such vdW attractions (and other nonlocal-correlation effects) also when they occur within materials fragments or between fragments with a finite density overlap [2–4, 10, 12, 45].

### C. Inclusion of GGA-type screening effects

In the following sections, supported by Appendix A, we explain how the vdW-DF method adapts the second susceptibility term $E_{\tilde{\boldsymbol{\alpha}}}^{\rm res}$ of Eq. (36) to secure a consistent description both of vdW interactions and of general screening effects, including vertex corrections.

We recall that the electron-gas is a many-particle system and that the bare Coulomb form, $|\mathbf{r}_1-\mathbf{r}_2|^{-1}$, of the interaction matrix element can only directly describe the situation with exactly two electrons at play. Vertex corrections quantify the extent that an actual interaction event (for example, defining the quasi-particle dynamics) differs from the bare Coulomb interaction [167, 169]. In MBPT, we handle a large part of all the interaction consequences by collecting these raw-Coulomb modifications into an effective form, described by an interaction vertex $\Gamma$, while keeping track of double counting.

The approach is exact when consistently implemented [89, 168, 169, 188, 189]. As we shall summarize in Sec. III, the extent of vertex corrections also directly controls the formal differences between constraint-based GGAs and LDA, when starting from an RPA-like model analysis [25, 91, 174, 190]. The top panel of Fig. 3 shows a Dyson expansion for the electron Green function where the fourth term is a simple example of a contribution containing an interaction vertex correction.

The handling of vertex corrections in the vdW-DF method works both in the HEG limit and in the presence of density gradients. For example, the recent (s)vdW-DF-cx version [9, 11] competes favorably with constraint-based GGAs, even in materials with a dense electron distribution. A high accuracy for (s)vdW-DF-cx (and for the related variants, vdW-DF-C09 and vdW-DF-optB86b [41, 42]), is documented both for oxide ferroelectrics [45, 191], for the structural and thermo-physical properties of metals [42, 94, 95], and for many other types of problems, as mentioned in the introduction.

The term $E^{\text{res}}_{\tilde{\boldsymbol{\alpha}}}$ is formally the sum of residuals at the poles of $\kappa_{\text{ACF}}(\omega)$, but that observation does not reflect how we put the material susceptibility to work in the vdW-DF method. Within a single-particle picture, such as the Hartree-Fock (HF) approximation, Mahan derived the simple form [158],

$$E^{\text{res,HF}}_{\tilde{\boldsymbol{\alpha}}} = -\sum_{\xi} \frac{\omega^{\text{sp}}_{\xi}}{2}\,, \tag{38}$$

where $\omega^{\text{sp}}_{\xi}$ denotes the single-particle excitation energies. Mahan's characterization of vdW interactions [158] provides an XC-energy determination, via Eq. (36), that is consistent with the random phase approximation (RPA) for spatially inhomogeneous systems [10]. The RPA is a special limit of the vdW-DF method, Ref. 10, but the vdW-DF method aims to go beyond RPA in keeping vertex corrections.[1]

In the vdW-DF method, we use a bootstrap approach to nucleate the nonlocal-correlation functionals around GGA ideas [4, 8, 10]. This is done instead of the RPA reliance of a lowest-order approximation for the local-field response function. DFT wisdom suggests keeping local exchange and correlation together [25] as done, e.g., in the screened-exchange description for local correlations in LDA [21, 22]. Vertex corrections are of central relevance for the success of LDA [19, 20, 25, 84]. However, vertex corrections are also essential for an accurate description of screening and electron correlations in cases with a pronounced density gradient. In a discussion of the quasi-particle dynamics, they can be succinctly described in cases with recoil-less interactions [85, 86, 163, 164], and we seek to port that insight.

Specifically, we rely on an effective model component, $E_{\text{xc},\alpha} \neq E^{\text{res,HF}}_{\tilde{\boldsymbol{\alpha}}}$, that reflects the non-vdW parts of a GGA-like XC functional. It does that in terms of a simple, scalar model susceptibility form $\alpha(\omega)$. We note that a local-field susceptibility, like $\tilde{\boldsymbol{\alpha}}$ or $\alpha$, naturally contains nonlocal-correlation effects also when they arise within a density fragment. Such fragments can be molecules or traditional bulk matter, cases where a GGA can be expected to work. In practice, we nucleate this vdW-DF start around a so-called internal functional $E^{\text{in}}_{\text{xc}}$ that retains all of the vertex correction effects entering the LDA specification. The vdW-DF method then tries to capture additional gradient-corrected vertex corrections though an exponential re-summation [86, 163, 164] in $E_{\text{xc},\alpha}$.

Collective excitations are expected to contribute significantly also in the GGA-type description $E_{\text{xc},\alpha}$. A proper GGA definition already exhausts most of the plasmon term $E^{\text{ZPE}}_{\text{pl}}$ in the implicit XC functional specification, Eq. (36). We can assume that most of the screening effects, including vertex corrections [85, 86, 163, 164], are already included in the GGA. It is primarily the pure vdW interaction term, denoted $E^{\text{nl}}_{\text{c,vdW}}$ (and given by the logic described in subsections II.A and II.B,) that is missing from the GGA account when the GGA is used to described sparse-matter systems. The formal structure of Eq. (36) is highly useful since the contour integral tracks all screening effects in the electron gas [10, 158]. It allows us to isolate the pure vdW-type interactions in a term with an explicit longitudinal projection, as discussed in Sec. II.B.

In this paper we focus on the class of consistent vdW-DF versions (like the vdW-DF-cx version [9]), that arise with an indirect strategy for XC-functional design (as defined and detailed below) and that also adhere to the Dyson/Lindhard screening logic [10, 12]. The design approach can be seen as simply trusting the ACF guiding principle for the vdW-DF method [4, 7, 10, 12, 14, 45]. What consistent vdW-DF versions do, in practice, is to renormalize the physics content of the Eq. (36) contributions:

$$E^{\text{res}}_{\tilde{\boldsymbol{\alpha}}} - E_{\text{self}} \longrightarrow E_{\text{xc},\alpha}\,, \tag{39}$$

$$E^{\text{ZPE}}_{\text{pl}} \longrightarrow E^{\text{nl}}_{\text{c,vdW}}\,. \tag{40}$$

This leads to a revised functional specification

$$E^{\text{DF}}_{\text{xc}} = E_{\text{xc},\alpha} + E^{\text{nl}}_{\text{c,vdW}}\,. \tag{41}$$

What we have effectively implemented in the recent consistent-exchange (spin-)vdW-DF-cx version [9–11, 45], is thus a pure vdW term $E^{\text{nl}}_{\text{c,vdW}}$ which aims to supplement the non-vdW parts of a GGA-type description (in $E_{\text{xc},\alpha}$).

The reorganization, Eqs. (36), (39), and (40), reflects the interpretation of the vdW-DF method as a mutual electrodynamics coupling of GGA-type XC holes [10].

[1] We note that both RPA and the vdW-DF method evaluate the XC energy as a formal contour-integral counting of ZPE shifts [10, 158]. The formal structures of the XC specification in RPA and the vdW-DF method are identical, just given by different models of the longitudinal dielectric functions [10].

The same GGA-like screening logic [10] (defined by formal input $E_{\rm xc}^{\rm in}$) enters in both terms of Eq. (41). The form of $E_{\rm c,vdW}^{\rm nl}$, expressed in terms of the physics content of $E_{\rm xc}^{\rm in}$, is another key vdW-DF step [4, 7, 11, 12, 45], that will be summarized later.

## III. SCREENING IN FORMAL DFT

Here we present a more systematic discussion of the role of screening in formulations of local, semi-local, and nonlocal XC energy functionals. We motivate the design logic of the vdW-DF method in general. By stressing the role of an effective Dyson equation and a current-conservation criterion, we also prepare for our subsequent presentation (in Section IV) of the consistent-exchange vdW-DF-cx version and for an analysis of the screening nature of the vdW-DF-cx components.

### A. Response in the electron gas

The exact XC energy functional for DFT calculations is given by formal response theory through the ACF [22, 25, 46], Eq. (1). To detail the discussion, it is convenient to introduce the longitudinal Lindhard dielectric function [61–63]:

$$\kappa_\lambda(\omega) = (1 + V_\lambda \chi_\lambda(\omega))^{-1} \,. \tag{42}$$

Here $\chi_\lambda(\omega)$ denotes the Fourier transform of Eq. (13). It is also convenient to define a screened or effective many-body interaction

$$W_\lambda(\omega) = \kappa_\lambda(\omega)^{-1} V_\lambda \,. \tag{43}$$

given by the integral equation

$$W_\lambda(\omega) = V_\lambda + W_\lambda(\omega)\tilde{\chi}_\lambda(\omega) V_\lambda \,. \tag{44}$$

This screened interaction is independent of the spin of scattering electrons [169].

The Lindhard analysis of screening gives

$$\tilde{\chi}_\lambda(\omega) = \chi_\lambda(\omega)\kappa_\lambda(\omega) \,, \tag{45}$$

$$\kappa_\lambda(\omega) = 1 - V_\lambda \tilde{\chi}_\lambda(\omega) \,, \tag{46}$$

reflecting a Dyson equation for the density-density correlation function

$$\begin{aligned} \chi_\lambda(\omega) &= \tilde{\chi}_\lambda(\omega) + \tilde{\chi}_\lambda(\omega) V_\lambda \tilde{\chi}_\lambda(\omega) + \dots \\ &= \tilde{\chi}_\lambda(\omega) + \tilde{\chi}_\lambda(\omega) V_\lambda \chi_\lambda(\omega) \,. \end{aligned} \tag{47}$$

The interaction-kernel function $\tilde{\chi}_\lambda(\omega)$ of Eq. (47) describes the local-field density response [169, 188]. The density change $\delta n_\lambda^\omega$, Eq. (16), induced by an external potential $\delta\Phi_{\rm ext}^\omega$, can equivalently be expressed

$$\delta n_\lambda^\omega = \tilde{\chi}_\lambda(\omega)\delta\Phi_{\rm loc}^\omega \,, \tag{48}$$

with $\Phi_{\rm loc}^\omega \equiv \kappa_\lambda^{-1}(\omega)\delta\Phi_{\rm ext}^\omega$.

Physics insight on the local-field response $\tilde{\chi}_\lambda$ (or equivalently on screening) leads to guidelines for the design of density functional specifications. For example, a Hubbard-type analysis of the HEG density response [19, 20, 25, 84, 190, 192] to a potential perturbation $\delta\Phi_{\rm loc}^\omega(\mathbf{q})$ (defined by frequency $\omega$ and wavevector $\mathbf{q}$) modifies the RPA by inclusion of a vertex-correction function $\gamma(q = |\mathbf{q}|)$:

$$\tilde{\chi}_\lambda(\mathbf{q},\omega) = \frac{\tilde{\chi}_{\lambda=0}(\mathbf{q},\omega)}{1 + V(q)\gamma(q)\tilde{\chi}_{\lambda=0}} \,. \tag{49}$$

Here $V(q)$ denotes the Fourier transform of the electron-electron interaction matrix element

$$V(q = |\mathbf{q}|) \equiv \int d\mathbf{r}\, \frac{e^{i\mathbf{q}\cdot(\mathbf{r}-\mathbf{r}')}}{|\mathbf{r}-\mathbf{r}'|} = \frac{4\pi}{q^2} \,. \tag{50}$$

The function $\gamma(q)$ approximates the interaction vertex $\Gamma$ at conditions that holds for the HEG. The Hubbard-type response approximation corresponds to the density-density correlation function [25]:

$$\chi_\lambda(\mathbf{q},\omega) = \frac{\tilde{\chi}_{\lambda=0}(\mathbf{q},\omega)}{1 - \lambda b(\mathbf{q},\omega)} \,, \tag{51}$$

$$b(\mathbf{q},\omega) = V(q)[1-\gamma(q)]\tilde{\chi}_{\lambda=0}(\mathbf{q},\omega) \,. \tag{52}$$

The wavevector arguments reflect a Fourier transform in $(\mathbf{r}-\mathbf{r}')$ of, for example, $\langle \mathbf{r}|\chi_\lambda(\omega)|\mathbf{r}'\rangle$. The Hubbard approximation for the HEG response, Eqs. (51) and (52), makes it possible to provide an analytical evaluation of the coupling-constant integral. The result is a formal, Hubbard-based specification of the HEG XC energy [25]:

$$E_{\rm xc}^{\rm Hub} = \int \frac{du}{2\pi}\, \frac{d\mathbf{q}}{(2\pi)^3}\, \frac{\ln(a(\mathbf{q},\omega=iu))}{[1-\gamma(q)]} - E_{\rm self} \,, \tag{53}$$

$$a(\mathbf{q},\omega) \equiv [1 - b(\mathbf{q},\omega)] \,. \tag{54}$$

The form Eq. (53) reduces to the RPA result for the homogeneous gas in the limit $\gamma(|\mathbf{q}|) \to 0$. The longitudinal projections, Eqs. (32) and (34), are implicit in the Hubbard-type specification of the HEG XC energy, Eq. (53). This follows because the translational invariance ensures that $a(\mathbf{q},\omega)$ is diagonal in $q$ space.

The success of DFT for descriptions of traditional, dense material can be seen as a consequence of including vertex corrections, that is, taking $\gamma(|\mathbf{q}|) \neq 0$. The RPA description fails but a MBPT characterization of response in the near-HEG limit, and of the role of the $V(q)\gamma(|\mathbf{q}|)$ variation at $\mathbf{q} \to 0$ in particular, formally guides the design of the LDA and GGA XC energy functionals [25, 26, 89, 91, 168, 189]. In practice, the effects of the important $V(q=0)\gamma(|\mathbf{q}|=0)$ value is absorbed into the modern QMC-based specification [23] of the LDA XC energy functional $E_{\rm xc}^{\rm LDA}$, Ref. 25 and 91.

Going beyond LDA, the initial analysis work concentrated on setting a quadratic nonlocal form [1, 65, 89, 173, 193]:

$$E_{\rm xc} \approx E_{\rm xc}^{\rm LDA} - \frac{1}{2}\int_{\mathbf{r}}\int_{\mathbf{r}'} K_{\rm xc}[n](\mathbf{r},\mathbf{r}')\,(n(\mathbf{r}) - n(\mathbf{r}'))^2 \,. \tag{55}$$

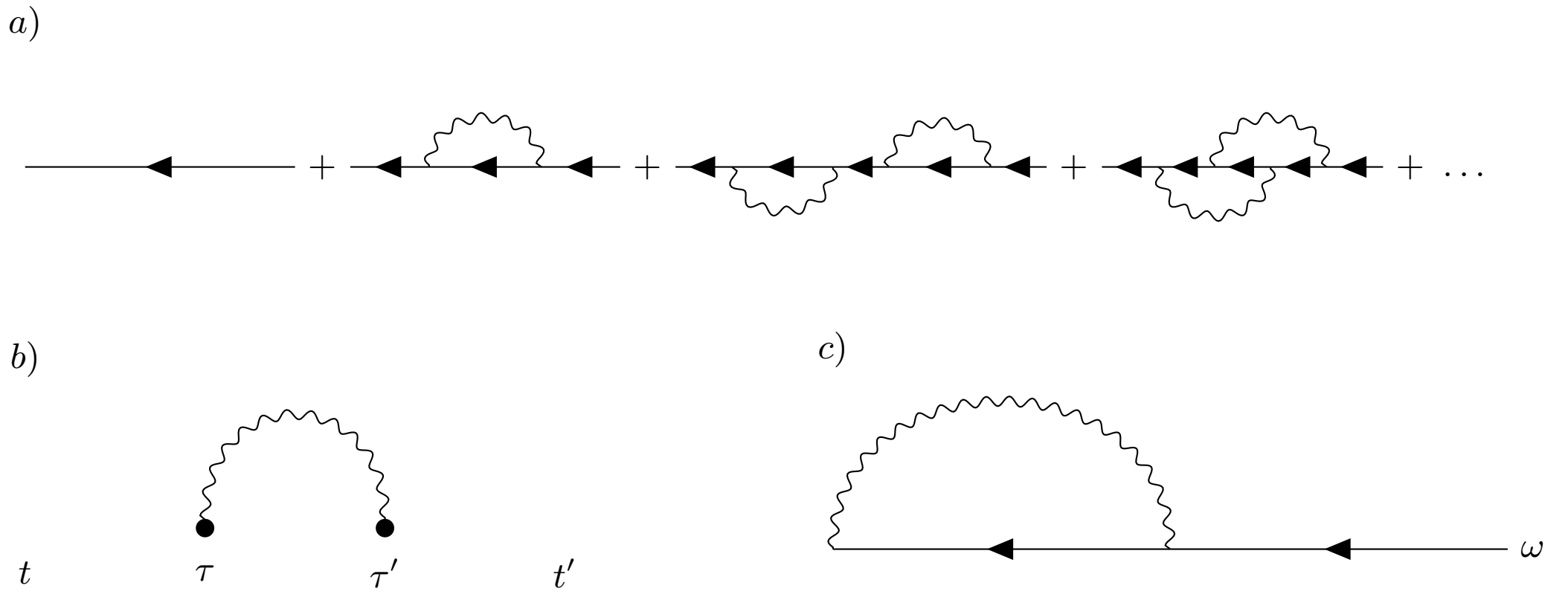

FIG. 3. Feynman-diagram solutions for the quasi-particle dynamics of a level having recoil-less interactions with a surrounding itinerant electron gas, described by the ZPE dynamics of plasmons $\omega_\eta$ (shown as wiggly lines). The level can be a core level or an extended state near the Fermi surface [85, 86, 163, 164]. The (core-)level Green function $\mathcal{G}^0$ is shown by arrow lines. Panel a) illustrates a traditional Dyson expansion for the interacting or screened Green function $\mathcal{G}$, with the fourth term illustrating the nature of vertex corrections. Panel b) illustrates the elegance and nature of the linked-cluster expansion or cumulant solution [86, 159–162] for $\mathcal{G}(t-t')$; In the time domain it is sufficient to consider a single connected diagram as the exponential re-summation provides an automatic inclusion of, for example, all vertex contributions (in the absence of recoil [86, 163, 164]). Panel c) illustrates that the $\mathcal{G}^0W$ approximation for the electron self energy underpins the corresponding frequency-domain formulation [164] of this cumulant expansion, Eq. (67).

For DFT calculations, XC functional specifications can be obtained by considering the global density variation $n(\mathbf{r})$ and enforcing XC hole conservation, as done in the averaged/weighted density approximations (ADA/WDA) [173]. They can also be obtained in a perturbation analysis, considering small density fluctuations $n(\mathbf{r}) = n_0 + \delta n(\mathbf{r})$ around a constant (average) density $n_0$. In the near-HEG limit, the kernel $K_{\rm xc}$ is translational invariant, set by the average electron density, and given by the Fourier components

$$K_{\rm xc}(\mathbf{q}) \equiv \int d\mathbf{r}\, e^{i\mathbf{q}\cdot(\mathbf{r}-\mathbf{r}')}\, K_{\rm xc}(\mathbf{r}-\mathbf{r}')\,. \tag{56}$$

For density perturbations, $\delta n_{\mathbf{q}}(\mathbf{r}) = \delta n_{\mathbf{q}} e^{i\mathbf{q}\cdot\mathbf{r}}$ of a given wavevector $\mathbf{q} \neq 0$, the near-HEG results Eq. (55) becomes [91]:

$$E_{\rm xc} \approx E_{\rm xc}^{\rm LDA} + K_{\rm xc}[n](\mathbf{q})\,|\delta n_{\rm q}|^2\,. \tag{57}$$

Using the Hubbard response description the MBPT result for the near-HEG kernel is [91]

$$K_{\rm xc}(\mathbf{q}) = -2V(q)\gamma(q=|\mathbf{q}|)\,, \tag{58}$$

and thus formally set by the $\mathbf{q}$ variation of the Hubbard-vertex function $\gamma(\mathbf{q})$. The kernel momentum dependence is expressed [47, 91, 174]:

$$K_{\rm xc}(q=|\mathbf{q}|) = K_{\rm xc}(0) + \frac{\pi}{8k_{\rm F}^4} Z(q) q^2\,. \tag{59}$$

The $\mathbf{q}=0$ contribution can be ignored since it is already part of the LDA specification [91]. The corrections for semi- and nonlocal XC functionals are reflected in the dimensionless quantity $Z(q)$.

The nature of the local-field response, and hence screening can, in principle, be computed from a diagram analysis [169]; In practice, such analysis is made primarily to extract design guidelines, for example, Refs 89, 91, 168, 174, and 189. The local-field response can be further separated into spin components of the interaction kernel [188],

$$\tilde{\chi}_\lambda(\omega) \equiv \sum_{\mu,\nu} P_\lambda^{\mu,\nu}(\omega)\,. \tag{60}$$

Here $P_\lambda^{\mu,\nu}(\mathbf{r}_1,\mathbf{r}_2;\omega)$ describes how the $\mu$-spin density at position $\mathbf{r}_1$ is affected by a local potential acting on the $\nu$-spin density at position $\mathbf{r}_2$. The vdW-type nonlocal-correlation diagram [64, 65, 91] – Fig. "1c" in Ref. 189 and Fig. "4c" in 91 – has diagonal as well as off-diagonal components.

The semilocal (or GGA) XC functionals can, in general, be expressed by the specification of a local energy-per-particle variations $\varepsilon_{\rm xc}^0(\mathbf{r})$:

$$E_{\rm xc}^0 = \int_{\mathbf{r}} n(\mathbf{r})\, \varepsilon_{\rm xc}^0(n(\mathbf{r}), s(\mathbf{r}))\,. \tag{61}$$

A closely related representation is also useful for a discussion of vdW-DF-cx and other nonlocal-correlation XC functional versions [14, 55, 57],

$$E_{\rm xc} = \int_{\mathbf{r}} n(\mathbf{r})\, \varepsilon_{\rm xc}[n](\mathbf{r})\,. \tag{62}$$

This latter form merely reflects an expression of the energy density for the XC functional; The energy-per-particle variation is then, as indicated, a functional of the electron density variation. We shall mostly use the energy-per-particle concept for local and semilocal functionals, as indicated by the superscript '0' in Eq. (61).

In the LDA, the $\varepsilon_{\rm x/c}^{\rm LDA}$ variation is entirely specified by the local value of the Fermi vector $k_F(\mathbf{r}) = (2\pi n(\mathbf{r}))^{1/3}$. In the near-HEG limit, relevant for the GGA design, the value of $\varepsilon_{\rm xc}^0$ also depends on the scaled form of the local density gradient [4, 7, 25],

$$s(\mathbf{r}) = \frac{|\nabla n(\mathbf{r})|}{[2n(\mathbf{r})k_F(\mathbf{r})]} \,. \tag{63}$$

For small periodic density variations $\delta n_{\mathbf{q}}(\mathbf{r}) = \delta n_{\mathbf{q}} e^{i\mathbf{q}\cdot\mathbf{r}}$, the length of the density gradient is itself proportional to $q = |\mathbf{q}|$, and the same applies to the value of the scaled gradient $s$, Eq. (63). Following Refs. 7, 47, 91, and 174 and using Eq. (55), we extract the perturbation limit

$$\varepsilon_{\rm xc}^0 - \varepsilon_{\rm xc}^{\rm LDA} \to -\varepsilon_{\rm x}^{\rm LDA} \frac{Z_{\rm xc}}{9}\, s^2 \,. \tag{64}$$

Here $Z_{\rm xc} = Z(q \to 0)$ is asserted from exchange-like $P^{\mu=\nu}(\omega)$ diagram contributions [7, 89, 91, 174]. Since $Z_{\rm xc}$ is given by the $\gamma(q \to 0)$ limit, vertex corrections also guide the GGA designs [91, 174].

For a motivation of the design choices made in the vdW-DF method, we will, below, consider a relation between the quasi-particle dynamics, expressed for spin $\nu$, and $\sum_\mu W_\lambda(\omega) P_\lambda^{\mu,\nu}(\omega)$. The effective interaction kernel is here

$$P_\lambda(\omega) = \sum_\mu P_\lambda^{\mu,\nu}(\omega) \tag{65}$$

and we suppress the spin dependence for simplicity in the discussion; In the standard class of spin-balanced problems, we have $P_\lambda(\omega) = \tilde{\chi}_\lambda(\omega)/2$.

### B. Screening and exponential re-summation

Exponential re-summation [86, 159, 160, 167] and canonical transformation are related solution strategies for the interacting electron gas problem. The central idea is that the many-particle interactions are effectively screened. For example, the use of a canonical transformation [159, 194–196] $H_\lambda \to \tilde{H}_\lambda = \hat{U} H_\lambda \hat{U}^\dagger$, can produce a significant cancellation of the leading many-particle interaction terms (in $\tilde{H}_\lambda$) with a suitable choice of the unitary operator $U$ [159]. The corresponding transformation of the many-particle wavefunction is often expressed

$$\tilde{\Psi}(\mathbf{r}_1, \ldots \mathbf{r}_N) = \exp(\tilde{J})\Psi(\mathbf{r}_1, \ldots \mathbf{r}_N) \,, \tag{66}$$

in terms of a so-called Jastrow factor [194, 195] $\tilde{J}$. This factor is chosen to partly reflect electron correlation and suppress $\tilde{\Psi}$ values whenever $|\mathbf{r}_i - \mathbf{r}_j|$ values are small [194]. The implied screening simplifies the numerical evaluation of the interaction effects, as used, for example, in Quantum Monte Carlo calculations [195, 196].

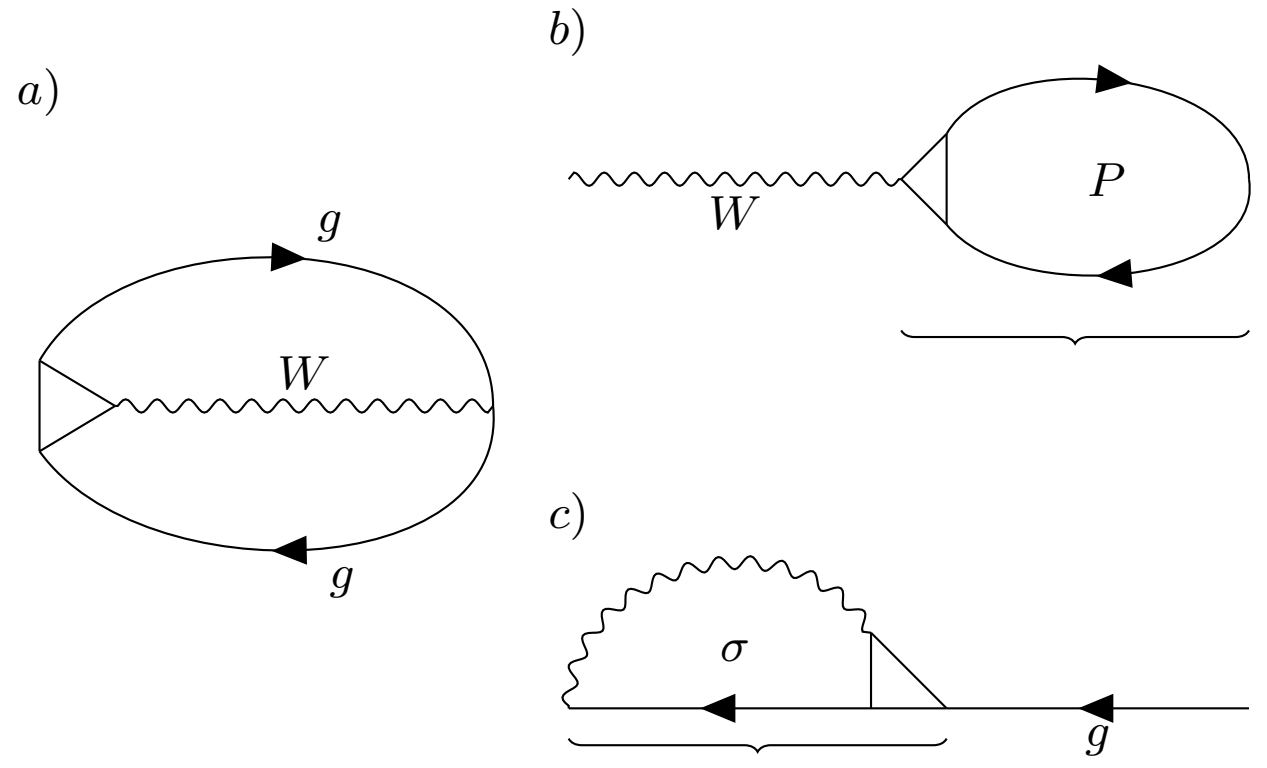

FIG. 4. Formal linked-cluster result for the total energy in the interacting electron gas (panel a). We suppress spin indices on electron Green functions $g$ and on the screened electron-electron interaction $W$; The triangle denotes the full vertex function. The same vertex function also enters the formal specifications of the spin-resolved local-field response function $P(\omega)$ and of the electron self-energy $\sigma(\omega)$. The cluster expansion result leads to a relation between the factor $W(\omega)P(\omega)$ (panel b) and the Dyson-correction factor $\sigma(\omega)g(\omega)$ (panel c). The former screens and thus defines the effective electron-electron interaction [188], while the latter guides a cumulant-expansion approximation for the quasi-particle dynamics [86, 163, 164].

To set the stage for our discussion of screening, we consider the Green function description of the dynamics of a quasi-particle

$$\begin{aligned} g_\lambda(\omega) &= g_0(\omega)\,[1 + \sigma_\lambda(\omega) g_\lambda(\omega)] \\ &\equiv g_0(\omega)\, \exp(J_\lambda(\omega)) \,, \end{aligned} \tag{67}$$

in the presence of an electron-electron interaction and screening. The Dyson correction factor, in the square brackets, is here cast as a cluster or cumulant formulation [86, 163, 164], defined by an exponential factor $J_\lambda(\omega)$. This factor is seen as analogous to the Jastrow factor, although in the present MBPT context. The Green function $g(\omega)$ reflects the single-electron excitations to vacuum [169], and the spatial variation defines corresponding orbitals of the quasi-particle dynamics.

Figure 3 shows related Feynman-diagram solutions for the so-called plasmaron model [85, 86, 163, 164], assuming recoil-less interactions. The dynamics of each quasi-particle level, or orbital, is described by an orbital-specific Green function $\mathcal{G}$. In a single-particle description it is described by $\mathcal{G}_0$, represented by arrows; The coupling to other quasi-particle levels is ignored. The wiggly lines represent approximations to the screened interactions, for example, given by the so-called plasmon propagator

$$S_\lambda(\omega) \equiv -V_\lambda \chi_\lambda(\omega) \,, \tag{68}$$

that tracks $W_\lambda(\omega) - V_\lambda$, Refs. 21, 22, 83, 85, and 86; The full Green function solution $\mathcal{G}$ then captures the characteristic screening (that affects given quasi-particle level) by the surrounding electron gas, as described by the ZPE dynamics of plasmons [85, 86, 163]. The same formal expansion can also be used to describe the dynamics of a level interacting with virtual vibrational excitations [164]. The top panel of Fig. 3 shows the form of a traditional Dyson expansion subject to assumption of recoil-less interactions, making it clear that vertex corrections plays an important role in setting the quasi-particle dynamics.

The bottom left panel of Fig. 3 shows the elegance of the cumulant approach for solving the time evolution $\mathcal{G}(t)$ of a specific orbital, again under the assumption of recoil-less interactions. This cumulant approach adapts the ideas of the linked-cluster expansion [167] to the description of the quasi-particle dynamics [86, 160, 161]. The point is that the linked-cluster expansion reduces to just a single connected diagram for which there exists a complete analytical solution [86, 163, 164].

Importantly, the cumulant expansion provides an automatic inclusion of all screening and vertex correction effects, relative to a stated approximation for the plasmon (or phonon) propagator. The underlying assumption of recoil-less interactions holds, for example, for a description of core levels [85, 86] but also for quasi-particle states near the Fermi level [163]. More generally, the exponential-re-summation idea can be used to extract potentially highly accurate solutions from MBPT, focusing on simple diagrams [86, 163, 164].

The bottom-left panel of Fig. 3 shows the cumulant-expansion idea as represented instead in the frequency domain [164], highlighting a formal similarity with the GW approximation [163, 188]. In frequency space, the expansion of the quasi-particle dynamics in the electron-plasmon coupling yields [164]:

$$\begin{aligned}\mathcal{G}_0(\omega)e^{J_\lambda(\omega)} &= \mathcal{G}_0(\omega)[1 + J_\lambda(\omega) + \ldots] \\ &= \mathcal{G}_0(\omega)[1 + \sigma^{(1)}_{\mathcal{G}W_\lambda}(\omega)\mathcal{G}^0(\omega) + \ldots]\,, \end{aligned} \quad (69)$$

$$\sigma^{(1)}_{\mathcal{G}W_\lambda}(\omega) = i\int_{\omega'} \mathcal{G}_0(\omega-\omega')W_\lambda(\omega')\,, \quad (70)$$

again under assumption of recoil-less interactions. Setting the exponential re-summation factor,

$$J_\lambda(\omega) = \sigma^{(1)}_{\mathcal{G}W_\lambda}(\omega)\mathcal{G}_0(\omega)\,, \quad (71)$$

allows us to capture vertex corrections implicitly. The use of this exponential re-summation factor allows a $\mathcal{G}W^{(1)}$-based account to be meaningful for metals, at least near the Fermi level [163, 164].

The vdW-DF method aims to capture the vdW forces without loosing the response and screening logic that underpins the GGA success. The electron-gas response provides a formal definition of the XC hole, Eq. (17), and, in turn, the ACF functional specification, Eq. (5). The screening, including vertex-correction effects, enters in the specifications of the local-field response [6, 7, 25, 89, 91, 174, 189] $\tilde{\chi}_\lambda(\omega)$ as well as in the Dyson-equation specification Eq. (47) of $\chi_\lambda(\omega)$, and hence in $\kappa_\lambda^{-1}(\omega)$.

The cumulant expansion [86, 163], analyzed in frequency space [164], is a guide for understanding the vdW-DF design logic. Screening in vdW-DF is described by constructing a scalar, but fully nonlocal, model for a local-field susceptibility $\alpha(\omega)$ that reflects a GGA-type response behavior.

Figure 4 shows, in panels a) through c), a compact formulation of the complete Feynman-diagram evaluation of the electron thermo-dynamical potential [167], the screening $WP$ of the effective interaction [188], and the quasi-particle dynamics [169], respectively. The triangle represents the general electron-electron interaction vertex form, denoted $\Gamma$, while arrows here depict the fully interacting electron Green functions $g$. The wiggly lines represent the screened interaction $W$. All internal coordinates (at interaction vertices) are integrated out.

Importantly, the linked-cluster result (panel a), also defines full specifications of both the interaction screening $W_\lambda(\omega)P_\lambda(\omega)$ (panel b) and of $\sigma_\lambda(\omega)g_\lambda(\omega) \approx J_\lambda(\omega)$ (panel c); They arise by pulling out $W$ and $g$, respectively. The common-origin argument motivates a mutual relation

$$\begin{aligned}\sigma_\lambda(\omega)g_\lambda(\omega) &\sim -W_\lambda(\omega)P_\lambda(\omega) \\ &= -V_\lambda\chi_\lambda(\omega)/2\,, \end{aligned} \quad (72)$$

where we focus on the quasi-particle dynamics for a given spin, and where the last line holds when $P_\lambda(\omega) = \tilde{\chi}_\lambda(\omega)/2$. The Hedin equations [188] prove Eq. (72) when expressed in a form with full frequency integration, Appendix A.

We see Eq. (72), and the usefulness [86, 163, 164] of the cumulant expansion idea, Eqs. (69)-(71), as a guide to construct $\tilde{\chi}_{\mathrm{ACF}}$ approximations, for use in the vdW-DF method. By relying on an exponential re-summation, we can build from a simple model local-field susceptibility $\alpha(\omega)$, as long as we also enforce the projection Eq. (32). A GGA-like internal functional is a good place to nucleate this response modeling [4, 7, 8, 10].

The vdW-DF design strategy for truly nonlocal functionals thus consists of a sequence of steps. First we use

$$\begin{aligned}\kappa_{\mathrm{int}} &= 1 - \tilde{\chi_{\mathrm{int}}}V \\ &\sim 1 + 4\pi\alpha(\omega) \equiv \epsilon(\omega)\,, \end{aligned} \quad (73)$$

to implicitly define an internal (or lower-level) dielectric-function approximation $\epsilon(\omega)$; This follows by partial integration since $\nabla^2 V = -4\pi$. Second, we use

$$\begin{aligned}\epsilon(\omega) &= \exp(S_{\mathrm{xc}}(\omega)) \\ &= 1 + S_{\mathrm{xc}}(\omega) + \ldots\,, \end{aligned} \quad (74)$$

to cast the associated scalar-model susceptibility $\alpha(\omega)$ – and hence $\tilde{\chi}_{\mathrm{ACF}}(\omega)$ – through an exponential re-summation,

$$4\pi\alpha(\omega) = S_{\mathrm{xc}}(\omega) + \sum_{j=2}^{\infty}\frac{1}{j!}(S_{\mathrm{xc}}(\omega))^j\,, \quad (75)$$

in an effective plasmon propagator $S_{\rm xc}$. As indicated by the subscript, this effective plasmon propagator is set so that it is consistent with the energy-per-particle variation of an internal semilocal XC functional [8, 10, 12], as summarized in Appendix B. Finally, we use Eq. (32) together with an effective Dyson equation (30) to complete the specification of the integrand $\kappa_{\rm ACF}(\omega) \approx \kappa_{\rm long}(\omega)$ in the ACF recast, Eq. (18). These steps describe a full implementation of the vdW-DF method; For the design of the popular general-geometry vdW-DF versions, the resulting description is also expanded in $S_{\rm xc}$, Refs. 4, 7, and 12.

We note that the implied combination of an exponential re-summation, Eq. (74) and a Dyson expansion, Eq. (30), captures higher-order and truly nonlocal correlation effects [1, 12, 65, 91]. This internal semilocal XC functional has no correlations beyond those reflected in LDA [4, 7]. In other words, all gradient-corrected and truly nonlocal-correlation effects must emerge from the third of the above-listed steps. Still, the summation in $S_{\rm xc}(\omega)$ provides the proper framework for capturing screening and vertex effects in $\alpha(\omega)$. Meanwhile, the combination of Eqs. (30) and (32) ensures a longitudinal projection in $\kappa^{-1}_{\rm ACF}(\omega) = 1 + V\chi_{\rm ACF}(\omega)$. This projection is, as discussed in Sec. II.B, sufficient to ensure that Eq. (18) also captures the asymptotic vdW interactions.

For a more detailed motivation of the vdW-DF design strategy, we consider an exponential re-summation for the Lindhard screening function

$$\begin{aligned} \kappa_\lambda^{-1}(\omega) &\equiv \exp(-F_\lambda(\omega)) && (76)\\ &= 1 + V_\lambda\chi_\lambda(\omega) = 1 - S_\lambda(\omega)\,. && (77) \end{aligned}$$

For a particular coupling constant $\lambda$, we have a direct specification of the screening

$$\begin{aligned} F_\lambda(\omega) &\equiv \ln\left(\kappa_\lambda(\omega)\right) \\ &= (1-\kappa_\lambda^{-1}) + \frac{1}{2}(1-\kappa_\lambda^{-1})^2 + \ldots \\ &= (-V_\lambda\chi_\lambda) + \frac{1}{2}(-V_\lambda\chi_\lambda)^2 + \ldots \\ &= S_\lambda(\omega) + \frac{1}{2}S_\lambda(\omega)^2 + \ldots\,. \end{aligned} \tag{78}$$

The re-summation can be expected to converge fast, $F_\lambda \approx S_\lambda$ as it describes the electrodynamics directly in terms of the screened response $\chi_\lambda(\omega)$. We also note that

$$\begin{aligned} \kappa_\lambda^{-1}(\omega) &= 1 + V_\lambda\chi_\lambda(\omega) \\ &= 1 + W_\lambda\tilde{\chi}_\lambda(\omega) \\ &\sim 1 - 2\sigma_\lambda(\omega)g_\lambda(\omega) \end{aligned} \tag{79}$$

follows from Eq. (72). Again, the factor of 2 arises from summation over spin contributions in the quasi-particle description. Equation (79) is consistent with a classic MBPT text-book demonstration [169] that the electron-electron interaction energy can be computed both from knowledge of the external-field response $\chi(\omega)$ and from the $\sigma(\omega)g(\omega)$ product.

Taking Eqs. (78) and (79) together suggests that we can set the exponential factor in the dielectric function

$$F_\lambda(\omega) \sim 2\sigma_\lambda(\omega)g_\lambda(\omega)\,. \tag{80}$$

This identification should be compared with the re-summation factor $J_\lambda(\omega)$ that characterizes the quasi-particle dynamics, Eq. (71), and the Jastrow factor $\tilde{J}$ in Eq. (66).

We observe that the vdW-DF reliance on exponential re-summations of dielectric functions, Eqs. (19), (76), and (80), provides the correct formal structure for capturing a range of important vertex effects in the electron-gas response. We recall that the cumulant formulation, Eq. (76), for $\kappa_\lambda^{-1}(\omega)$ was, in fact, originally set up to capture the range of vertex-correction and hence screening effects for a core electron interacting with the itinerant electron gas [86]. This was expressed through the quasi-particle dynamics in the plasmaron model [85, 86]. Hedin as well as Gunnarsson and co-workers noted and demonstrated that the model ability to capture the range of vertex corrections for recoil-less interactions has broader use. It works, for example, for descriptions of quasi-particles near the Fermi surface [163, 164]. Finally, we find that the Hedin equations, Appendix A, suggests the connection Eq. (72) between the quasi-particle dynamics and the cumulant or exponential re-summation, Eq. (76), for $\kappa_{\rm ACF}(\omega)$.

Of course, for the ACF specification Eqs. (18) and (19) we seek a $\lambda$-averaged evaluation of the response. By construction [10, 47] we have

$$-\int_0^1 d\lambda\, V\chi_\lambda(\omega) = \ln(\kappa_{\rm ACF}(\omega))\,, \tag{81}$$

while Eq. (78) gives

$$\begin{aligned} \int_0^1 \frac{d\lambda}{\lambda}\,\ln(\kappa_\lambda(\omega)) &\approx -\int_0^1 \frac{d\lambda}{\lambda}\,V_\lambda\chi_\lambda(\omega) \\ &= -\int_0^1 d\lambda\, V\chi_\lambda(\omega)\,. \end{aligned} \tag{82}$$

That is, we can see $\ln(\kappa_{\rm ACF}(\omega))$ as a mean-value evaluation given a characteristic exponent

$$F_{\rm ACF}(\omega) \equiv \ln(\kappa_{\rm ACF}(\omega)) \sim F_{\lambda_{\rm eff}(\omega)}\,. \tag{83}$$

We finally note that treating $\kappa_{\rm ACF}(\omega)$ like an actual Lindhard dielectric function $\kappa_{\lambda_{\rm eff}}(\omega)$ is fully consistent with the logic of the coupling-constant analysis of XC functionals [14, 48–51, 90, 197]. For example, based on the coupling-constant analysis of the consistent-exchange vdW-DF-cx version [14, 15], we find that the actual XC functional $E_{\rm xc}^{\rm DF}$ should be seen as a suitable average of the $\lambda = 0$ (all-exchange) limit and $\lambda \to 1$ (strongly correlated) limit [15]. This observation reflects the behavior in the associated XC holes [14]. The corresponding response description differs from that of the physical system but it is a still defined by a long-range particle interaction, $V_{\rm ACF} \sim V_{\lambda_{\rm eff}}$.

Overall, we are led to trust a lower-level internal response description for a characterization of $F_{\rm int}(\omega) \approx \ln(\kappa_{\rm int}(\omega)) \sim S_{\rm xc}(\omega)$, precisely because we rely on exponential re-summation. We note that a direct use of Eqs. (69)-(71) leads to an explicit specification,

$$F_{\lambda_{\rm eff}}(\omega) \sim 2\sigma^{(1)}_{GW_{\lambda_{\rm eff}}}(\omega) g_0(\omega)\,? \tag{84}$$

However, this would be an uncontrolled approximation because we do not know the extent that we have thus respected the longitudinal projection. We also do not know the value of $\lambda_{\rm eff}$. Instead, in the vdW-DF method, we seek to retain the elegance of the cumulant-expansion approach [86, 163, 164] with an implicit construction of $\alpha$ from $S_{\rm xc}$. We do that by setting $S_{\rm xc} = \ln(\epsilon)$, where $\epsilon \approx \kappa_{\rm int}$ denotes an approximation set by an internal functional.

### C. Electrodynamics nature of XC functionals

We make 3 observations to further motivate and detail the logic of the vdW-DF method and the formulation of the consistent vdW-DF versions. The exact vdW-DF framework contains the constraint-based semilocal functionals as a limit [4, 7, 10, 12]. This makes it possible to also discuss the connection between the semilocal and the truly nonlocal functionals in some detail.

*First,* the identification of the ACF recast, Eqs. (18) and (19), with the screening at some effective coupling constant ($0 < \lambda_{\rm eff} < 1$), Eq. (83), implies that we should leverage insight from the theory of light-matter interactions and screening [63] in the vdW-DF designs.

It is, for example, interesting to use Eq. (79) to discuss optical excitation of a quasi-particle orbital of energy $\epsilon_r$. We assume that we are at a frequency $\omega$ far from collective excitations and that there is only a limited amount of actual transitions, meaning $\Im\sigma(\omega) \to 0$. The imaginary part of the right-hand side of Eq. (79) is then, for $\omega \approx \epsilon_r$, set by the spectral density, i.e., set by the relevant inelastic excitation from $\epsilon_r$ up to the vacuum level [169]. Meanwhile, the imaginary part of the left-hand side of Eq. (79) is given by $\Im\chi(\omega)$, and effectively set as an evaluation of Fermi's golden rule for inelastic transitions [63]. The transition rate must, at $\omega \approx \epsilon_r$, have significant contributions from such optical excitations to the vacuum level. It is gratifying that the real part of the model dielectric function $\kappa_\lambda(\omega \approx \epsilon_r)$ define the screening of such excitations.

Moreover, since we have build both $\tilde{\chi}_{\rm ACF}(\omega)$ and $g_{\lambda_{\rm eff}}(\omega)$ from exponential re-summations, we expect the two expansions to be related. We note that the cumulant approach for the quasi-particle dynamics, given by Eq. (72), has accuracy near the Fermi level (and for localized orbitals). This suggest that our exponential resummation for $\epsilon(\omega)$ has the formal structure to be accurate for screening at corresponding frequencies.

For a practical design framework, we set the exponential re-summation [8, 12, 47]

$$\epsilon(\omega) = e^{S_{\rm xc}(\omega)}\,. \tag{85}$$

from an internal semi-local XC functional

$$\begin{aligned} E^{\rm in}_{\rm xc} &= \int_0^\infty \frac{du}{2\pi}\,{\rm Tr}\{\ln(\epsilon(iu))\} - E_{\rm self} \\ &= \int_0^\infty \frac{du}{2\pi}\,{\rm Tr}\{S_{\rm xc}(iu)\} - E_{\rm self}\,. \end{aligned} \tag{86}$$

The formal structure is the same as that used for a direct design of GGA functionals, Eq. (89), as discussed below. This allow us to use Eq. (86) to set the details of $S_{\rm xc}(\omega)$ approximation, as summarized in Appendix B and elsewhere [4, 7, 9, 12].

*Second,* approximations for the ACF electrodynamics and for $\kappa_{\rm ACF}(\omega)$ should be build in compliance with constraints and with the Lindhard screening logic [63].

We should seek approximations that adhere to the Dyson equation [61, 62], Eq. (30) for the effective ACF electrodynamics response. This follows because we should think of the ACF recast as reflecting an actual electrodynamics, given by a long-ranged interaction $V_{\rm ACF} \sim V_{\lambda_{\rm eff}}$.

Also, approximation for this ACF electrodynamics must be made subject to a current-conservation criterion [10], that is, $\kappa_{\rm ACF}(\omega) \approx \kappa_{\rm long}(\omega)$. We are constructing a ground-state XC functional that reflects an electrodynamics coupling of the ZPE dynamics of XC holes, Eq. (36). However, zero-point vibrations, like plasmons, involve electron currents and we should view vdW-inclusive XC functionals (for ground state DFT) as a limit [198, 199] of time-dependent DFT [200] or of current-density functional theory [201, 202].

The longitudinal projection, Eq. (33), is hardwired into the resulting vdW-DF functional description of nonlocal-correlation effects, Eq. (2), as explained below. As detailed in Sec. IV, the consistent-exchange vdW-DF-cx version seeks to enforce both the Dyson criterion, Eq. (30), and the longitudinal projection in the full nonlocal-functional specification.

*Third,* the physics of the ACF electrodynamics and dielectric function $\kappa_{\rm ACF}(\omega) \approx \kappa_{\rm long}(\omega)$ is at the same time partly known and incompletely explored. Equations (21) and (22) provide a direct link between the formally exact XC hole, Eq. (17), and $\kappa_{\rm ACF}(\omega)$. On the one hand, we know a lot about this ACF dielectric function when it is used to describe semilocal functionals via Eq. (22). On the other hand, the Mahan and Aschroft analysis of the nature of vdW forces [64, 65, 158], as well as the Anderson-Langreth-Lundqvist-Dobson launching of the vdW-DFs [1, 2, 166, 203–205], suggest that we must seek truly nonlocal formulations of $n_{\rm xc}$ and $\kappa_{\rm ACF}(\omega)$.

In the vdW-DF method we seek to port the GGA experience in the former to enhance the quality of the latter.

### D. Nonlocal and semilocal XC functionals

Interestingly, our analysis of the screening nature of functionals leaves us with two suggestions for actual designs, a direct and an indirect formulation. They differ in how we connect the physics content of approximations like Eq. (34) to the underlying formally exact ($\lambda$-averaged) response description contained in Eq. (1).

*The direct approach* is using Eq. (77) to just recoup the standard ACF

$$E_{\rm xc} \equiv \int_0^\infty \frac{du}{2\pi}\,{\rm Tr}\{\langle(1-\kappa_\lambda^{-1}(iu))\rangle_\lambda\} - E_{\rm self}$$
$$= \int_0^\infty \frac{du}{2\pi}\,{\rm Tr}\{\langle S_\lambda(iu)\rangle_\lambda\} - E_{\rm self}\,, \qquad (87)$$

where we have used

$$\langle A_\lambda\rangle_\lambda \equiv \int_0^1 \frac{d\lambda}{\lambda}\,A_\lambda\,. \qquad (88)$$

to define the $\lambda$ averaging.

The direct-design approach can also formally be seen as a result of both truncating the exponential re-summation Eq. (19) and the logarithm in Eq. (18) to lowest relevant order in $V\chi_\lambda(\omega)$. Since Eq. (87) is, in principle, exact by itself, it is clear that there are important cancellations in ACF recasting and in the exponential summation. Use of a direct approach (relying directly on the original ACF formulation), Eq. (87) is a highly motivated design strategy, as long as we are actually able to directly assert a good, robust approximation for $\langle(1-\kappa_\lambda^{-1}(iu))\rangle_\lambda = -\int_0^1 d\lambda V\chi_\lambda(iu)$.

In practice, the direct design strategy rests on making systematic approximations that are based on formal MBPT for the HEG and for the weakly-perturbed electron gas [4, 7, 10, 11, 13–15, 21, 23, 25, 26, 28, 29, 48, 50, 64, 65, 89, 91, 168, 174]. The electron response is dominated by plasmons, i.e., collective excitations defined as zeros of the dielectric function of the system. We thus expect the semilocal (GGA) functional descriptions to be fairly approximated by a simple model for a plasmon propagator $S'_{\rm xc}$. We can, for example, use the formulation inspired by a formal MBPT gradient expansion [47, 87],

$$E_{\rm xc} \approx E^0_{\rm xc} \equiv \int_0^\infty \frac{du}{2\pi}\,{\rm Tr}\{S'_{\rm xc}(iu)\} - E_{\rm self}\,, \qquad (89)$$

Appendix B. The specification (89) is equivalent to LDA/GGA formulations given the semilocal XC hole

$$n^0_{\rm xc}(\mathbf{r};\mathbf{r}'-\mathbf{r}) = -\delta(\mathbf{r}-\mathbf{r}')$$
$$-\frac{1}{2\pi n(\mathbf{r})}\quad\int_0^\infty \frac{du}{2\pi}\nabla^2_{\mathbf{r}}\langle\mathbf{r}|S'_{\rm xc}(iu))|\mathbf{r}'\rangle\,. \qquad (90)$$

It is a problem that use of even a double-pole plasmon approximation for $S'_{\rm xc}$ leads to a GGA-type functional in the direct-design approach. As summarized in Appendix B and in Refs. 4, 7, 10, and 12, the local energy-per-particle variation $\varepsilon^{\rm in}_{\rm xc}(\mathbf{r})$ of the semi-local internal function $E^{\rm in}_{\rm xc}$ specifies the local plasmon dispersion [4, 7, 25], and invariably reduces the XC functional specification Eq. (90) to that of a semilocal form. The semilocal XC functional form has no clear mechanism for reflecting truly nonlocal correlation effects, including pure vdW attraction [64, 65, 91], Sec. II.

*The indirect, or vdW-DF, approach,* is to instead exploit Eqs. (18) and (19) for higher-order expansions in the plasmon propagator, building from Eq. (85). The exponential re-summation $\epsilon(\omega) = \exp(S_{\rm xc}(\omega))$ is, by itself, exclusively used to reflect a GGA behavior minus pure vdW forces.

However, the indirect design provides a balanced account of general interactions by also explicitly enforcing a longitudinal projection [4, 10], That is, in the vdW-DF method, we use Eq. (34), with $\epsilon(\omega)$ inserted for $\boldsymbol{\epsilon}(\omega)$, to define the XC functional

$$E_{\rm xc} \approx E^{\rm DF}_{\rm xc} \equiv \int_0^\infty \frac{du}{2\pi}\,{\rm Tr}\{\ln(\kappa_{\rm long}(iu))\} - E_{\rm self}\,. \quad (91)$$

The implicit ACF approximation, $\kappa_{\rm ACF} \approx \kappa_{\rm long}$, secures a strict enforcement of current conservation in the description of the electrodynamics response [10].

The vdW-DF method is, to our knowledge, the first example of this indirect (Dyson-based) functional design approach. There is seamless integration with the underlying GGA formulation, given by $\epsilon(\omega) = \exp(S_{\rm xc}(\omega))$, as long as we properly balance exchange and correlation terms, as discussed in Sec. IV.

To define specific versions in this vdW-DF method we assert the local field response, given by $\tilde\chi_{\rm ACF}(\omega)V$, from a trusted MBPT analysis [4, 7, 91]. It is sufficient to formulate the input in terms of an effective scalar local-field susceptibility, given by $\boldsymbol{\alpha} = \alpha\mathbf{I}$ and Eq. (75), where $\mathbf{I}$ is the unit matrix in spatial coordinates; The associated (internal-functional) approximation for the scalar dielectric function is given $\boldsymbol{\epsilon} = \epsilon\mathbf{I}$ with $\epsilon(\omega) = 1 + 4\pi\alpha(\omega)$. We use the MBPT analysis to set the gradient-corrected exchange of the internal semilocal XC functional, Eq. (86), and thus the details of the approximation for the vdW-DF plasmon propagator $S_{\rm xc}(\omega)$, Appendix B. The description of $\tilde\chi_{\rm ACF}(\omega)V$ then follows by the exponential re-summation, Eq. (75).

The lowest-order contribution in Eq. (94) is just the internal functional itself. The vdW-DF method has a leading term given as

$$E^{\rm DF}_{{\rm xc},n=1} = \int_0^\infty \frac{du}{2\pi}\,{\rm Tr}\{\nabla S_{\rm xc}(iu)\cdot\nabla G\} - E_{\rm self}$$
$$= E^{\rm in}_{\rm xc}\,, \qquad (92)$$

where $G = -V/4\pi$. The last line follows since $\nabla^2 G = 1$.

The vdW-DF method yields a nonlocal-correlation energy approximation [2–4],

$$E^{\rm nl}_{\rm c} = \int \frac{du}{2\pi}\,{\rm Tr}\{\ln(\nabla\epsilon(iu)\cdot\nabla G) - \ln(\epsilon(iu))\}\,, \qquad (93)$$

that is set entirely by the density variation [1, 6, 32, 53, 59, 165, 180, 205–211] and hence by occupied orbitals [31].

The vdW-DF version specifications [4, 8, 9, 11, 12] are given by Eqs. (91), (34), (86) and (93). They are, in principle, strictly nonempirical [4, 10, 12, 30, 45] since the evaluation of Eq. (93) is defined by a systematic expansion in a known approximation for the plasmon propagator $S_{\rm xc}$. In practice, the resulting functional specification, Eq. (3), normally includes a cross-over term, $\delta E_{\rm x}^0$.

Moreover, in the popular general-geometry vdW-DF versions [4, 5, 7–9, 12, 40–43, 212] the formal nonlocal-correlation energy specification Eq. (93) is expanded to second order in $S_{\rm xc}$, using $G = -V/4\pi$ and

$$\chi_{\rm ACF} V \approx -\nabla \left[ S_{\rm xc} + \frac{1}{2}\, S_{\rm xc}^2 \right] \cdot \nabla G + (\nabla S_{\rm xc} \cdot \nabla G)^2 \,. \quad (94)$$

The two terms grouped in the square bracket in Eq. (94) are part of an expansion for $\tilde{\chi}_{\rm ACF}$, while the third term is the lowest-order contribution to the screening implied in Eq. (30). The combination of the second and third terms in Eq. (92) defines Eq. (2). However, we may also retain more or all steps in the expansion, Eq. (94), as done in the immediate precursor, namely the layered-geometry vdW-DF0 version [2, 3, 6].

Importantly, when expanding to order $n = 2$ (or more), we arrive at a truly nonlocal form for the description of nonlocal correlations. The vdW-DF method comes with a universal-kernel evaluation [4, 5] of the nonlocal-correlation term

$$E_{\rm c}^{\rm nl} = \frac{1}{2} \int_{\mathbf{r}} \int_{\mathbf{r}'} n(\mathbf{r})\, \phi(n(\mathbf{r}), n(\mathbf{r}'), s(\mathbf{r}), s(\mathbf{r}'))\, n(\mathbf{r}') \,. \quad (95)$$

The universal-kernel evaluation, the definition of associated effective potential, and its use in efficient DFT calculations are detailed elsewhere [7, 12, 17, 18, 47, 60, 171].

Also, the nonlocal-correlation evaluation Eq. (95) has seamless integration [4, 5] with LDA and it is consistent with Eq. (55). This follows both because it nominally involves a double spatial integral over the density distribution and because Eq. (2) is easily shown to vanish in the HEG limit [4]. At the same time, neither Eq. (95) nor Eq. (55) should be seen as just a two-density-point summation, for example, as discussed in Ref. 10. It captures the mutual electrodynamics coupling of two electron-XC-hole systems (that are centered at positions $\mathbf{r}$ and $\mathbf{r}'$).

It should be noted that the input $\boldsymbol{\epsilon}(\omega)$ to any such vdW-DF version or variant, reflects the screening that exists in the GGA dielectric function $\epsilon(\omega) = \exp(S_{\rm xc}(\omega))$. As such, the vdW-DF method provides a systematic extension that captures vdW forces as already screened by, for example, itinerant electrons [10, 64, 65].

Finally, we mention that care must be taken when building the full vdW-DF functional designs from an understanding of screening contained in a GGA functional. We nucleate the description around an internal semi-local functional $E_{\rm xc}^{\rm in}$. Appendix B documents that it is therefore sufficient to approximate $\ln(\epsilon(\omega)) = S_{\rm xc}(\omega)$ by a double-plasmon-pole approximation $S_{\rm xc}(\omega)$ that reflects the logic of a gradient expansion [4, 47, 87]. However, the nonlocal-correlation effects cannot be limited to the vdW binding mechanism discussed in Sec. II, Refs. 10, 64, and 65. The second term of Eq. (36) must also reflect the exponential re-summation implied in asserting, $S_{\rm xc}(\omega) = \ln(\epsilon)$.

The second term $E_{{\rm xc},\alpha}$ in Eq. (41) reflects the poles of $\alpha(\omega) = (\epsilon(\omega) - 1)/4\pi$. We keep local XC effects together in $E_{\rm xc}^{\rm in}$, and hence in $S_{\rm xc}(\omega)$, so clearly $E_{{\rm xc},\alpha} \neq E_{\boldsymbol{\alpha}}^{\rm res,HF} - E_{\rm self}$. By the cumulant expansion [86, 163, 164], the electron dynamics (in $g_0(\omega)^{-1} g_{\lambda_{\rm eff}}(\omega)$) also reflects terms that are quadratic in $S_{\rm xc}$ and thus truly nonlocal (as we detail in Appendices B through D). If the inner functional $E_{\rm xc}^{\rm in}$ already reflected gradient-corrected correlation, we could be double counting.

The vdW-DF solution strategy is that of the Occam's razor, that is, to simply set the semilocal inner functional as LDA plus gradient-corrected exchange. The idea is to let the vertex correction effects in GGA correlation emerge in $E_{{\rm xc},\alpha}$, rather than dealing with a potential double counting. However, this Occam solution strategy is only partly motivated by the analysis in Refs. 86, 163, and 164. Important semilocal-correlation effects, for example, reflecting vertex corrections might be missing and we must validate that vdW-DF designs also works for traditional dense-matter challenges, for example, Refs. 45, 95, and 191.

## IV. SCREENING EFFECTS IN CONSISTENT VDW-DF VERSIONS

This section summarizes the rationale for the consistent-exchange vdW-DF-cx version. It does so in terms of illustrations of accuracy in computing the density variation and based on formal MBPT arguments. The section furthermore presents a detailed analysis of the underlying electrodynamics description. This section therefore discusses a different splitting of vdW-DF-cx terms than what is presented in Eq. (3). Appendix C explains how the electrodynamics nature of the vdW-DF method suggests an ordering of general nonlocal-correlation-energy terms, going beyond the truncated expansion of the vdW-DF nonlocal-correlation energy.

Finally, to make it possible to compute and thus explore the electrodynamics analysis in practice, this section presents universal-kernel calculations of the individual pure-vdW and cumulant components. They describe the functional contributions to quadratic order in $S_{\rm xc}$ and Appendix D provides details of the universal-kernel derivation and evaluation.

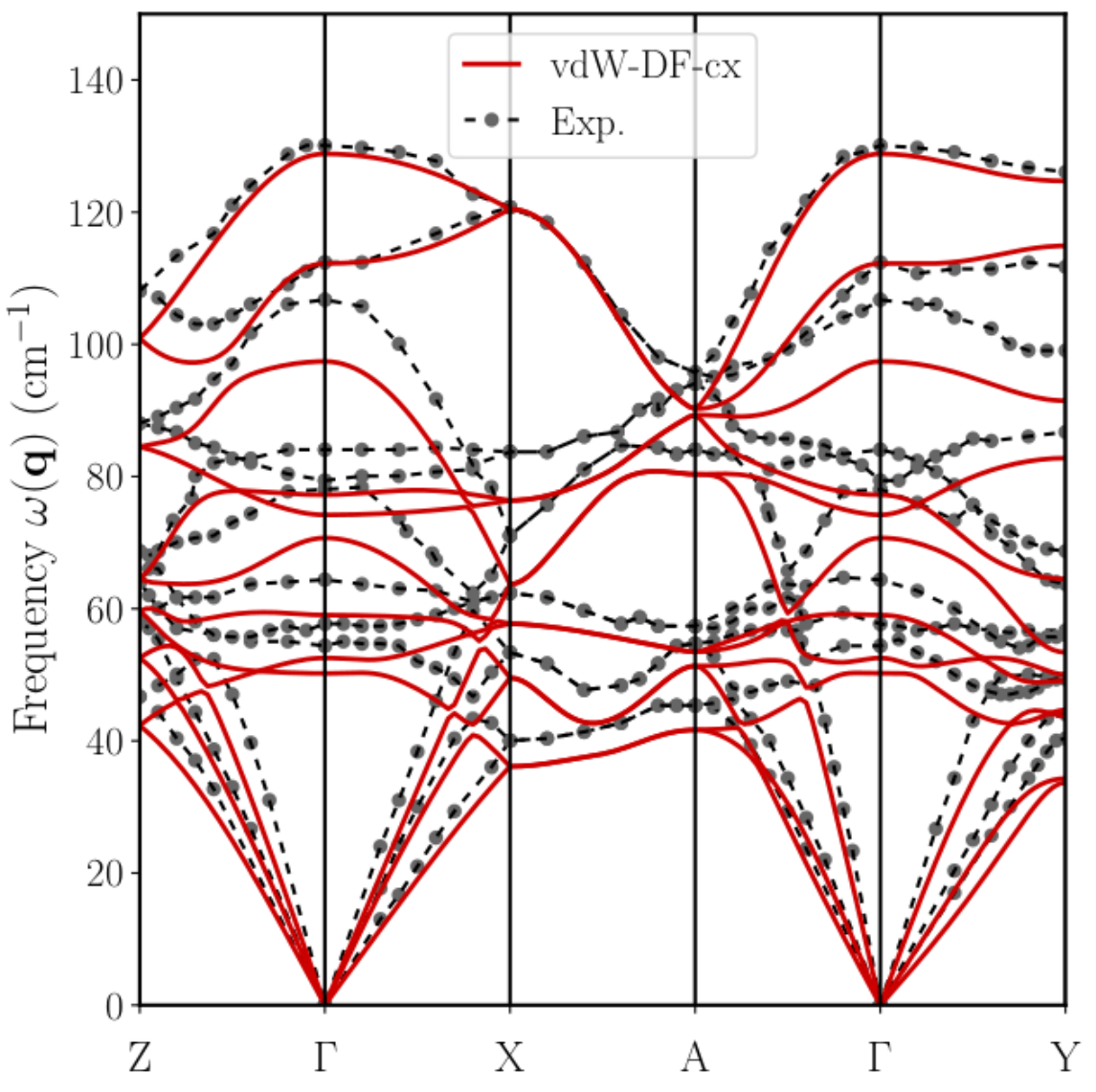

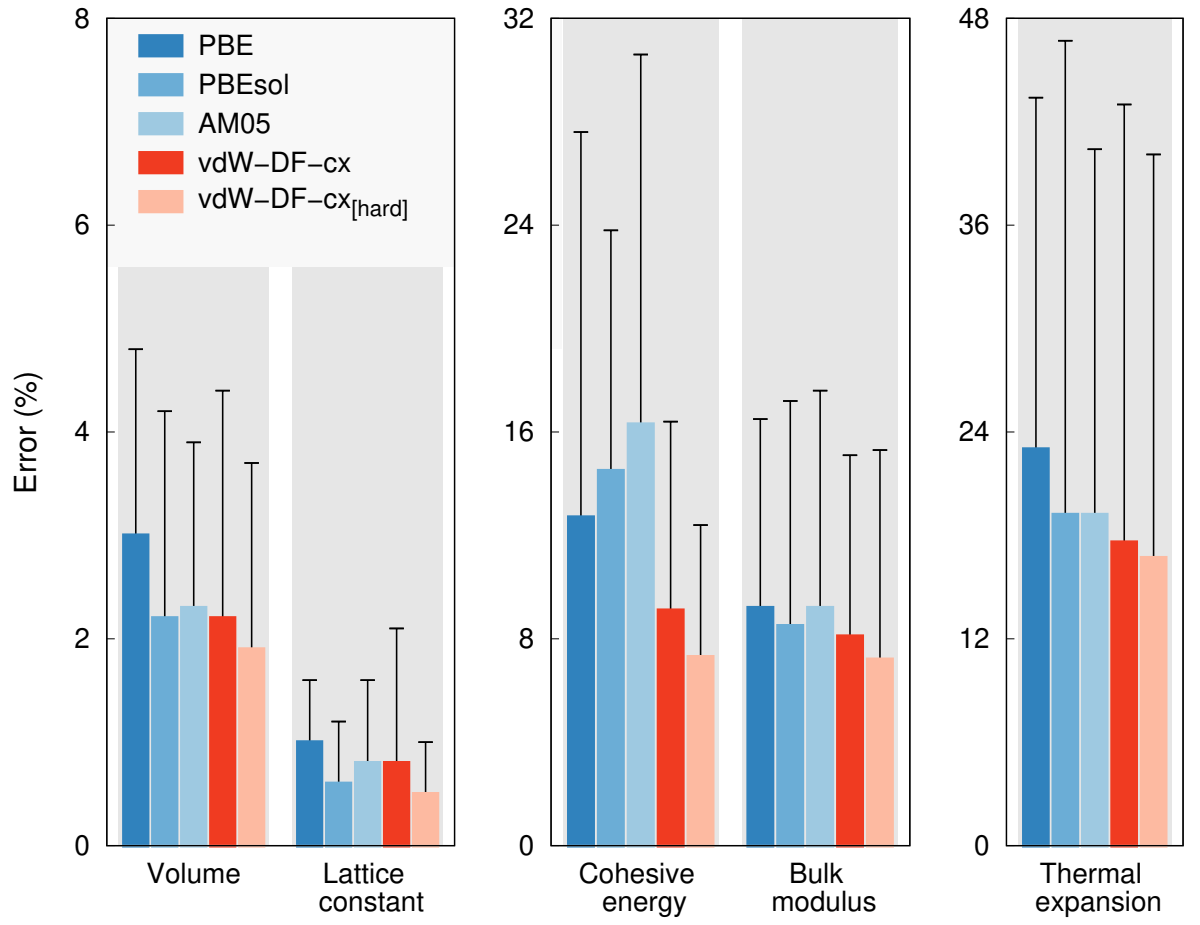

FIG. 5. *Top panel:* Phonon dispersion for the set of intermolecular (libration and rocking) modes in naphthalene, contrasting neutron measurements [213] (black) and vdW-DF-cx calculations (red) obtained with full stress and atomic-position relaxations [99]. The phonons are computed at the vdW-DF-cx result for the lattice constants and structure angles. *Bottom panel:* Comparison of vdW-DF-cx and PBE, PBEsol, and AM05 [214] performance for nonmagnetic transition metals. The panel summarizes calculations in ground-state DFT but corrected using the quasi-harmonic approximation [215] that is evaluated from full phonon calculations [216] in each of the XC functionals. These calculations were performed in the plane-wave code VASP [217, 218] using normal or hard PAW setups (as noted by the legend). The atomic reference energies are, however, corrected by a study of the atomic spin polarization, as obtained using the proper spin vdW-DF-cx implementation that is available in Quantum Espresso [11, 36, 95, 219]. The panel compares, among the functionals, the percentage errors (bars) as well as standard deviation (vertical lines) for the set of thermo-physical properties in a comparison with room-temperature measurements. The panels are adapted with permission from Refs. 99 and 95, respectively. Copyright (2016 and 2017) by the American Physical Society.

## A. Look how good the densities are: molecular libration modes and hard-matter thermo-physics

The key DFT concept is the electron density variation. There are concerns that while many of the recent XC density functional developments may improve the description of energies, they tend to destroy the description of the electron densities [220]. We share the opinion that the real test of a functional usefulness and robustness lies in its ability to accurately predict the electron density variation in general systems. Of course, it is also relevant to test the accuracy in descriptions of reaction and process energies, for example, as in Fig. 1. However, to be useful as a general-purpose tool [45], DFT with the vdW-DF-cx [9] approximation must be accurate on the electron density from soft molecular matter [98, 99], where there are important sparse regions [32], and to hard traditional matter [45, 95], where the electron distribution is dense.

Good discriminators for quality in the density description are highly accurate predictions of atomic structure, of atomic vibrations and phonons, as well as of surface-specific work functions. The relevance of the first discriminator follows from the Hohenberg-Kohn theorem [221]: the structure-optimized nuclei positions $\mathbf{R}_i$ must uniquely reflect the ground-state density variation $n(\mathbf{r})$. The relevance of the third follows since the work function reflects the variation (across the surface) in the full electrostatic potential $\Phi(\mathbf{r})$ that acts on the electrons; The work functions and work-function shifts (computed for specific surfaces) is an established indicator of quality of XC functionals, for example, as discussed and used in Refs. 25, 89, 222–226.

The relevance of finding accurate vibrations follows since we work in the standard Born-Oppenheimer or adiabatic formulation of DFT. We do not treat the nuclei dynamics directly, since the problem only depends parametrically on the nuclei positions, denoted $\mathbf{R}_i$. Instead we compute DFT results for another (related) electrostatic potential $\Phi_{\mathbf{R}_i}(\mathbf{r})$, that influences a given nuclei $i$. It contains the Hartree potential from the electron density as well as the electrostatic-potential contributions that are produced at the rest of the nuclear positions, $\mathbf{R}_{j\neq i}$. For an infinitesimal displacement, the restoring force on a nuclei $i$ must be given exclusively by $\Phi_{\mathbf{R}_i}(\mathbf{r})$, as discussed, for example, in Ref. 7. These restoring forces are a direct reflection of the electron-density variation, $n(\mathbf{r})$, and they define the phonons, and vibrations in general [227].

Figure 5 highlights the vdW-DF-cx performance on density quality. The figure summarizes two sets of vdW-DF-cx studies of the oligoacenes molecular crystals and of thermo-physical properties in the full set of nonmagnetic transition metals [95, 98, 99]. Computational details are given in the cited papers. Both panels assess the density-quality performance in term of the first (structure) and second (vibration) discriminators. We shall return to a discussion of a quality assessment of vdW-DF-cx in terms of the surface work function discriminator in Sec. VI.

The top panel documents that the vdW-DF-cx results for the weak libration and rocking modes of the naphthalene crystals are in exceptional agreement with low-temperature measurements [99, 213]. It is important to note that this vdW-DF-cx phonon prediction is computed free of all experimental input, on structure or otherwise. The phonon dispersion is computed for the lattice constants and structure angles that results with full stress and atomic relaxations in vdW-DF-cx. The vdW-DF-cx structure predictions are, in turn, in close agreement with low-temperature measurements across the set of oligoacene structures [98]. The vdW-DF-cx is found able to predict the molecular structure and vibrational response ahead of synthesis.

We also demonstrated that vdW-DF-cx structure predictions are accurate enough to, in principle, motivate (computationally costly) GW and Bethe-Salpeter calculations of the molecular-crystal optical response ahead of experimental input [98]. The ability to genuinely predict, for example, solar-cell function before actual synthesis has potential for accelerating, for example, molecular-based green-technology developments. Here, we take the set of oligoacene results [98, 99] as a validation that consistent vdW-DF-cx functional is accurate on the electron-density variation in sparse, molecular matter.

The bottom panel of Fig. 5 illustrates what vdW-DF-cx can do for characterizing transition-metal structure and thermo-physical properties at room temperature. As discussed in Refs. 95, 113, and 121, the vdW-DF-cx not only works well at describing the structure at a Born-Oppenheimer level, in raw DFT, but it has a robust description of the phonons that emerge with variations of both cubic and hcp lattices, for bulk and layered structures. At least for some of the cubic nonmagnetic-transitions metal cases, we could have just compared our raw-DFT vdW-DF-cx structure characterization with measurements of structure and cohesive energy that are also back-corrected for ZPE and thermal expansion effects [228–230]. However, the back correction introduces a further dependence on experimental input. Instead we chose to combine the robust phonon description of vdW-DF-cx (and of highly successful GGA descriptions: PBE [28], PBEsol [29], and AM05 [214]) with the quasi-harmonic approximation [215, 216] for a full (or native) DFT determination of ZPE and thermal-expansion effects as a function of temperature [95].

The panel shows a comparison of performance for the set of thermo-physical properties: volume, main lattice constant (often denoted $a$, extracted for both cubic and hcc lattice structures), cohesive energy, bulk modulus and thermal-expansion coefficient. These are computed at room temperature and can be compared directly with room-temperature measurements. Also, since the data set of nonmagnetic transition metal is sufficiently large, we can extract meaningful statistical averages of deviations (denoted 'Errors' in Ref. 95 and in the panel) for this important class of systems with a dense electron distribution. In the vdW-DF-cx case, we provide this hard-matter benchmark not only for the normal PAW set up of VASP [217, 218] but also (as noted by a subscript) with the 'hard' PAW setup to test convergence of the $E_{\mathrm{c}}^{\mathrm{nl}}$ evaluation [18]. Also, for the study of cohesive energies, we relied on the Quantum Espresso implementation of spin vdW-DF-cx [11] to describe the atom reference energy, finding a systematic improvement [95].

The bottom panel Fig. 5 suggests that vdW-DF-cx is highly accurate on structure and vibrations and therefore on the electron-density description in dense matter. The transition-metal benchmarking result supplements other findings of good vdW-DF-cx performance on structure (and vibrations). Besides low-temperature comparisons with oligoacene and polyethylene measurements [98, 99, 103, 107], we mention that the vdW-DF-cx is found accurate on describing polarization of $PbTiO_3$ and vdW-bound layered materials [9, 45, 68, 113, 121, 230].

### B. Charge and current conservation in vdW-DF

We believe that the vdW-DF-cx robustness advantages follow from the emphasis on conservation laws in the full vdW-DF method. We explicitly enforce current conservation in the response descriptions, Eq. (34), and this also ensures an automatic compliance with charge conservation for the vdW-DF XC hole [10].

Adapting Eq. (20), the full vdW-DF XC hole is

$$n_{\mathrm{xc}}^{\mathrm{DF}}(\mathbf{r};\mathbf{r}'-\mathbf{r}) = -\delta(\mathbf{r}-\mathbf{r}') - \frac{1}{2\pi n(\mathbf{r})} \int_0^\infty \frac{du}{2\pi} \nabla_{\mathbf{r}}^2 \langle \mathbf{r}| \ln(\kappa_{\mathrm{long}}(iu))|\mathbf{r}'\rangle \,. \quad (96)$$

The Fourier transform of this XC hole is

$$n_{\mathrm{xc}}^{\mathrm{DF}}(\mathbf{r};\mathbf{q}') = \int_{\mathbf{u}} e^{i\mathbf{q}'\cdot\mathbf{u}}\, n_{\mathrm{xc}}^{\mathrm{DF}}(\mathbf{r};\mathbf{u}=\mathbf{r}'-\mathbf{r}) \,. \quad (97)$$

Overall XC hole conservation, Eq. (4), mandates that $n_{\mathrm{xc}}^{\mathrm{DF}}(\mathbf{r};\mathbf{q}'\to 0) = -1$. This is equivalent to demanding $\Delta n_{\mathrm{xc}}^{\mathrm{DF}}(\mathbf{r};\mathbf{q}'\to 0) = 0$ where $\Delta n_{\mathrm{xc}}^{\mathrm{DF}}$ is used to identify all XC hole contributions that originate from the $\ln(\kappa_{\mathrm{long}})$ part of the vdW-DF specification.

For our discussion it is convenient to explore the formal connections to a characterization of light-matter interactions [47]. Because we use a scalar representation of the model susceptibility, the current-conservation criterion, Eq. (32), reduces to

$$\tilde{\chi}_{\mathrm{ACF}}(\omega)V = \nabla\alpha(\omega)\cdot\nabla V \,. \quad (98)$$

However, the external-field susceptibility, given by Eq. (31), remains a tensor given by the analysis behind Eq. (28). To make the identification we simply expand $\boldsymbol{\alpha}_{\mathrm{ext}}$ in terms of $\tilde{\chi}$ (and thus $\alpha$), sorting the contributions according to the number $m$ of times that $\tilde{\chi}$ enters in the associated Dyson equation (30):

$$\boldsymbol{\alpha}_{\mathrm{ext}} = \sum_{m=1}^{\infty} \boldsymbol{\alpha}_m \,. \quad (99)$$

We note that

$$\chi_{\rm ACF} V = \sum_{m=1}^{\infty} (\nabla \boldsymbol{\alpha} \cdot \nabla V)^m \,, \tag{100}$$

and it follows that $\boldsymbol{\alpha}_1 = \alpha \mathbf{I}$ while

$$\boldsymbol{\alpha}_{m\geq 2} = \alpha(\nabla V)(\nabla\alpha \cdot \nabla V)^{m-2}(\nabla\alpha) \tag{101}$$

are matrix contributions defined by the outer product of the first $\nabla V$ and the last $\nabla\alpha$.

The evaluation of the vdW-DF method relies on expanding every instance of $\alpha(iu)$ (entering in Eq. (99)), in terms of $S_{\rm xc}$ to some order given $j_k$, using Eq. (75). The implied effects on the optical-response description can be sorted

$$\boldsymbol{\alpha}_{\rm ext} = \sum_{n=1}^{\infty} \sum_{m\geq n}^{\infty} \sum_i \boldsymbol{\alpha}^{(i)}_{m,n} \,, \tag{102}$$

according to both $m$ and to the total number $n = j_1 + j_2 + \ldots + j_m$ of $S_{\rm xc}$ factors. An index $i$ is used to keep track of the different ways that a multiple expansion complies with the $m$ and $n$'s. Those details are not written out here as they have no relevance for the present discussion.

Each of the susceptibility terms of Eq. (102) corresponds to the XC hole contribution

$$\Delta n^{{\rm DF},(i)}_{{\rm xc},m,n} = \nabla^2_{\mathbf{r}'} \int_{\mathbf{r}''} \nabla_{\mathbf{r}} \cdot \mathbf{A}^{(i)}_{m,n}(\mathbf{r},\mathbf{r}'') \cdot \nabla_{\mathbf{r}''} V(\mathbf{r}'' - \mathbf{r}') \,, \tag{103}$$

where

$$\mathbf{A}^{(i)}_{m,n}(\mathbf{r},\mathbf{r}'') = \frac{1}{2\pi n(\mathbf{r}) m} \int_0^{\infty} \frac{du}{2\pi} \, \boldsymbol{\alpha}^{(i)}_{m,n}(\mathbf{r},\mathbf{r}''; iu) \,. \tag{104}$$

Evaluating the Laplacian and performing a partial integration yields

$$\Delta n^{{\rm DF},(i)}_{{\rm xc},m,n} = 4\pi \nabla_{\mathbf{r}} \cdot \nabla_{\mathbf{r}'} \mathbf{A}_m(\mathbf{r},\mathbf{r}') \,, \tag{105}$$

where

$$\nabla_{\mathbf{r}} \cdot \nabla_{\mathbf{r}'} \equiv \left\{ \frac{\partial}{\partial x}\frac{\partial}{\partial x'} + \frac{\partial}{\partial y}\frac{\partial}{\partial y'} + \frac{\partial}{\partial z}\frac{\partial}{\partial z'} \right\} . \tag{106}$$

The longitudinal projection leading to the form Eq. (105) is thus sufficient to ensure explicit charge conservation for each partial XC-hole contribution, $\Delta n^{{\rm DF},(i)}_{{\rm xc},m,n}(\mathbf{r};\mathbf{q}' \to 0) = 0$.

### C. vdW-DF versions and variants

The original general-geometry vdW-DF1 version [4] did not fully rely on the logic of the vdW-DF method (as summarized above); Instead it relied on Eqs. (2) and (3), that is, using a formulation in which the semilocal functional component $E^0_{\rm xc}$ is not firmly linked to the internal functional $E^{\rm in}_{\rm xc}$. In effect, there is a presence of a cross-over component $\delta E^0_{\rm x} = E^0_{\rm xc} - E^{\rm in}_{\rm xc}$. The same is, in principle, true for all other present general-geometry vdW-DF versions and variants [8, 9, 16, 40–43]. It is also true for the precursor, the layered geometry vdW-DF0 version [3, 6].

The reason for originally tolerating a cross-over term $\delta E^0_{\rm x}$ is a build-in frustration in the present vdW-DF designs. On the one hand, the exchange content of $E^{\rm in}_{\rm xc}$ must help curb nonlocal correlation contributions from low-density regions [1, 6, 9, 10, 12, 65, 171]. This motivated picking a LV form [7, 91, 174] for the exchange enhancement factor $F_{\rm x}(s) = 1 + \mu_{\rm LV} s^2$ in the internal functional. On the other hand, the LV exchange [91] is not, in itself, a good exchange description and leads to poor descriptions of, for example, atomization energies of molecules.

In the original layered- and general-geometry vdW-DF versions [3, 4], we picked revPBE as the functional exchange choice, entering in $E^0_{\rm xc}$. The idea was simply to minimize potential for double counting [6] and the argument was given by demonstrating that this original exchange choice for vdW-DF essentially eliminates all binding from exchange in the case of layered materials and noble gas atoms [3, 4, 6].

The emphasis on plasmon consistency in (spin) vdW-DF-cx means that the actual vdW-DF-cx exchange (in $E^0_{\rm x}$) is also given by the LV analysis [7, 91, 174], at scaled gradients $s < 2 - 3$. At small $s$, this LV behavior is similar to the PBEsol gradient expansion [29]. However vdW-DF-cx (PBEsol) builds from the LV handling of a screened-exchange (from a pure-exchange) analysis of the $\mathbf{q} \to 0$ response limit [7, 174]. Since the PBEsol exchange enhancement (at small $s$) is in turn similar to that used in vdW-DF-C09 [41] and vdW-DF-optB86b [42], these variants are thus related to the consistent-exchange vdW-DF-cx [9].

The exchange choice in other vdW-DF variants and in vdW-DF2 [8] is picked from other considerations.

The vdW-DF family of XC functionals comes with a highly effective Roman-Soler algorithm [17] for evaluating $E^{\rm nl}_{\rm c}$. It also comes with an open-source library, termed LIBVDWXC, for massively-parallel computations of this scheme [18]. While based on (parallel) fast-Fourier transforms, the scheme (and the library) can easily be adapted to all-electron calculations on radial grids [231]. Generally, the $E^{\rm nl}_{\rm c}$ evaluation has excellent scaling [18] and it is never a speed hindrance compared to, for example, a GGA in a plane-wave code, for a given choice of the wavefunction energy cut off.

A full, first-principle nonlocal-DFT characterization of truly large organic and even biological systems is today possible. The biophysics potential is illustrated by a recent subsystem and linear-scaling but first-principle DFT study of structure and molecular dynamics in a tobacco mosaic virus in an explicit aqueous solution [232].

The virus study used the related rVV10 nonlocal-correlation functional [66, 67] as implemented in CP2K [233] and a modern TIER0 supercomputer [232]. However, in computational terms, rVV10 and vdW-DF differs essentially by looking up different universal-kernel files [4, 7, 67]; In Quantum Espresso [36, 219], the rVV10 implementation is, in part, adapted from the subroutines for vdW-DF calculations [11, 36]. The rVV10 and vdW-DF share the Roman-Soler [17] evaluation scheme which seems to present no computational bottleneck up to at least a million atoms [232].

### D. Rationale for consistent-exchange vdW-DF-cx

A simple argument makes it clear that the Dyson equation, Eq. (30), uniquely specifies the correct balance between vdW-DF exchange and correlation terms. The exchange exclusively resides in $\tilde{\chi}_{\rm ACF}$ (as it must be independent of $\lambda$) while the rest of the Dyson-equation terms can exclusively contribute to the nonlocal-correlation term. However, as the ACF effectively reflects an actual long-range interaction, $V_{\rm ACF}$, the Dyson equation (and Lindhard screening logic) leaves no wiggle room between local- and external-field response components [62, 63].

The recent consistent-exchange vdW-DF-cx version [9, 45] seeks to restore compliance with this electrodynamics guide for the vdW-DF method. Effectively, this means that we must use $E_{\rm xc}^{\rm in}$ to specify the exchange component. A complete elimination of the cross-over term $\delta E_{\rm x}^{0}$ is not possible in present vdW-DF versions, given by Eq. (3). However, the exchange component of vdW-DF-cx is chosen to effectively eliminate the adverse effects of $\delta E_{\rm x}^{0} \neq 0$ in the description of binding.

Part of the motivation for taking the consistent-exchange vdW-DF-cx path is a concern about conservation laws. The effect of using a traditional or direct GGA-type specification for the semilocal functional component $E_{\rm xc}^{0} = E_{\rm xc}^{\rm in} + \delta E_{\rm x}^{0}$ can be seen as shifting to a direct-design approach, $S_{\rm xc} \to S'_{\rm xc}$, exclusively in the leading Eq. (94). This corresponds to an adjusted vdW-DF method framework:

$$\begin{aligned} E_{\rm xc}^{\rm DF'} &\equiv \int_0^\infty \frac{du}{2\pi} {\rm Tr}\{\ln(\kappa_{\rm long}(iu) + \delta S(iu))\}\,, \\ &= E_{\rm xc}^{\rm vdW-DF} \\ &\quad + \int_0^\infty \frac{du}{2\pi} {\rm Tr}\{\ln(1 + \kappa_{\rm long}^{-1}(iu)\,\delta S(iu)\}\,, \end{aligned} \tag{107}$$

where $\delta S \equiv S'_{\rm xc} - \nabla S \cdot \nabla G$. It is not clear when the second term of Eq. (107) complies with the current-conservation logic of the vdW-DF method [9]. Also, if we assume some other approach to enforce charge conservation in $E_{\rm xc}^{0}$, it is not clear how that input can easily be merged with the conservation constraints in the present vdW-DF versions. The Occam razor approach of consistent vdW-DF implementations is to try to avoid such potential complications.

Over and above that, we recommend the vdW-DF-cx for general-purpose use because it aims to be true to the underlying MBPT logic. It does not break the Dyson/Lindhard logic of screening, and avoids an uncontrolled approximation [63]. Also, it exclusively relies on analysis of quantum Monte Carlo (QMC) data (in the LDA correlation [23] part of $E_{\rm xc}^{\rm in}$) and of MBPT (in the screened-exchange [7, 91, 174] part of $E_{\rm xc}^{\rm in}$). Of course, the vdW-DF-cx alignment with the full vdW-DF method idea is only perfect when the materials binding arises in regions with small to moderate values of the scaled gradients $s < 2 - 3$, as discussed elsewhere [9, 10]. That criterion is, however, expected to hold in bulk problems and for covalent binding in molecules. It often holds also for intermolecular binding cases [10].

Finally, sticking to the vdW-DF-cx logic holds another advantage: since all functional components are set directly in terms of $S_{\rm xc}$, it makes it easier to seek systematic improvements in nonlocal-correlation functionals [9, 10, 12].

### E. Spin vdW-DF-cx: a systematic extension

Spin effects enter the vdW-DF description because spin polarization adjusts the plasmon dispersion [22], and hence the plasmon propagation [11]. The vdW-DF propagator form $S_{\rm xc}$ is set by a semilocal functional $E_{\rm xc}^{\rm in}$ and it is thus entirely given by LDA correlation and by gradient-corrected exchange, Appendix B. Spin effects on the latter is exactly specified by an exact spin-scaling result [27]. Spin effects on the former is believed to be well described by the analysis in Refs. 23 and 24. In the vdW-DF method, with the present choice of plasmon-propagator model, there is only one possible specification of spin effects on $S_{\rm xc}$.

The implication is that there exists a proper spin extension of the vdW-DF method, Ref. 11. The extension is uniquely defined for any given choice of the internal functional if we strictly follow the Occam design strategy: excluding gradient-corrected correlation in $E_{\rm xc}^{\rm in}$ and sticking with the simple, gradient-expanded $S_{\rm xc}$, Appendix B and Refs. 4 and 12. The natural spin extension (of a given vdW-DF version) results by simply continuing to use $S_{\rm xc}$ to determine all terms of the full vdW-DF method specification, Eq. (91). While the standard vdW-DF versions, given by Eq. (3), also contain an exchange cross-over term, $\delta E_{\rm x}^{0}$, the spin extention is still uniquely determined since the exact spin-scaling criterion [27] also defines spin-polarization effects on $\delta E_{\rm x}^{0}$.

Figure 6 summarizes a study of the weak chemisorption of graphene. We use a 6 layer unit-cell slab and a lateral unit-cell choice that is indicated by the white frame in the top panel. It was chosen to also permit a comparison with RPA calculations [235, 236]. The formulation of spin vdW-DF [11] permitted us to complete the bottom-panel comparison of adsorption of graphene on Ni(111). Here we focus on a comparison with an experimental observation of the optimal adsorption distance [234] (indicated by the vertical dashed line).

Contrasting the graphene binding-energy curves we find that vdW-DF-cx differs from vdW-DF1 [4], from vdW-DF2 [8], from rVV10 [66, 67], and from PBE [28] in predicting both a physisorption minima and a weak-chemisorption minima at a shorter graphene-surface separation. The RPA results yield no clear preference for a chemisorption configuration at a 2.2 Å binding separation [235, 236]. Meanwhile, vdW-DF-cx correctly predicts the stability of the weak chemisorption minima, occuring at a graphene-surface separation 2.1 Å that aligns with the measured Ni(111) value $2.11 \pm 0.07$ Å, [234].

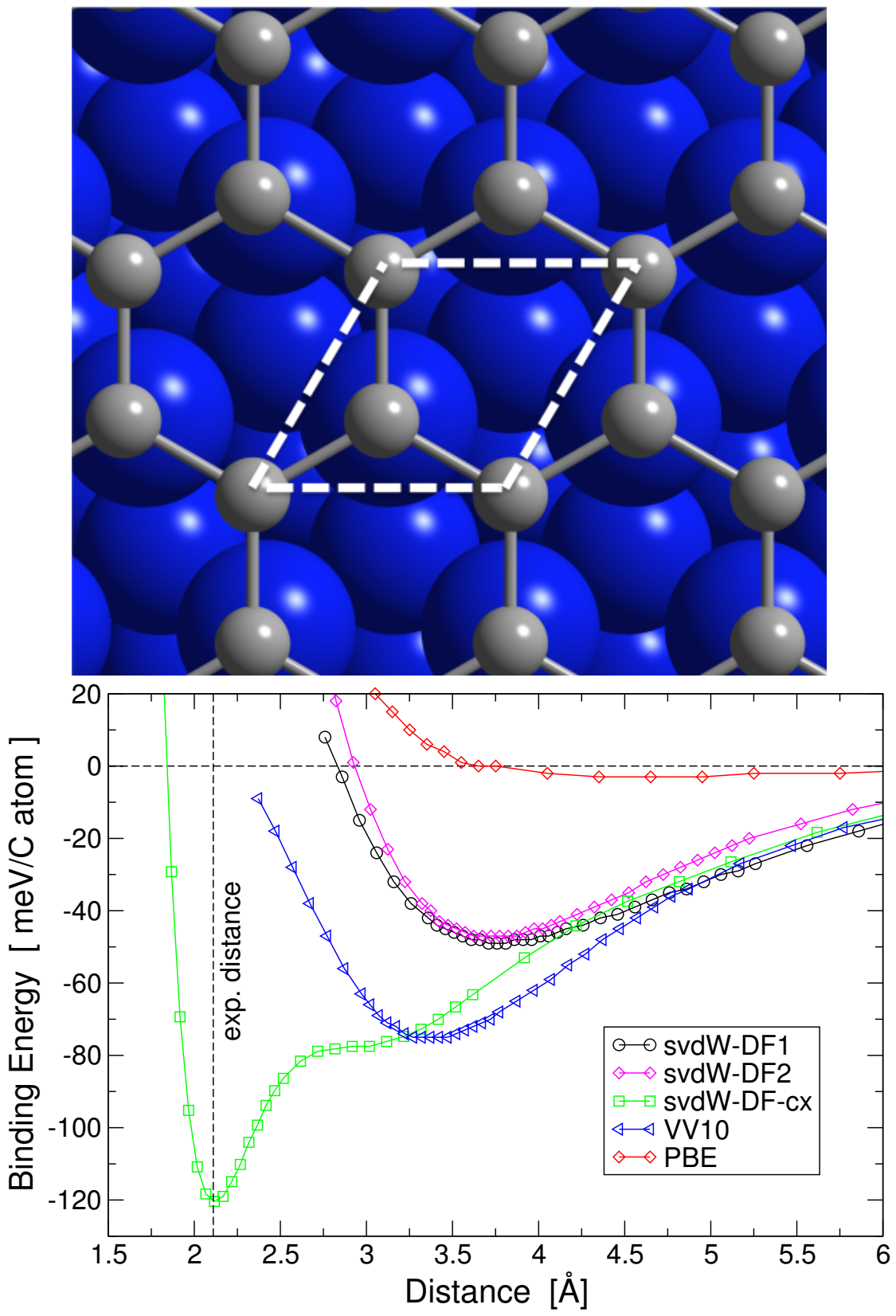

FIG. 6. Weak chemisorption of graphene on Ni (top panel) as described using proper spin extensions of the vdW-DF method (bottom panel). Blue and grey spheres identify Ni and C atoms, respectively. The vertical dashed line in the bottom panel shows the experimental results for the adsorption minimum distance $2.11 \pm 0.07$ Å [234]. The panel contrasts the results for the adsorption energy variation as obtained in the spin extension of the original vdW-DF1 (black circles), of vdW-DF2 (purple diamonds), and of vdW-DF-cx (green squares). For completeness the panels also shows the results for PBE and VV10 [66, 67]. The panels are reproduced, with permission, from the supplementary materials of Ref. 11. Copyright (2015) by the American Physical Society.

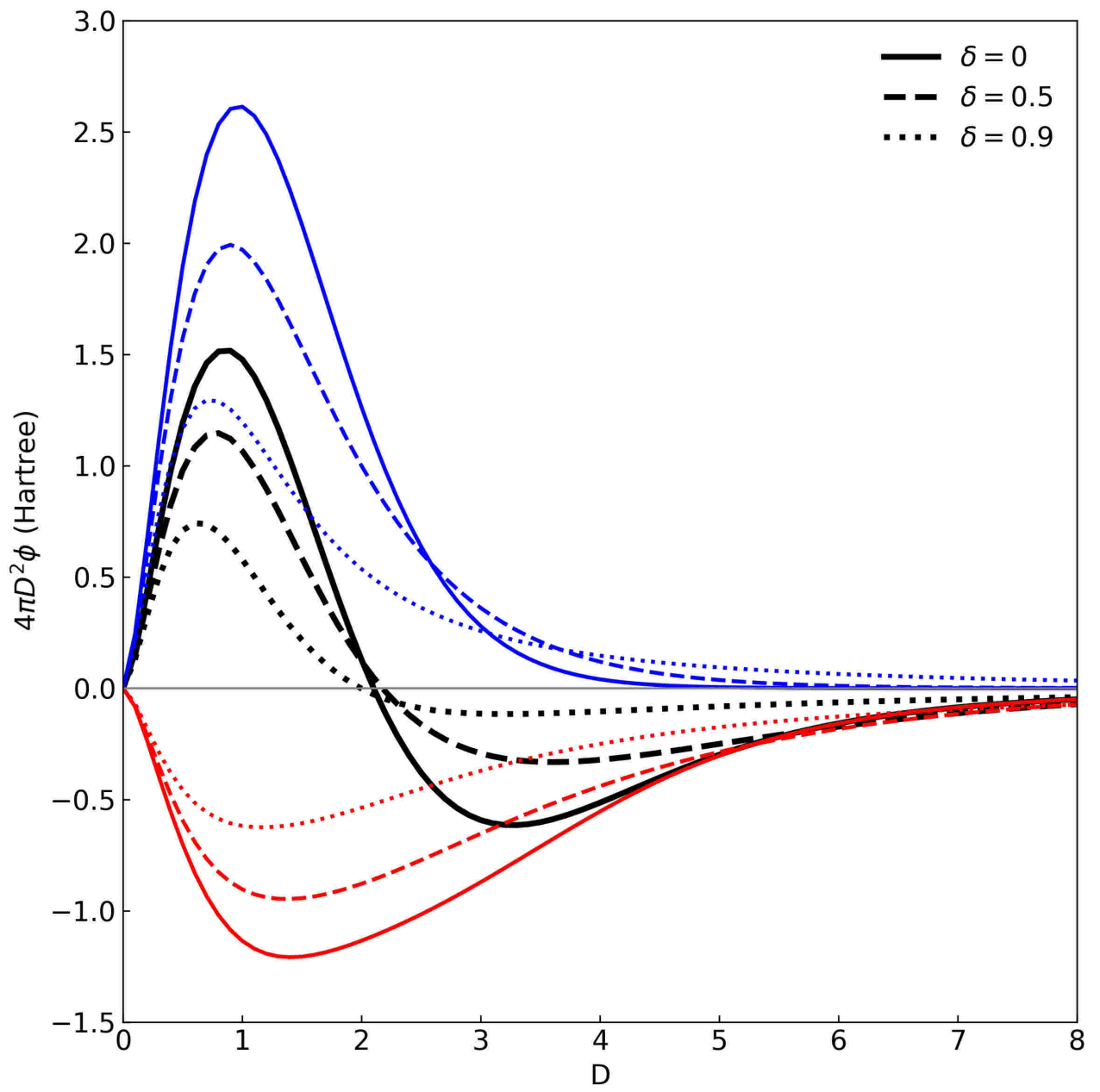

FIG. 7. Universal nonlocal-correlation kernels for the full $E_c^{nl}$ behavior (black curves) and for the vdW and cumulant components (red and blue curves respectively). The kernels reflect nonlocal-energy contributions from an electrodynamics coupling among GGA-type XC holes, centered at two different positions $\mathbf{r}$ and $\mathbf{r}'$, Ref. 10. The kernel behavior can, however, be fully represented by an effective distance $D$ and a parameter $\delta$ which reflects how different these internal GGA-type XC holes are at positions $\mathbf{r}$ and $\mathbf{r}'$. It is exclusively the set of $\delta = 0$ curves that are relevant for the homogeneous electron gas (HEG) limit; Seamless integration with LDA follows in the HEG because the solid-black curve integrates to zero [4, 5].

In molecular problems [35], and especially for atomization energies [13, 15, 237–242], a correct handling of spin effects is important. The same was also found to be true for transition-metal cohesive energies in cases where the atom is in a high-spin state [95]. It is necessary to eventually seek a non-collinear spin implementation of vdW-DF-cx [44] (and of other vdW-DF versions and variants) to address some triangular (or other frustrated) molecular problems [243] as well as for some so-called vdW crystals [44]. However, such a formulation is not yet publicly available.

For a general treatment of magnetic effects and molecular spin it is also possible that the vdW-DF ability to reflect spin effects may be improved by permitting more flexible formulations, that merge the vdW-DF-cx logic with the spin handling that is used in PBE [244, 245]. However, we are presently sticking with the above-summarized Occam design strategy for vdW-DF-cx [11]. This is because simplicity in the design focus (using just $E_{xc}^{in}$ as the nucleation point in spin vdW-DF-cx) makes it easier to check the impact of fundamental design choices, for example, regarding the plasmon dispersion [9, 45].

### F. Electrodynamics interpretation of consistent vdW-DF designs

Below, we discuss the formal nature of general consistent vdW-DF designs. That is, we consider the class of vdW-DF versions that comply with the screening laws of the ACF and of the vdW-DF framework, and that aim to implement the full logic of the vdW-DF method for XC functional designs.

In the consistent-exchange vdW-DF designs, the leading term must effectively be set by $E_{\rm xc}^{\rm in}$. It is thus restricted to GGA exchange and LDA correlation. The nonlocal-correlation energy of the vdW-DF method can formally be written

$$E_{\rm c}^{\rm nl} = E_{{\rm c},\alpha}^{\rm nl} + E_{\rm c,vdW}^{\rm nl} \,. \tag{108}$$

The splitting is here made depending on whether the nonlocal-correlation effects originate from $\tilde{\chi}_{\rm ACF}$ or from the $m \geq 2$ components of the Dyson expansion, Eq. (99). Taken together, the terms of Eq. (108) must serve as the consistent-vdW-DF replacement for the GGA formulation of gradient-corrected correlation. Appendices C and D discuss the details in a general such GGA extension, here we summarize the analysis in an electrodynamics context.

*Electrodynamics-response terms with* $m = 1$ define one important subset of contributions

$$\begin{aligned} E_{{\rm xc},\alpha} &\equiv \int_0^\infty \frac{du}{2\pi}\, {\rm Tr}\{4\pi\alpha(iu)\} \\ &= E_{\rm xc}^{\rm in} + E_{{\rm c},\alpha}^{\rm nl} \,. \end{aligned} \tag{109}$$

These are naturally ordered by the number $n$ of $S_{\rm xc}$ factors:

$$E_{{\rm c},\alpha}^{\rm nl} = \sum_{n=2}^{\infty} E_{{\rm c},\alpha}^{n} \,, \tag{110}$$

$$\begin{aligned} E_{{\rm c},\alpha}^{n} &= (n!)^{-1} \int_0^\infty \frac{du}{2\pi}\, {\rm Tr}\{\nabla (S_{\rm xc})^n \cdot \nabla G\} \\ &= (n!)^{-1} \int_0^\infty \frac{du}{2\pi}\, {\rm Tr}\{(S_{\rm xc})^n\} \,. \end{aligned} \tag{111}$$

Here (as in Appendix C) we retain the notation introduced in Eq. (102), using $n$ to track the number of $S_{\rm xc}$ factors. The $m = n = 1$ component is just the internal functional, as implied in Eq. (109).

The exponential-re-summation term of $E_{\rm xc}^{\rm DF}$, that is, Eq. (110), is set up to track general screening, including vertex corrections, when used together with the leading $E_{\rm xc}^{\rm in}$ term. This follows by the discussion in III.B and Appendix B. However, in the context of describing vdW interactions, 'screening' is often taken to imply a moderation of the dispersion forces [81, 202]. To avoid confusion, we therefore choose to denote Eq. (110) instead as a 'cumulant' term, in our discussions below.

We note that the cumulant term captures vertex corrections even if the expansion in Eq. (111) is truncated at second order, at $n = 2$. The $n = 1$ term, i.e., the internal functional $E_{\rm xc}^{\rm in}$, is designed to carry the vertex corrections that are in LDA correlation directly [4, 7]. When truncating the expansion of $E_{{\rm c},\alpha}^{\rm nl}$ at $n = 2$, as done in popular versions defined by Eq. (94), we cannot fully leverage the expontial re-summation that underpins the vdW-DF method. However, we retain a re-summation character since we always keep at least two expansion terms in $E_{{\rm c},\alpha}^{\rm nl}$.

*The remaining nonlocal-correlation part*

$$E_{\rm c,vdW}^{\rm nl} = -\sum_{m=2}^{\infty} \int_0^\infty \frac{du}{2\pi m}\, {\rm Tr}\{(\nabla\alpha \cdot \nabla V)^m\} \,, \tag{112}$$

$$E_{\rm c,vdW}^{\rm nl} \equiv \sum_{n=2}^{\infty} \sum_{m\geq n} \sum_{i} E_{{\rm xc},m,n}^{{\rm DF},(i)} \,, \tag{113}$$

primarily reflects pure or asymptotic vdW effects. The association is directly motivated for the second-order expansion form

$$E_{\rm c,vdW}^{\rm nl,m=n=2} = -\int \frac{du}{4\pi}\, {\rm Tr}\{\nabla\cdot\boldsymbol{\alpha}_1(iu)\cdot\nabla V \nabla\cdot\boldsymbol{\alpha}_1(iu)\cdot\nabla V\} \tag{114}$$

This is always describing an attraction, appendix D, and it reflects the form for the vdW interaction that holds for two disjunct fragments, Sec. IV and Ref. 10.

More generally, the full specification Eq. (112) contains factors that can always rewritten

$$\frac{1}{m}(\nabla\alpha\cdot\nabla V)^m = \frac{1}{m(m-1)} \sum_{\nu=1}^{m-1} (\nabla\alpha\cdot\nabla V)^{m-\nu}\, (\nabla\alpha\cdot\nabla V)^{\nu} \,. \tag{115}$$

For partly separated fragments, the interaction is still dominated by terms where the implied repeated spatial integrations will only involve two Coulomb factors $V(\mathbf{r}-\mathbf{r}')$ connecting these fragments. Apart from the weighting in Eq. (115), we have again terms that reflects the asymptotic vdW form, Sec. II.

We need the full vdW-DF machinery, given by the interplay of the vdW term Eq. (112) and of the cumulant term Eq. (111), in general. The full account sorts out the weighting between nonlocal-correlation effects in the more interesting general cases, when density fragments can no longer be considered disjunct.

*Interpretation of terms in the HEG limit* merits a separate discussion. We note that, as we approach the HEG, we will no longer have system fragments and the original premise of interpreting Eq. (112) as pure vdW interactions eventually breaks down; Similarly, we should explain the nature of Eq. (110) in the HEG limit [4].

In the HEG, there cannot be any actual vdW forces nor can there be any beyond-LDA vertex corrections. A vdW-DF-cx calculation is fully consistent with those observations: We explicitly comply with the criteria $E_{\rm c}^{\rm nl} \equiv 0$ and $E_{\rm xc}^{\rm in} \to E_{\rm xc}^{\rm LDA}$ in the HEG limit [4, 12]. The question is only one of *interpretation*.

We find it motivated to always view Eq. (112) as reflecting pure vdW interactions, i.e., forces arising from the Ashcroft-ZPE coupling mechanism [1, 10, 64, 65, 158]. This picture, Fig. 2, is valid in the HEG limit (even if it has no consequences on material binding then). Yes, these ZPE electron-correlation effects are already baked in into the LDA description of the HEG [25]. However, it is still instructive to view the relevance of the pure vdW interactions in isolation, as an evaluation of Eq. (112) allows us to do.

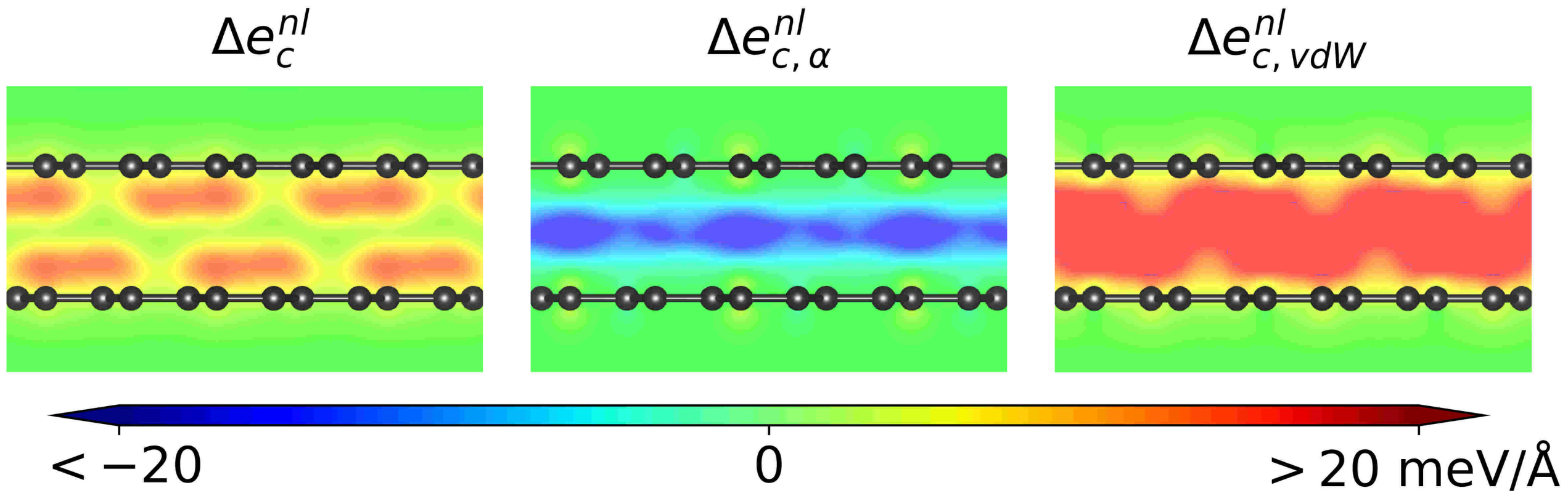

FIG. 8. Nonlocal correlation contributions to binding in bilayer graphene. The left panel shows the spatial variation in the total nonlocal correlations computed in vdW-DF-cx. The middle and right panels depict the cumulant (or vertex-correction) and the pure vdW interaction components, respectively.

We view Eq. (110) as representing cumulant (i.e., vertex and other GGA-type screening) effects in general, again even in the HEG limit. Since Eq. (112) represents the Ashcroft-ZPE mechanism, there is also a need (even in the HEG) to explicitly treat a compensating effect, namely as provided by the cumulant term. This term must therefore be extracted from the LDA correlation that we have originally inserted in $E_{\rm xc}^{\rm in}$. Equation (110) specifies that subtraction exactly, ensuring that the vdW-DF method has seamless integration with LDA in the HEG limit [4].

*Electrodynamics interpretation of terms in the standard, truncated-expansion vdW-DF versions.* The leading ($m = n = 2$) components of Eq. (108) define the physics content of the (spin) vdW-DF-cx version [9, 11]. It is worth exploring in detail.

Appendix D shows that those terms can be formulated in terms of a universal kernel evaluation

$$E_{\rm c,\alpha}^{{\rm nl},n=2} = \frac{1}{2}\int d^3\mathbf{r} \int d^3\mathbf{r}'\, n(\mathbf{r})\, \phi^{\alpha}(d,d')\, n(\mathbf{r}')\,, \quad (116)$$

$$E_{\rm c,vdW}^{{\rm nl},n=2} = \frac{1}{2}\int d^3\mathbf{r} \int d^3\mathbf{r}'\, n(\mathbf{r})\, \phi^{\rm vdW}(d,d')\, n(\mathbf{r}')\,, \quad (117)$$

where $d$ and $d'$ denote suitable ways to scale the distance between $\mathbf{r}$ and $\mathbf{r}'$.

Figure 7 summarizes our numerical evaluation of the (universal-)kernel components, $\phi_\alpha$ and $\phi_{\rm vdW}$. The variation is given in terms of scaled differences, as detailed in Appendix D and elsewhere [4, 5, 60]. The sum

$$\phi(d,d') = \phi^{\alpha}(d,d') + \phi^{\rm vdW}(d,d')\,, \quad (118)$$

defines the full $E_{\rm c}^{\rm nl}$ kernel, and both of these kernels reflect the general two-density-point structure of Eq. (55). However, the universal-kernel forms, $\Phi_0^{\alpha}(d,d')$ and $\Phi_0^{\rm vdW}(d,d')$, are found to always remain positive and negative, respectively.

In the vdW-DF-cx version, we have an expansion given by Eqs. (41) and (109). The beyond-$E_{\rm xc}^{\rm in}$ are terms given by Eqs. (108), (116) and (117) and we make the following observations:

The first nonlocal-correlation part, Eq. (116), is, in itself, of direct interest for understanding the physics content of our electrodynamics modeling. This follows because it can be combined with $E_{\rm xc}^{\rm in}$ to yield a mapping of the model susceptibility $\alpha(\omega)$. The associated local energy contribution

$$e_{\rm c,\alpha}^{\rm nl}(\mathbf{r}) = n(\mathbf{r}) \int_{\mathbf{r}'} \phi^{\alpha}(d(\mathbf{r},\mathbf{r}'), d'(\mathbf{r},\mathbf{r}'))\, n(\mathbf{r}')\,, \quad (119)$$

provides a mapping of where we can expect cumulant effects (including nonlocal vertex corrections), in Eq. (109), to play a larger role in material binding. Being based on an exponential re-summation, it has a formal structure that allows a low-level approximation to be accurate on the quasi-particle dynamics, at least for core levels and near the Fermi surface [86, 163, 164]. We are motivated to extract this mapping also broadly, since we are documenting a strong general performance for vdW-DF-cx, Refs. 9–11, 15, 45, and 95 and Sec. VI.

The second nonlocal part of Eq. (108) represents pure vdW interactions [1, 10, 64, 65, 158], as they emerge in the presence of screening by the surrounding electron gas [10, 12, 64]. The associated local energy contribution

$$e_{\rm c,vdW}^{\rm nl}(\mathbf{r}) = n(\mathbf{r}) \int_{\mathbf{r}'} \phi^{\rm vdW}(d(\mathbf{r},\mathbf{r}'), d'(\mathbf{r},\mathbf{r}'))\, n(\mathbf{r}')\,, \quad (120)$$

provides a mapping of where pure vdW binding contributes significantly to materials binding.

*Seamless integration requires treating the cumulant and the pure-vdW terms together.* There are in general compensations between the $e_{\rm c,\alpha}^{\rm nl}$ and $e_{\rm c,vdW}^{\rm nl}$ contributions, while $e_{\rm c,vdW}^{\rm nl}$ dominates at larger separations. However, both parts are needed to secure seamless integration of vdW-DF-cx with LDA, i.e., compliance with Eq. (55) in the near-HEG limit. Inserting a constant density, $n_0$, yields $E_{\mathrm{c},\alpha}^{\mathrm{nl}} < 0$. One needs the combined nonlocal-correlation effects, reflected in Eqs. (95) and (118), to ensure seamless integration, as discussed elsewhere [4, 12].

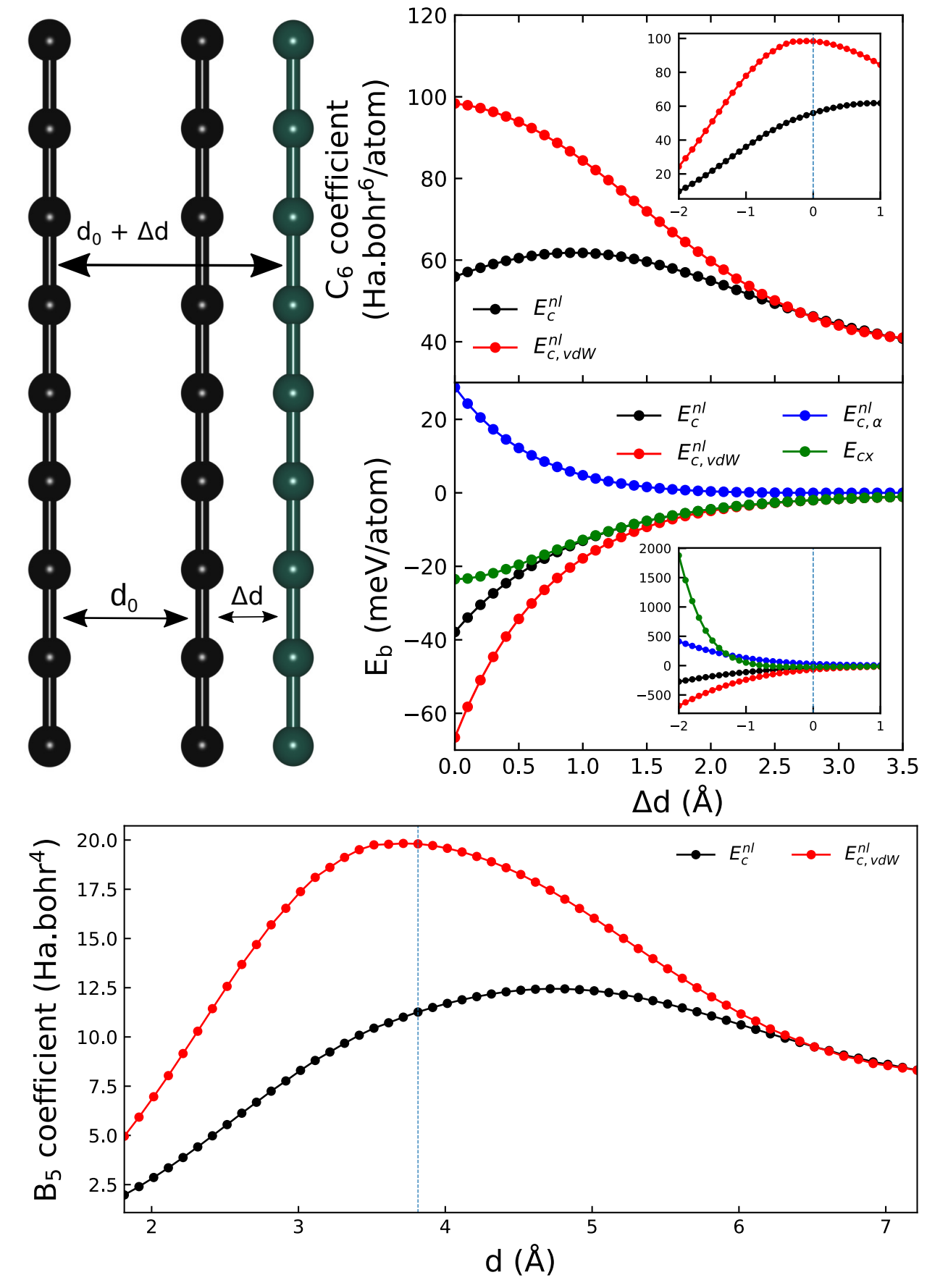

FIG. 9. Scaling in the vdW-DF-cx results for binding between a pair of carbyne wires. The left panel shows the geometry, identifying also a focus on tracking the mutual attraction per unit length as a function of the changes in wire distance $d = d_0 + \Delta$, where $d_0 = 3.81$ Å is the vdW-DF-cx results for the equilibrium wire separation. The middle right panel shows the computed variation in the total wire binding energy (per carbon atom in a wire), in the full nonlocal-correlation binding contribution (black curve) and in the binding contributions from pure vdW interactions (red curve) and cumulant corrections (blue curve). The top right and bottom panels show the corresponding scaling of an interpretation of the nonlocal-correlation binding in terms of an atom-based-vdW-pair interaction (with effective atom-centered-vdW $C_6$ coefficients) and in terms of a wire-based-vdW interaction form (with effective wire-vdW $B_5$ coefficients), respectively. The vertical line in the bottom panel positions the optimal separation $d_0$ relative to the $B_5(d)$ enhancements.

*The universal-kernel descriptions simplify an analysis of the nature of binding.* Figure 8 illustrates how having efficient determinations of nonlocal-correlation contributions Eqs. (119) and (120) can provide a spatial mapping of the nature of binding, here explored for the case of a graphene bilayer. The left panel shows a mapping of the full $E_{\mathrm{c}}^{\mathrm{nl}}$ binding contribution; Such mapping was introduced and discussed in Refs. 55, 57, and 58 and can be supplemented by a coupling-constant scaling analysis that isolates the kinetic-correlation energy binding component [14].

The middle and right panels of Fig. 8 shows a here-explored separation of $E_{\mathrm{c}}^{\mathrm{nl}}$ binding contributions into cumulant effects (nonlocal-vertex corrections) and pure vdW attraction, respectively. We see that the vdW attraction is larger and even more widely distributed in the trough region *between* the layers than what is apparent from the $E_{\mathrm{c}}^{\mathrm{nl}}$ description [9, 10, 14, 59]. However, this pure vdW binding is also offset by a repulsion in $E_{\mathrm{c},\alpha}^{\mathrm{nl}}$, in the central region.

Figure 9 summarizes a vdW-DF-cx calculation and our pure-vdW/cumulant-resolved analysis of the binding between two carbyne wires. That is, we consider the attraction between two parallel one-dimensional wires of pure carbon as a function of their separation $d$, as illustrated in the top left panel.

This is a system where vdW-DF-cx has problems at describing the mutal binding energy (that is, the energy difference between the combined system and those of the components) per unit length at large wire distances. Depending of the intra-wire carbon-carbon separation $l_{C-C}$, this problem allows a simple tuning between an insulating nature (with dimerization) and a conducting nature (with no dimerization) within a tight-binding DFT model, by simply varying the intra-wire carbon-carbon separation [246]. This leads in turn to an elegant analysis of the impact of 1-D conduction on the long-range component of the vdW attraction [105, 202, 246–254]. Ref. 246 provides the analysis within a TS-based Many-Body-Dispersion (MBD) method [80, 81], denoted TS-MBD. It tracks the power-law behavior of the mutual vdW interaction for variations of the wire separations $d$. The TS-MBD method is found to correctly produce a $d^{-5}$ asymptotic scaling for insulating wires and an expected significant difference in the exponent for metallic wires [202, 247, 250].

For vdW-DF method implementations that rely on the universal-kernel formulation (as vdW-DF1 and vdW-DF-cx do,) the interaction enhancement can stretch out to about a nanometer due to multipole or image effects. This is documented, for example, in a previous study of parallel semi-conducting nanotubes [53]. The enhancement in vdW-DF-cx attraction cannot stretch as far as in TS-MBD because the kernel contribution from an electron-XC-hole pair, at $\mathbf{r}_1$ and $\mathbf{r}_2$, *eventually* approaches a form decaying like $|\mathbf{r}_1 - \mathbf{r}_2|^{-6}$, Refs. 4 and 105. The consequence is that vdW-DF-cx and vdW-DF always have an asymptotic vdW interaction energy that decays like $d^{-5}$ for parallel wires. The vdW-DF-cx predicts an absence of dimerization and a conduting nature of the wire, but it fails to predict the associated enhancement and non-integer power-law interaction be-havior that should therefore result at large separations [246, 250].

However, the nature of vdW attraction at and near binding must necessarily be markedly different from the asymptotic form [105]. The vdW interaction among nanostructures is indeed also documented [53] to be different than what is in the asymptotics [10, 69, 105, 246, 253, 254]. We and others have therefore argued that even non-MBD XC functionals can still be useful and accurate for characterizations on the mutual attraction at binding separations [10, 69, 105].

The middle-right panel of Fig. 9 reports our vdW-DF-cx calculation of the mutual binding (per carbon atom in a wire) described by differences in the total energy (in green, with $E_{\rm cx}$ label) and in the nonlocal-correlation energy (in black, with $E_{\rm c}^{\rm nl}$ label). We have furthermore resolved the latter in pure-vdW (red) and cumulant (blue) contributions. The results are obtained for the fully relaxed intra-wire carbon-carbon separation $l_{C-C} = 1.28$ Å. The interaction results are shown as a function of $\Delta d = d - d_0$, where $d_0 = 3.81$ Å denotes the equilibrium vdW-DF-cx value for the wire separation. Computational details are given in Sec. V.

The bottom panel of Fig. 9 also shows that pure-vdW term dominates the total nonlocal-correlation term as the distance increases, but the repulsive $E_{{\rm c},\alpha}^{\rm nl}$-based contributions significantly compensate this attraction at and near binding separations. There is also a saturation directly in $E_{\rm c,vdW}^{\rm nl}$ (not shown in this panel). The sum of these nonlocal-correlation contributions scales away in the high-density region, to provide a seamless integration with LDA [4, 5, 7, 10]. The sum of these contributions eventually vanishes when the distance $d$ decreases.

Differences between the pure vdW attraction ($E_{\rm c,vdW}^{\rm nl}$) and the binding arising from the sum of nonlocal correlations (in $E_{\rm c}^{\rm nl}$) are certainly the correct behavior in weak-chemisorption binding [59] where competition among forces are responsible for producing density changes [12, 45, 255, 256]. It is also, at least qualitatively, the correct behavior for pure physisorption [175, 176, 181, 182] since both nonlocal-correlation components are needed to reflect the vdW-multipole [10] as well as other screening effects [10, 105]. The screening produces, for example, large image-plane corrections in vdW binding at surfaces [157, 179, 182, 184, 185] and among extended molecules [53]. The image-plane effects enhance the interaction description at close-to-intermediate separations from the asymptotic $d^{-5}$ interaction behavior [53].

We shall return to a more detailed discussion of this interaction. Here we note that in vdW-DF-cx the description at and near materials binding separations is guided by the Lindhard/Dyson screening logic and have a formal reason for being robust towards system-specific changes in the electron-density variation [45, 56, 59].

### G. Open questions for consistent vdW-DF

The presentation of the vdW-DF method and of consistent-exchange vdW-DF versions (like vdW-DF-cx and svdW-DF-cx) leave us at this stage with a set of open questions:

*GGA-level vertex corrections?* Consistent vdW-DF relies on an exponential re-summation to capture effects similar to nonlocal vertex corrections. However, it is not clear if the re-summation in $\alpha(\omega)$ is enough, if it can, in practice, benefit from the vertex advantages that exist for cumulant expansion for the quasi-particle dynamics [86, 163, 164]. The constraint-based semilocal functionals, like PBE and PBEsol, set a high standard. Transition metals and their surfaces are often considered a challenge for XC functionals [25, 56, 89, 95, 133, 181, 222, 223, 225, 226, 257]. A particular questions is then: Does the indirect design logic of consistent vdW-DF retain enough semilocal correlation to work even here?

*Non-additivity in vdW interactions?* In the general-geometry vdW-DF formulation [4, 12], we truncate the expansion of nonlocal-correlation effects in the plasmon-propagator $S_{\rm xc}$ to second order. This is not enough to capture so-called many-body dispersive interactions in the asymptotic limit [10, 64, 81, 105, 202, 252]. However, two questions remain: 1) does the reliance of the vdW-DF method on a GGA-type screening [10] provide non-additivity at binding separations? 2) Since not all fragments of a molecular crystal reside at the optimal mutual separation, do we get the right description of molecular crystals at binding?

*All-round performance?* We want nonlocal-correlation XC functionals that work for concurrent descriptions of both dense and sparse matter components. In consistent vdW-DF, we leverage both Dyson/Lindhard screening logic and current conservation. An important questions is therefore: Does the general-purpose capability [45] survive in the resulting vdW-DF-cx version?

*Role nonlocal vertex effects?* The GGAs and meta-GGAs are expected to reflect a mixture of nonlocal vertex effects and short-to-immediate-range vdW interactions [69], but offer no detailed characterization of the balance. This review explains that vdW-DF-cx calculations can be used to track such screening effects in isolation. It is therefore natural to inquire, where do we find important cumulant (vertex-correction) effects in general materials binding?

## V. COMPUTATIONAL DETAILS

Some of the open questions for consistent vdW-DF can be settled by analyzing published results, for example, vdW-DF-cx studies for molecules as well as for bulk and layered matter. For computational details of those studies we refer directly to the cited papers.

Other questions for consistent vdW-DF can be addressed by validation studies that we also provide in this paper. The results of some of these test have already been partly presented in Figs. 1, 5, 6, 8, and 9. However, the discussion will generally be expanded in the following validation section and we have collected the computational details in this section.

For a discussion of molecular and bulk binding processes (like binding and reactions) we generally investigate a compound system 'AB' as well as its relevant constitutes, for example, 'A' and 'B'. We compute in vdW-DF-cx, or in a general functional 'DF', both the total energy (in this general discussion denoted $E_{\rm DF}$) and XC functional terms, like $E_{\rm c}^{\rm nl}$. These are sometimes computed with full atomic relexations and sometimes at so-called reference geometries, when given in benchmark suites like the GMTKN55 [35]. The energy values of interest are the energy differences for the investigated process 'p', like

$$E_{\rm p} \equiv \Delta E_{\rm DF} = E_{\rm DF}^{\rm A} + E_{\rm DF}^{\rm B} - E_{\rm DF}^{\rm AB}\,, \qquad (121)$$

$$E_{\rm p,c}^{\rm nl} \equiv \Delta E_{\rm c}^{\rm nl} = E_{\rm c}^{\rm nl,A} + E_{\rm c}^{\rm nl,B} - E_{\rm c}^{\rm nl,AB}\,. \qquad (122)$$

The process energies may at times reflect a binding or a bulk cohesion, and they are then denoted $E_{\rm b}$ or $E_{\rm coh}$ to have better consistency with papers and surveys that are cited. Similar structures of energy differences, Eqs. (121) and (122), hold when the properties 'p' of interest are isomerization energies and molecular unfolding energies, reaction barriers, or charge transfers (including ionization or electron-affinity processes).

We systematically use the plane-wave QUANTUM ESPRESSO DFT code package [36, 219]. Like our c-based library LIBVDWXC [18], this code suite has highly efficient, native implementations of both the fast-Fourier-transform (universal-kernel) evaluation [17] of the $E_{\rm c}^{\rm nl}$. It has the vdW-DF family of functions, including the consistent-exchange vdW-DF-cx version [9], and it has the proper vdW-DF spin extension [11]. It also permits stress relaxations of nonlocal-correlation functionals. Core electrons are represented via pseudopotentials since we use QUANTUM ESPRESSO [36, 219].

### A. Molecular-system benchmarking

To obtain statistics on functional performance, we complete a full functional survey of molecular properties as tested in the extensive GMTKN55 suite of benchmarks sets [35]. The GMTKN55 not only has a wide range of benchmark sets for intra- and inter-molecular noncovalent-interaction energies, as already discussed in Fig. 1. However, the GMTKN55 suite also has groups of benchmark sets probing the performance for small- and large-molecular properties, and a group of benchmarks probing barriers for molecular reactions [35]. Taken together, the results on performance give a fair impression of the robustness and transferability of semi- and truly nonlocal density functionals, when used in the molecular realm.

We perform and report this comprehensive survey for both the consistent-exchange vdW-DF-cx version [9] and for the original vdW-DF1 [4, 5, 7].

For the molecular benchmarking we use Troullier-Martins norm-conserving pseudo potentials of the ABINIT package [38], with a 80 Ry wavefunction cutoff. We and collaborators have previously used this pseudopotential and cut off choice for benchmarking performance on molecular interactions, in connection with launching hybrid vdW-DF(-cx) versions [13, 15] and while testing the dependence on unit-cell sizes for plane-wave studies of molecular systems [37].

We use the Gamma point $k$-point for our periodic unit-cell representation of the molecular problems. Within each of the 55 different molecular benchmark sets, we identified the largest systems and then added 10 Å of vacuum to have a standard unit-cell size for all comparisons (among the individual processes and among functionals). This approach yields a setup that for the S22 benchmark set corresponds to the convergence recommendations made in Ref. 37. Several benchmark sets involve comparing system of different charge states. For these we used the Makov-Payne correction [39] to compensate for spurious electrostatic interactions that arise because we use a periodic unit-cell representation for molecules.

The relevant comparison of (spin) vdW-DF-cx accuracy should be made on the so-called second (GGA-based) rung of dispersion-corrected DFTs [33]. This is because vdW-DF-cx, while a consistent implementation of the vdW-DF method, is still limited to a semi-local choice of the exchange formulation [59]. According to the full GMTKN55 comparison [35] of dispersion-corrected DFTs [33], one of two best performing second-rung DFT-D3 formulations is revPBE-D3 [33, 34]. This functional is also available to us (with the very same pesudopotentials) in QUANTUM ESPRESSO [36, 219]. It is an interesting comparison for revPBE-D3 has the very same exchange choice as do both vdW-DF1 [4, 7] and vdW-DF0 [3]. We therefore include a full QUANTUM ESPRESSO [36, 219] survey of the revPBE-D3 as well as of PBE [28].

We note that a comparison of vdW-DF1, revPBE-D3 and vdW-DF-cx allows us to highlight the role of balancing exchange and correlation contributions, as already discussed in the introduction in connection with Fig. 1.

We have not made our own comparison of vdW-DF-cx molecular performance versus the best-performing version on the third (meta-GGA) rung of the DFT-D3s, namely SCAN-D3 [33, 35, 258]. Expanding the survey is beyond the present scope and it would distract from our present focus. Here the purpose is to explore if use of Lindhard/Dyson screening logic to balance exchange and correlation significantly enhances the robustness and transferability of XC functional designs in the vdW-DF method. A comparison with SCAN-D3 [33, 35, 258] and SCAN+rVV10 [69] is relevant for a possible, future meta-vdW-DF, a version that may perhaps use a meta-GGA exchange as the internal functional.

We do, however, find that the vdW-DF-cx perfor-mance (as we assert in QuantumEspresso) generally approaches literature reports for SCAN+rVV10 [69] performance, for here-investigated systems. This holds for SCAN-D3 studies for molecules, computed in Turbomole [259, 260] and reported in the GMTKN55 functional survey [35]. It also holds for SCAN+rVV10 studies of surface properties [223] computed in VASP [217, 218]. We mention also these SCAN-D3/+rVV10 results in a vdW-DF-cx comparison when relevant.

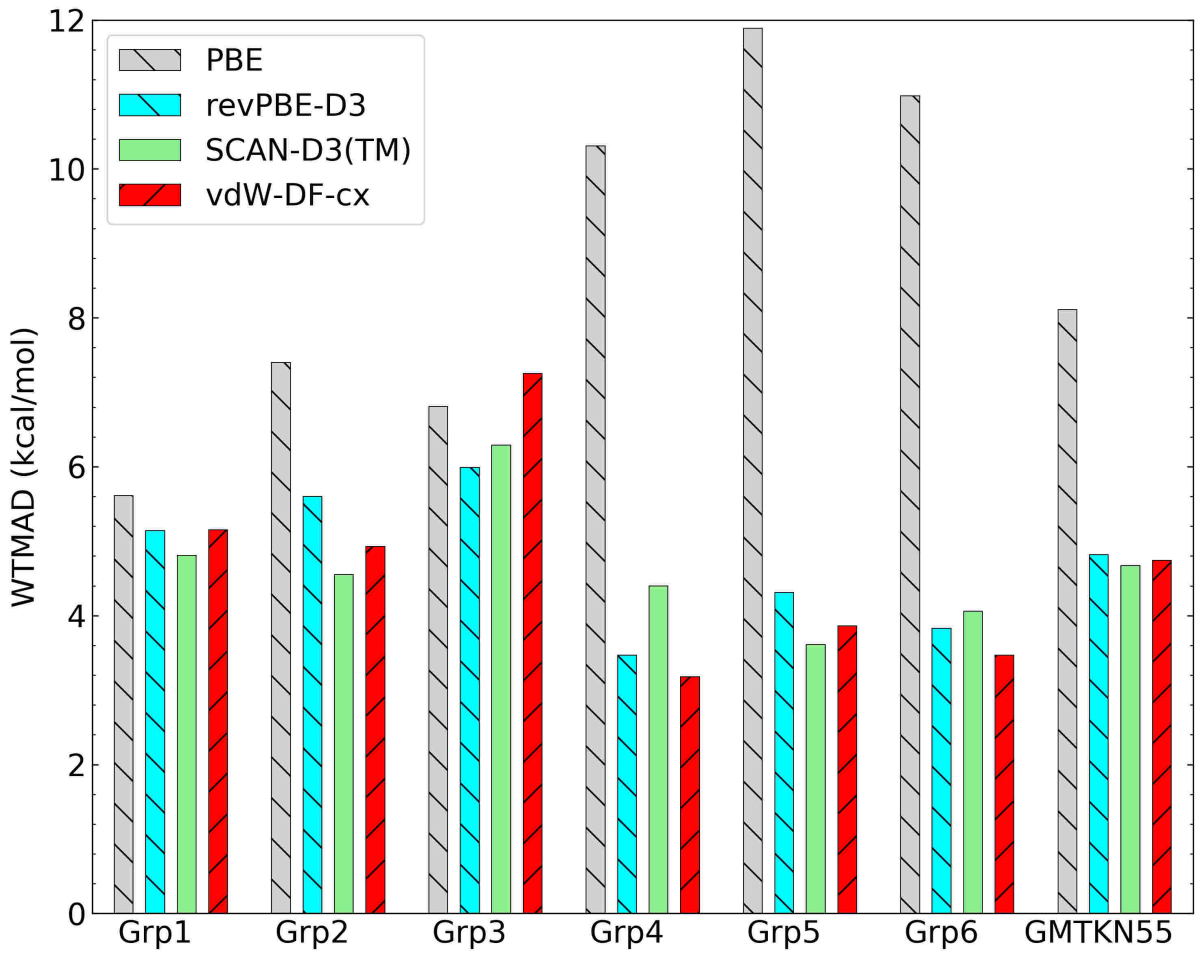

FIG. 10. Full survey of vdW-DF-cx performance across the GMTKN55 suite [35] of benchmarks sets, as asserted in the plane-wave code Quantum Espresso [36, 219] and using a weighted total mean absolute deviation measure 'WTMAD' defined in Ref. 35 and in the text. For comparison, the figure also includes corresponding results for PBE and revPBE-D3 (also in Quantum Espresso calculations) and those obtained with Turbomole [259, 260] for SCAN-D3 in Ref. 35. Benchmark groups 4, 5 correspond to a set of benchmarks on inter- and intra-molecular noncovalent interactions. Group 6 is the union of group 4 and 5, that is, the set of general noncovalent interaction benchmarks that was detailed in Fig. 1. Group 1, 2, and 3 are collections of benchmark sets reflecting small-molecule properties, large-molecule properties, and transition barriers, respectively [35].

At the same time, we note that care must be taken in comparing our plane-wave/pseudopotential benchmarks with those of quantum-chemistry DFT codes (that are typically used for the surveys in Ref. 35). There are advantages and disadvantages of both approaches [37]. However, it is clear that converging negatively charged systems in a plane-wave code is generally difficult, at least with the presently used pesudopotential choice. Furthermore, our Quantum Espresso testing is complicated by the fact that we do not have access to a non-collinear spin formulation [44] for vdW-DF-cx.

To provide a check on our molecular GMTKN55 survey we therefore first discuss differences and similarities between our revPBE-D3 benchmarking in Quantum Espresso and that reported in Ref. 35. We find that sometimes our plane-wave revPBE-D3 results are closer to reference values (available in Ref. 35) and sometimes the quantum-chemistry results [35] are closer. However, in almost all cases the agreement is good in terms of percentages and in terms of absolute MAD values.

For all of the benchmark sets involving noncovalent interactions, the agreement between the here-reported revPBE-D3 benchmarking and that of Ref. 35 is typically within one tenth of a kcal/mole and within two tenths of a kcal/mol for benchmarks containing charged systems, marked with an asterisk in Fig. 1. The alignment is also strong with regards to RMSD values and in terms of percentage errors. This robustness is important for our analysis: Outliers indicate that something is not covered by the logic and our aim is to draw general conclusions about the logic in consistent vdW-DF versions [9].

The results for the WATER27 benchmark set of the GMTKN55 [35] are, however, an exception with over one kcal/mol difference on a total 3.5 kcal/mol MAD value reported in Ref. 35. We expect that we can improve our Quantum Espresso benchmarking for water with a better pseudopotential choice, and that may hold for plane-wave benchmarking in both revPBE-D3 and vdW-DF-cx. The net result on the performance statistics of any one such benchmark is very low.

The GMTKN55 suite also has other benchmarks. There are group 1, 2, and 3 of benchmarks for small-molecule properties, for large-molecule properties, and for transition barriers, respectively [35]. Here the differences between our plane-wave benchmarking and the quantum-chemistry approach of Ref. 35 are larger for revPBE-D3, but not out of proportion to the overall deviations from the reference energies that exist here. The differences in MAD values for the revPBE-D3 benchmarking are generally well within 1 kcal/mol. The differences are, however, about 2 kcal/mol for the MB16 mindless-benchmarking set and for the DC13 benchmark set of difficult-for-DFT tests [35].

Interestingly, we find that we are 2 kcal/mol closer to reference values for the W4-11 set of atomizations energies than what is reported for revPBE-D3 in Ref. 35. However, for vdW-DF-cx we find a MAD value of almost 11 kcal/mol, one of the few benchmarks where vdW-DF-cx fares substantially worse than revPBE-D3. Other benchmarks were vdW-DF-cx generally fares worse than revPBE-D3 are in the GMTKN55 group 3. This will be analyzed in a separate paper.

We observe that bad plane-wave performance (of vdW-DF-cx/revPBE-D3) does result in some charged-system benchmarking problems. It also results when Quantum Espresso finds a spin configuration that is frustrated, for example, in triangular molecules like $ClO_2$ and even in the linear CCH system. Some molecules are known to require a non-collinear treatment of spin [44, 243], but we lack access for vdW-DF and vdW-DF-cx. In our molecular benchmarking we do make sure that we provide the right value of the total molecular spin $M_{\mathrm{tot}}$, as given in the reference data [35]. That is, denoting the spin-

up(-down) density components as $n_{\uparrow(\downarrow)}(\mathbf{r})$, we enforce

$$M_{\rm tot} = \int d\,\mathbf{r}[n_\uparrow(\mathbf{r}) - n_\downarrow(\mathbf{r})]\,, \qquad (123)$$

while also converging the orbital occupation. However, in some molecular problems we find non-integer values of the so-called absolute magnetic moment

$$M_{\rm abs} = \int d\,\mathbf{r}|n_\uparrow(\mathbf{r}) - n_\downarrow(\mathbf{r})|\,. \qquad (124)$$

A non-integer value of $M_{\rm abs}$ arises, in our quantum-espresso characterization of vdW-DF-cx performance, for $ClO_2$ and CCH (part of the W4-11 atomization benchmark set), for $C_4H_4$ and a $C_7H_8$ case (in the RC21 benchmark set), for $C_3H_2N$, $C_3H_5$, $C_2H_2N$, $C_3H_3$, and $C_7H_7$ (in the RSE43 benchmark set), as well as in a $C_3H_7$ transition state. A non-integer $M_{\rm abs}$ value also arises in the charged $CO^+$, $CS^+$, $NH^-$, $O^-$, and $O_2^-$ states, affecting our plane-wave assessment of performance for the electron-affinity (G21EA) and electron-ionization (G21IP) benchmark sets, themselves part of group 1 of the GMTKN55 benchmark suit [35].

Nevertheless, we report our Quantum Espresso survey of vdW-DF-cx performance not only for all the noncovalent-interaction benchmarks, Fig. 1, but across all of the GMTKN55 groups, below. We do that noting that the present GMTKN55-type [35, 267] survey reflects what is presently available to a user that aims to pursue (plane-wave) vdW-DF-cx calculations.

Also, since we aim to check the overall robustness, accuracy and the general-purpose capability, we are in any case interested in statistically averaged values. We worry about outliers that we do not understand but we can accept specific issues. This is so because the percentage differences between MAD values in the plane-wave [36, 219] versus quantum-chemistry [35] benchmarking remain small in essentially all cases, the W4-11 atomization energies, MB13 and DC13 being exceptions. It is nice to see that there is perhaps a correlation between performance hits and known limitations in the calculation platform that is available to us. However, we live with the fact that we could have put a different, and perhaps even a better, foot forward had vdW-DF-cx already been implemented in codes that focus on descriptions of molecules.

### B. Nature of interactions in low-dimensional systems

All of the following studies are provided using the GBRV ultrasoft pseudopotentials [268] with a 50-Ry wavefunction cutoff.

For our study of the binding in the graphene bilayer, Fig. 8, we first relaxed the atomic coordinates using a slab setup and a $20\times20\times1$ $k$-point sampling. The slab height is 50 Å to ensure amble vacuum and to eliminate coupling between periodic images. From the fully relaxed electron density variation we subsequently extract the nonlocal-correlation contributions, relying on our ppACF code [14, 15]. This code is designed to split the total system energy $E_{\rm vdW-DF-cx}$ into the range of XC-energy (and other) components, using also a coupling-constant scaling of vdW-DF-cx [14]. In anticipation of this benchmarking and analysis effort, we designed ppACF with an ability to also read in new universal-kernel components, for which Appendix D provides details of the evaluation.

Our ppACF code is released into the Quantum Espresso post-processing tool set [36, 219].

The study of the nature of binding among carbyne wires, Fig. 9, is performed with vdW-DF-cx for a set of unit cells of dimensions $l_{C-C}\times$ 50 Å $\times$ 50 Å using an $80\times1\times1$ $k$-point sampling. The value of $l_{C-C}$ is varied for convergence and we find no dimerization for vdW-DF-cx at any pressure. Consistent-exchange vdW-DF-cx describes the carbyne wires as metallic.

For a study of the nature and non-additivity of interactions in the fullerene dimer & trimer, we use Gamma-point calculations. For a similar characterization of the nanotube-trimer bundle we use a $k$-point mesh grids of size $16\times1\times1$. Again we use ppACF to extract the spatial variation in the nonlocal-correlation energies for each specific system. We then compute the relevant binding-energy differences with the Quantum Espresso post-processing tool set [36, 219] following the procedure introduced in Ref. 14.

### C. Benchmarking bulk and surface properties

Our surface-energy results are extracted by computing and comparing the energies of slabs with thickness varying from 4 to 12 layers, extracting an average as in Ref. 223. The work functions were computed from slabs of 8 layers by tracking the electrostatic potential outside the surfaces. All slab systems were first allowed full relaxations as described by the XC functional in question.

For tests of bulk-system properties we sample the Brillouin zone using Gamma-centered k-mesh grids of size $16\times16\times16$, whereas for surface-slab systems we use $16\times16\times1$. We use the set of GBRV ultrasoft pseudopotentials [268] with a 50-Ry wavefunction cutoff.

For the metal and MgO slab geometries we add 15 Å of vacuum to compensate for possible Coulombic interaction between the actual surface and its periodic image. We also use a dipole corrections [269] to compensate for polar effects. The 15 Å vacuum region allows us to track the electrostatic potential and thus extract work-function values [181, 222, 223, 225].

We compute and extract the surface-energy and work function values for vdW-DF-cx, PBE, and PBEsol in Quantum Espresso. Some SIESTA- [270] and VASP-based [217, 218] vdW-DF-cx results for metal-surface work functions and surface energies have already appeared [224–226]. However, here we provide a broader

TABLE I. Comparison of PBE [28], PBE0 [261], PBE-XDM [77, 78], PBE-D3 [33], PBE-TS [80], PBE-TS-SCS [81], and vdW-DF-cx [9] (here abbreviated CX) performance for dispersion-dominated inter- and intra-molecular interactions. The table summarizes performance that is characterized both in terms of MAD and (when available) the mean absolute percentage deviation (MAPD) values. The comparisons of functional performance are obtained, mostly in plane-wave DFT, both at reference geometries (as denoted with a '-refG' hyphen) and, when available, with relaxations native to the functional (as denoted with a '-minG' hyphen). We include MAD and MAPD listings for the so-called new-S22 and intramolecular dispersion (IDISP) subsets of GMTKN55 [35], as well as for the L7 [262], A24 [263], and the so-called 'Blind set' [264]. We have also included a comparison defined by theory accuracy in matching low-temperature measurements for structure and cohesion in oligoacene crystals [98]. Cohesion-energy measurements for the oligoacenes are limited to the benzene, naphthalene, and anthracene crystals; We report MAD values that are averaged over these. Structure characterizations, extracted in the low-temperature limit, are available for oligoacenes up to hexacenes [98]; Here we report MAD and MAPD values that are obtained by averaging deviations between computed results and measurements of the $a$, $b$, and $c$ lattice constants [98]. In the benchmarking summary for the IDISP set at relaxed coordinates, we have omitted the $C_{22}H_{46}$ case, as a more plausible atomic configuration were obtained with full vdW-DF-cx relaxations. This alkane-unfolding problem deserves a separate discussion, see text.

| Test Set | Measure | PBE | PBE0 | PBE-XDM | PBE-D3 | PBE-TS | PBE-TS-SCS | CX |
|---|---|---|---|---|---|---|---|---|
| S22-refG $E_b$ | MAD (kcal/mol) | $2.55^a/2.72^{b*}$ | $2.37^a/2.54^{b*}$ | $0.59^b$ | $0.48^a/0.54^b$ | $0.34^b$ | - | $0.47^b$ |
| | MAPD (%) | $57^{b*}$ | $55^{b*}$ | $9.8^b$ | $11.6^b$ | $10.0^b/9.2^c$ | $5.4^c$ | $9.0^b$ |
| S22-minG $E_b$ | MAD (kcal/mol) | - | - | $0.54^b$ | $0.57^b$ | $0.49^b$ | - | $0.68^b$ |
| | MAPD (%) | - | - | $8.4^b$ | $11.1^b$ | $12.8^b$ | - | $9.4^b$ |
| S22-minG $d$ | MAD (Å) | - | - | $0.075^b$ | $0.073^b$ | $0.040^b$ | - | $0.068^b$ |
| IDISP-refG $E_b$ | MAD (kcal/mol) | $10.78^a$ | $9.51^a/9.8^b$ | - | $2.76^a$ | - | - | $2.4^b$ |
| | MAPD (%) | - | $242^b$ | - | - | - | - | $32^b$ |
| IDISP-minG $E_b$ | MAD (kcal/mol) | - | - | - | - | - | - | $2.7^b$ |
| | MAPD (%) | - | - | - | - | - | - | $30^b$ |
| L7-refG $E_b$ | MAD (kcal/mol) | $18.6^d$ | - | - | - | $2.92^{d,d*}$ | $3.79^{d,d*}$ | $1.58^{d,d*}$ |
| | MAPD (%) | $120^d$ | - | - | - | $17^d$ | $21^d$ | $14^d$ |
| A24-refG $E_b$ | MAD (kcal/mol) | $0.40^d$ | - | - | - | $0.35^d$ | $0.29^d$ | $0.14^d$ |
| | MAPD (%) | $42^d$ | - | - | - | $28^d$ | $22^d$ | $11^d$ |
| 'Blind set' $E_b$ | MAD (kcal/mol) | $3.00^d$ | - | - | - | $0.65^d$ | $0.79^d$ | $0.54^d$ |
| | MAPD (%) | $52^d$ | - | - | - | $11^d$ | $13^d$ | $10^d$ |
| Acenes-minG $E_b$ | MAD (kcal/mol) | $15.2^e$ | - | - | - | $5.1^{e,f}$ | - | $2.3^e$ |
| Acenes-minG $\langle a,b,c\rangle$ | MAD (Å) | $0.76^e$ | - | - | - | $0.09^{e,f}$ | - | $0.06^e$ |
| | MAPD (%) | $9^e$ | - | - | - | $1^{e,f}$ | - | $1^e$ |

[a] Ref. 35.

[b] Ref. 15.

[b*] Extracted from the computational data that we also used for Ref. 15.

[c] Ref. 81.

[d] Ref. 124.

[d*] With DLPNO-CCSD(T) reference data [265], these calculations [124] yield a 1.39/2.25/1.51 MAE for TS/TS-SCS/CX.

[e] Ref. 98.

[f] Ref. 266.

vdW-DF-cx (and PBE/PBEsol) survey. This gives us data to report a performance characterization based on the metal-surface benchmarking that is implicitly suggested in Ref. 223. We find essentially identical results using the plane-wave code Quantum Espresso [36, 219] for the PBE and PBEsol studies that we repeated here (checking for, but finding no code dependences). We include also these calculations to highlight that we provide a consistent comparison in our assessment of the vdW-DF-cx performance.

### D. Understanding binding

To understand how vdW-DF-cx works for dense (and sparse) matter, we compare vdW-DF-cx and PBE descriptions of XC-energy contributions to binding in bulk. We also provide this comparison for two fullerene molecules, while we refer to Ref. 14 for a corresponding comparison and analysis for smaller molecules.

We base this comparison and analysis on calculations of the energy-per-particle variation [14, 55, 57, 58]. The standard energy-per-particle representations of semilocal XC functionals is given by Eq. (61), while the general-ization Eq. (62) applies to all investigated functionals, including our truly nonlocal vdW-DF-cx.

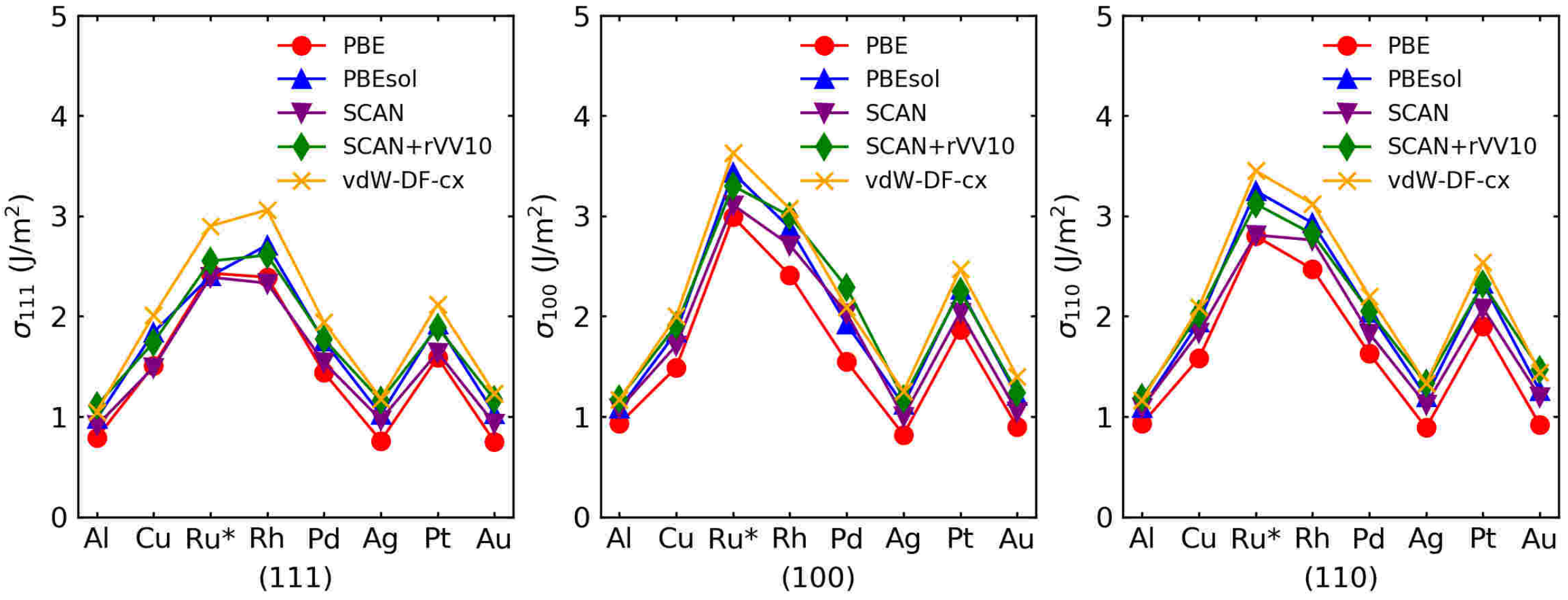

FIG. 11. Surface energies ($\bar{\sigma}$) of (111), (100), and (110) metal surfaces, as obtained in different functionals. We also compare our calculations for PBE [28], PBEsol [29], and vdW-DF-cx [9] with the SCAN [258] and SCAN+rVV10 [69] (that is, rVV10-corrected [66–68] SCAN) results reported in Ref. 223. As in that SCAN+rVV10 study, the high-temperature fcc (and not hcp) structure, labeled Ru*, is used for a computational study of Ru surface energies.

The general energy-per-particle energy variation $\epsilon^0_{\rm xc}[n](\mathbf{r})$ can be further split into exchange and correlation components. The spatial variation in the LDA/PBE correlation energy densities is simply $e_{\rm c}^{\rm LDA/PBE}(\mathbf{r}) = n(\mathbf{r})\epsilon_{\rm c}[n](\mathbf{r})$. For vdW-DF-cx it is natural to look at the nonlocal-correlation energy density [14, 55, 57, 58]:

$$\begin{aligned} e_{\rm c}^{\rm nl}(\mathbf{r}) &= \frac{1}{2}\int d\mathbf{r}'\, n(\mathbf{r})\,\phi(\mathbf{r},\mathbf{r}')\,n(\mathbf{r}') \\ &= e^{\rm nl}_{{\rm c},\alpha}(\mathbf{r}) + e^{\rm nl}_{\rm c,vdW}(\mathbf{r})\,. \end{aligned} \tag{125}$$

This vdW-DF-cx [9] characterization should be compared with the nonlocal, that is gradient-corrected, PBE correlation

$$e_{\rm c}^{\rm PBE,nl}(\mathbf{r}) \equiv e_{\rm c}^{\rm PBE}(\mathbf{r}) - e_{\rm c}^{\rm LDA}(\mathbf{r})\,. \tag{126}$$

It is also natural to track the spatial variation in the total nonlocal parts of the vdW-DF-cx and PBE XC energies, exploring differences in the nonlocal XC functional components

$$e_{\rm xc}^{\rm cx,nl}(\mathbf{r}) \equiv e_{\rm xc}^{\rm cx}(\mathbf{r}) - e_{\rm xc}^{\rm LDA}(\mathbf{r})\,, \tag{127}$$

$$e_{\rm xc}^{\rm PBE,nl}(\mathbf{r}) \equiv e_{\rm xc}^{\rm PBE}(\mathbf{r}) - e_{\rm xc}^{\rm LDA}(\mathbf{r})\,. \tag{128}$$

The relevant quantity for our discussion is the spatial variation in the binding contributions, for example,

$$\Delta e_{\rm c}^{\rm nl}(\mathbf{r}) = e_{\rm c}^{\rm nl,A}(\mathbf{r}) + e_{\rm c}^{\rm nl,B}(\mathbf{r}) - e_{\rm c}^{\rm nl,AB}(\mathbf{r})\,. \tag{129}$$

The integrals of these spatially resolved contributions yield binding-energy contributions like Eq. (122).

As we use the vdW-DF-cx version, we can also track the binding contributions arising from the (nonlocal-correlation part of) vertex corrections,

$$\Delta e_{{\rm c},\alpha}^{\rm nl}(\mathbf{r}) = e_{{\rm c},\alpha}^{\rm nl,A}(\mathbf{r}) + e_{{\rm c},\alpha}^{\rm nl,B}(\mathbf{r}) - e_{{\rm c},\alpha}^{\rm nl,AB}(\mathbf{r})\,. \tag{130}$$

A similar definition also allows us to track binding contributions arising from pure vdW interactions, i.e., given by $e^{\rm nl}_{\rm c,vdW}(\mathbf{r})$.

This set of analysis tools is available through our PPACF code [14, 15] and through the post-processing components of Quantum Espresso [36, 219]. We used these to map and contrast the nature binding of PBE and vdW-DF-cx for bulk Si and in metals as well as for a fullerene dimer.

For an analysis of the nature of binding in bulk, we first re-compute the electron density variation using Troullier-Martins norm-conserving pseudo potentials of the ABINIT package [38] with a 80 Ry wavefunction cutoff; For W bulk we also increase the $k$-point mesh to $24 \times 24 \times 24$. Furthermore, W is a heavy element with large gradients in the electron distribution around the individual atoms and we compute the electron-density variation of the atoms using an 800 Ry density cutoff. We apply a cubic-spline fit to the spatial variations of the XC energy components that we extract for each of the atoms. The high-accuracy characterizations of both bulk and atoms are finally used to determine binding contributions, using Eqs. (129) and (130) or using corresponding expressions for LDA and PBE characterizations. This set of steps reduces the numerical noise in the resulting mapping of vdW and other screening effects in bulk-crystal cohesion, below.

## VI. VALIDATION CHECKS ON VDW-DF-CX

The vdW-DF-cx compliance with the underlying screening logic of the ACF and current and with charge conservation is promising. We can expect that the consistent vdW-DF versions will be robust and likely transfer-able. However, usefulness as a general-purpose functional will only follow if vdW-DF-cx (and by extension consistent vdW-DF versions) continue to perform well in tests. Below we summarize some of those tests and we extend the testing to a broad set of molecular benchmarkings [35] and to metal surfaces [223].

As the vdW-DF-cx passes validation checks, we also propose to use the vdW-DF-cx Lindhard-screening logic to map interaction details, as illustrated below.

### A. All-round performance of consistent vdW-DF.

There are by now many studies for bulk, layered systems, and organics that document that vdW-DF-cx is robust and transferable, for example, Refs. 68, 95, 98, 99, 103, 107, and 113. Below we provide a brief overview of the vdW-DF-cx performance. We limit ourselves to a summary because we also want to provide specific validation checks on vdW-DF-cx ability to capture semilocal-correlation that plays an important role in the description of metal surfaces [29, 223]. Moreover, we want to illustrate the unique advantage that vdW-DF-cx calculations have in terms of mapping cumulant (screening and vertex-correction) effects in material binding.

We begin with a discussion of the asymptotic vdW interactions among atoms and molecules. The original vdW-DF1 [4] (and thus vdW-DF-cx [9]) nonlocal-correlation term leads to a good description of asymptotic $C_6$ vdW interaction coefficients for many atoms [9, 271]. The same is true for some molecular problems [210, 211], but not for hollow structures [105, 252], where screening causes dramatic changes in the lowest-energy collective mode [252]. A similar effect exists for metallic systems (including graphene sheets and conducting nanotubes) where the extended collective response causes fundamental changes in the nature of the interaction, again at asymptotic separations [202, 246, 251]. However, the asymptotic description is not the focus of the present implementations of the vdW-DF method [10].

Next we return to the broad survey of the vdW-DF-cx performance for molecular interactions, using the GMTKN55 suite of 55 benchmark sets on molecular properties [35]. The reference data for structure (including the molecular spin state) and energies (as well as per-functional/per-benchmark performance results for DFT-D3 versions) are available [267] and allow for an effective way of testing (spin) vdW-DF-cx relative to dispersion-corrected DFT. For individual benchmark sets, like G21IP, G2RC, S22, Al2X6, DARC, and Al2X6 of the GMTKN55 collection, we have previously compared the vdW-DF-cx performance relative to that of hybrids, Refs. 13 and 15. In those studies we checked if the good vdW-DF-cx performance arose from a lucky cancellation by also tracking the effects of relaxations (computed in vdW-DF-cx). Here instead, we safeguard the testing by going for the full benchmark suite [35].

To accomplish this we have written python scripts to download the data and submit Quantum Espresso calculations for the molecular problems, automatically implementing the computational strategy described in Sec. V. There are in total 6 GMTKN55 benchmark groups. Group 1 probes small-molecule properties (half of which involve charged systems), Group 2 probes large-molecule properties, Group 3 concerns barriers for molecular reactions, while Group 4 and Group 5 test intra- and inter-molecular noncovalent interactions, respectively. Group 6 is just the union of group 4 and 5. For each of the GMTKN55 benchmark groups, the scripts also compute the performance in the individual sets and extract values for the MAD, RMSD, and the mean absolute percentage deviation (MAPD) relative to the CCSD(T) reference data on process energies [35, 267]. We extract a weighted MAD average [35]

$$\mathrm{WTMAD} \equiv \frac{1}{55}\sum_{i=1}^{55} w_i \times \mathrm{MAD}_i \,. \tag{131}$$

The summation runs over the 55 different benchmark sets that make up the GMTKN55, with $\mathrm{MAD}_i$ denoting the individual-benchmark MAD value. The weighting factor $w_i$ are set to align benchmark with similar values of the reference energy values for the process [35]. The weighting is such that having 1 kcal/mol deviation on the weaker noncovalent interactions can cause as much as 100 times the WTMAD change that arises in benchmarks with large characteristic process energies. We furthermore extract and report a similar WTMAD value for each of the GMTKN55 benchmark groups [35].

In Fig. 1 we have already reported details of this benchmarking for all of the Group-6 benchmarks on noncovalent interactions in molecular systems. There we show the per-benchmark set MAD and RMSD values and contrast the vdW-DF-cx, vdW-DF, and revPBE-D3 performance. The vdW-DF-cx outperforms the original vdW-DF version and performs slightly better than revPBE-D3 on average. As discussed in the introduction, this fact is an argument for balancing the exchange and correlation terms using the vdW-DF screening logic. It is a motivation for using (spin) vdW-DF-cx.

Fig. 10 reports our full GMTKN55 survey, contrasting WTMAD values obtained from a full Quantum Espresso characterization in PBE, revPBE-D3, and vdW-DF-cx. We find that vdW-DF-cx performs significantly better than PBE, and at least matches revPBE-D3 performance on all but the Group 3 class of barrier problems, when compared on our plane-wave/pseudopotential platform. The comparison with revPBE-D3 is relevant because it is one of the two best-performing DFT-D3s that have a GGA-level exchange. We also note that on average vdW-DF-cx performs almost as good as SCAN-D3 [33, 258] even if vdW-DF-cx lacks a meta-GGA formulation of exchange.

As discussed in Sec. V, care should be taken in making the vdW-DF-cx comparison with these literature values for SCAN-D3, obtained in the orbital-based code, Turbomole [259, 260]. We are then comparing with results obtained in a different, non-plane-wave DFT framework. Some care must also be made when making comparison between our vdW-DF-cx and revPBE-D3 characterizations for Group 3, for Group 2, and, perhaps especially, for Group 1. This is because there are some small residual differences between what we obtain as the revPBE-D3 performance characteristics and what was reported in Ref. 35. The reason could in part be our representation of the molecular spin, as discussed in Sec. V. On the other hand, Fig. 10 does provide a present-day plane-wave (Quantum Espresso) user with an impression of what performance level is already available. We find that vdW-DF-cx is a realistic option even for a pure molecular problem.

TABLE II. Comparison of PBE [28], PBE0 [261], PBEsol [29], and vdW-DF-cx [9] performance (lattice constants, bulk moduli, and atomization energies) for ionic crystals and a few alkali metals. The reference values, denoted 'Exp.', are zero-point corrected experimental values, Ref. 29 and 272. Refs. 15, 45, 95, and 105 detail the vdW-DF-cx ability to characterize bulk-structure properties for other systems (graphene bilayers, $C_{60}$ crystals, Si and W bulk, and other metals) that are also further analyzed in this review.

| Bulk | property | PBE | PBE0 | vdW-DF-cx | Exp. |
|---|---|---|---|---|---|
| LiF | $a_0$ (Å) | 4.005 | 3.952 | 3.976 | 3.972 |
| | $E_a$ (eV/atom) | 4.289 | 4.224 | 4.417 | 4.46 |
| | $B_0$ (GPa) | 66.79 | 75.02 | 69.09 | 76.3 |
| NaCl | $a_0$ (Å) | 5.658 | 5.619 | 5.605 | 5.569 |
| | $E_a$ (eV/atom) | 3.146 | 3.106 | 3.270 | 3.34 |
| | $B_0$ (GPa) | 22.92 | 24.60 | 23.94 | 27.6 |
| MgO | $a_0$ (Å) | 4.227 | 4.186 | 4.202 | 4.189 |
| | $E_a$ (eV/atom) | 4.928 | 4.913 | 5.177 | 5.203 |
| | $B_0$ (GPa) | 147.8 | 164.6 | 155.5 | 169.8 |
| Li | $a_0$ (Å) | 3.405 | 3.429 | 3.407 | 3.453 |
| | $E_a$ (eV/atom) | 1.555 | 1.492 | 1.610 | 1.658 |
| | $B_0$ (GPa) | 13.6 | 13.7 | 13.3 | 13.9 |
| Na | $a_0$ (Å) | 4.196 | 4.237 | 4.183 | 4.214 |
| | $E_a$ (eV/atom) | 1.040 | 0.983 | 1.078 | 1.119 |
| | $B_0$ (GPa) | 7.6 | 7.3 | 8.3 | 7.9 |

We also find (not shown) a small but finite performance advantage of vdW-DF2-b86r [43] over vdW-DF-cx [9] for the broad molecular benchmarking that the GMTKN55 suite offers [35, 267]. The vdW-DF2-b86r (or rev-vdW-DF2) is a variant that Hamada launched to make vdW-DF2 applicable also for solids and general materials. It also has a GGA-level exchange and it is also highly successful. Completing a vdW-DF2-b86r survey analogous to Fig. 10, we provide the following vdW-DF-cx/vdW-DF2-b86r contrast of WTMAD performances (listed in kcal mol$^{-1}$): 5.15/5.18 (Group 1), 4.93/4.81 (Group 2), 7.25/6.68 (Group 3), 3.14/3.22 (Group 4), 3.86/3.61 (Group 5), 3.47/3.39 (Group 6), and 4.74/4.62 (Total GMTKN55).

We interpret this vdW-DF2-b86r advantage as a sign that the class of consistent vdW-DF versions has room to grow also beyond vdW-DF-cx. This is true even before going to a possible, future meta-vdW-DF formulation or a hybrid vdW-DF [13, 15]. We recall that for many problems the materials properties are often set by the low-to-medium values of the scaled density gradient $s$. In such cases, the vdW-DF-cx exchange description effectively resembles that of the Langreth-Vosko analysis [91], which is not ideal for molecules [92]. The vdW-DF-cx remains a relevant choice because it stands out in being consistent, in having a regular screening logic for balancing exchange and correlation. However, we can hope to eventually obtain a consistent vdW-DF design that has both a good exchange (as vdW-DF2-b86r has) and a good XC balance.

Table I summarizes performance comparisons in a broader context for the description of noncovalent interactions both within and between molecules. The table reflects systematic studies reported in Refs. 9, 13, 15, 98, 124, and 266. It includes a more detailed listing of the GMTKN55 comparison concerning both the S22 and the IDISP [35], and adds an overview of a benchmarking for the L7, A24, and so-called blind test cases [262–264]. Overall, the vdW-DF-cx is found to compete well with the class of dispersion-corrected DFT descriptions [33, 76, 78, 80, 81].

For the G1-set of molecular atomization energies [237, 238] the vdW-DF-cx version does perform worse than PBE (giving a MAD value of 9.7 kcal/mol as opposed to the PBE 7.5 kcal/mol, for fully relaxed structures) [13]. We ascribe this issue, in part, to the fact that the vdW-DF-cx elimination of the vdW-DF cross-over term $\delta E_{\mathrm{x}}^0$ is not perfect for atomic density profiles [9, 10]. We also note that the performance picture for the G1 set is reversed once we move to corresponding hybrid formulations, PBE0 and vdW-DF-cx0/-cx0p [13, 15].

Table I furthermore provides both direct and indirect evidence that vdW-DF-cx is accurate in its characterization of both the inter- and intra-molecular structure. Direct evidence is available from the excellent agreement between vdW-DF-cx results for lattice parameters (and to a lesser extent, for the fully relaxed binding energies) for the set of oligoacene molecular crystals [98]. The accuracy also extends to the characterization of the electron-density variation, as evident by the vdW-DF-cx ability to predict the soft intermolecular libration and rocking modes in the naphthalene [99], Fig. 5, and in the polyethylene [107] crystals.

Indirect evidence for vdW-DF-cx accuracy is also evident in Table I by looking at the effects of structural relaxations on the vdW-DF-cx performance on molecular binding/reaction energies. These results, at minimized geometries, are listed by the 'minG' entries (where available). For the tests on noncovalent and covalent binding properties (in the AL2X6, DARC, G2RC, G21IP subsets of GMTKN55 [35],) we find no significant changes in performance from that asserted at reference geometries (listed under 'refG'). There are small effects (improve-ments) on the binding-energy (and binding-separation) accuracy for the soft S22 [15].

TABLE III. Surface energies (in J/m$^2$) of the selected metals and MgO. For metals, the calculated surface energies are averaged over (111), (100), and (110) surfaces. Four- to twelve-layer slabs were used in the linear fit for the surface energy, as in Ref. 223. The headings 'SCVV' and 'CX' are abreviations for SCAN+rVV10 and vdW-DF-cx, respectively. We focus on the high-temperature fcc phase of Ru (as marked by an asterix), but also list results for the low-temperature hcp phase. We also report MAD and MAPD values obtained for the set of metal surfaces, for comparison with Ref. 223. For MgO, we concentrate on the nonpolar (110) surface.

| | PBE | PBEsol | SCVV$^a$ | CX | Exp. |
|---|---|---|---|---|---|
| Al | 0.88 | 1.05 | 1.16 | 1.13 | 1.14±0.2$^b$ |
| Cu | 1.52 | 1.87 | 1.89 | 2.03 | 1.79±0.19$^b$ |
| Ru*(fcc) | 2.74 | 3.02 | 2.99 | 3.32 | 3.04±0.33$^b$ |
| Ru(hcp) | 2.98$^c$ | 3.44$^c$ | – | 3.61$^c$ | – |
| Rh | 2.42 | 2.84 | 2.81 | 3.08 | 2.66±0.29$^b$ |
| Pd | 1.54 | 1.90 | 2.04 | 2.08 | 2.00±0.22$^b$ |
| Ag | 0.82 | 1.11 | 1.22 | 1.26 | 1.25±0.13$^b$ |
| Pt | 1.79 | 2.17 | 2.15 | 2.38 | 2.49±0.26$^b$ |
| Au | 0.86 | 1.17 | 1.29 | 1.36 | 1.51±0.16$^b$ |
| MAD | 0.41 | 0.16 | 0.12 | 0.16 | |
| MAPD (%) | 23 | 9 | 6 | 7 | |
| MgO | 0.89$^d$ | 1.01$^d$ | – | 1.18$^d$ | 1.04$^e$ |

$^a$ Ref. 223.
$^b$ From liquid metal surface tensions, Refs. 273 and 274.
$^c$ Surface energies are here averaged over the (0001), $(11\bar{2}1)$, $(10\bar{1}0)$, and $(11\bar{2}0)$ surfaces of hcp Ru.
$^d$ Nonpolar MgO(110) surface only.
$^e$ Ref. 275.

TABLE IV. Work functions $\phi$ (eV) for the selected metals. These values were extracted by tracking the electrostatic field outside fully relaxed slabs having 8 atomic layers. The headings 'SCVV' and 'CX' are abreviations for SCAN+rVV10 and vdW-DF-cx, respectively.

| Surface | PBE | PBEsol | SCVV$^a$ | CX | Exp.$^b$ |
|---|---|---|---|---|---|
| Al(111) | 4.10 | 4.18 | 4.23 | 4.26 | 4.32±0.06 |
| Al(100) | 4.19 | 4.25 | 4.42 | 4.29 | 4.32±0.06 |
| Al(110) | 4.15 | 4.18 | 4.00 | 4.24 | 4.23±0.13 |
| Cu(111) | 4.79 | 5.01 | 5.09 | 5.07 | 4.90±0.02 |
| Cu(100) | 4.51 | 4.63 | 4.54 | 4.69 | 4.73±0.1 |
| Cu(110) | 4.48 | 4.60 | 4.53 | 4.66 | 4.56±0.1 |
| Ru(11-21) | 4.48 | 4.61 | 4.65 | 4.66 | 4.71$^c$ |
| Ru(10-10) | 4.54 | 4.64 | 4.97 | 4.71 | 4.60±0.28 |
| Ru(11-20) | 4.36 | 4.46 | 4.72 | 4.54 | – |
| Ru(0001) | 4.95 | 5.10 | – | 5.17 | – |
| Ru*(111) | 5.08 | 5.22 | – | 5.29 | – |
| Ru*(100) | 4.89 | 5.03 | – | 5.09 | 5.40±0.11 |
| Ru*(110) | 4.46 | 4.56 | – | 4.64 | – |
| Rh(111) | 5.16 | 5.30 | 5.20 | 5.36 | 5.46±0.09 |
| Rh(100) | 5.08 | 5.22 | 5.37 | 5.29 | 5.30±0.15 |
| Rh(110) | 4.57 | 4.67 | 4.83 | 4.75 | 4.86±0.21 |
| Pd(111) | 5.26 | 5.38 | 5.47 | 5.45 | 5.67±0.12 |
| Pd(100) | 5.09 | 5.23 | 5.26 | 5.29 | 5.48±0.23 |
| Pd(110) | 4.72 | 4.84 | 5.09 | 4.91 | 5.07±0.2 |
| Ag(111) | 4.39 | 4.53 | 4.63 | 4.57 | 4.53±0.07 |
| Ag(100) | 4.22 | 4.38 | 4.37 | 4.43 | 4.36±0.05 |
| Ag(110) | 4.22 | 4.36 | 4.26 | 4.40 | 4.10±0.15 |
| Pt(111) | 5.68 | 5.80 | 5.97 | 5.87 | 5.91±0.08 |
| Pt(100) | 5.68 | 5.81 | 6.01 | 5.87 | 5.75±0.13 |
| Pt(110) | 5.31 | 5.42 | 5.36 | 5.51 | 5.53±0.13 |
| Au(111) | 5.05 | 5.20 | 5.41 | 5.24 | 5.33±0.06 |
| Au(100) | 5.09 | 5.18 | 5.28 | 5.26 | 5.22±0.31 |
| Au(110) | 5.03 | 5.11 | 5.30 | 5.20 | 5.16±0.22 |

$^a$ Ref. 223
$^b$ Ref. 276
$^c$ Ref. 277 for polycrystalline Ru is used as reference.

One of these benchmarks, namely the IDISP subset of the GMTKN55 [35], deserves a special discussion concerning the vdW-DF-cx structural characterization. Interestingly, there are no discernible relaxations for all but the $C_{22}H_{46}$ case of alkane unfolding in the IDISP set [15]. The consistent-exchange vdW-DF-cx version performs well also for describing the $C_{22}H_{46}$ unfolding energy at the reference geometry [15, 35]. However, at the reference geometry, the CCSD(T) result suggests an exothermal unfolding process, in contrast to the results of a fully relaxed vdW-DF-cx study [15]. Since an energy release for unfolding is in contrast with expectations from past investigations of alkane interactions [278], we have excluded the $C_{22}H_{46}$ case in our summary of relaxed-coordinate performances for the IDISP set in Table I.

The soundness and robustness of vdW-DF-cx for molecular relaxations can be further tested by analyzing the $C_{22}H_{46}$ unfolding case in detail, by instead comparing CCSD(T) and vdW-DF-cx energy results for fully relaxed vdW-DF-cx structures. As will be detailed elsewhere, this is done using the GMTKN55 procedure, described in Ref. 35, for the CCSD(T) calculations. The new CCSD(T) results reflect the expected endothermal behavior of $C_{22}H_{46}$ unfolding. In fact, at the vdW-DF-cx structure, the CCSD(T) result, 1.08 kcal/mol, for the unfolding energy cost aligns perfectly with the self-consistent, fully relaxed, vdW-DF-cx description, 1.06 kcal/mol.

A recent study provides a comparison of performance for extended system for many of nonlocal-correlation functionals [230], based on all-electron evaluations [231]. That study includes the full vdW-DF family of versions and variants as well as SCAN+rVV10 [69] and rVV10 [66, 67]. They find that vdW-DF-cx [9] and vdW-DF-C09 [41] comes out best statistically in the vdW-DF family for strongly bound bulk problems (close in performance to SCAN-rVV10), vdW-DF2-b86r [43] and vdW-

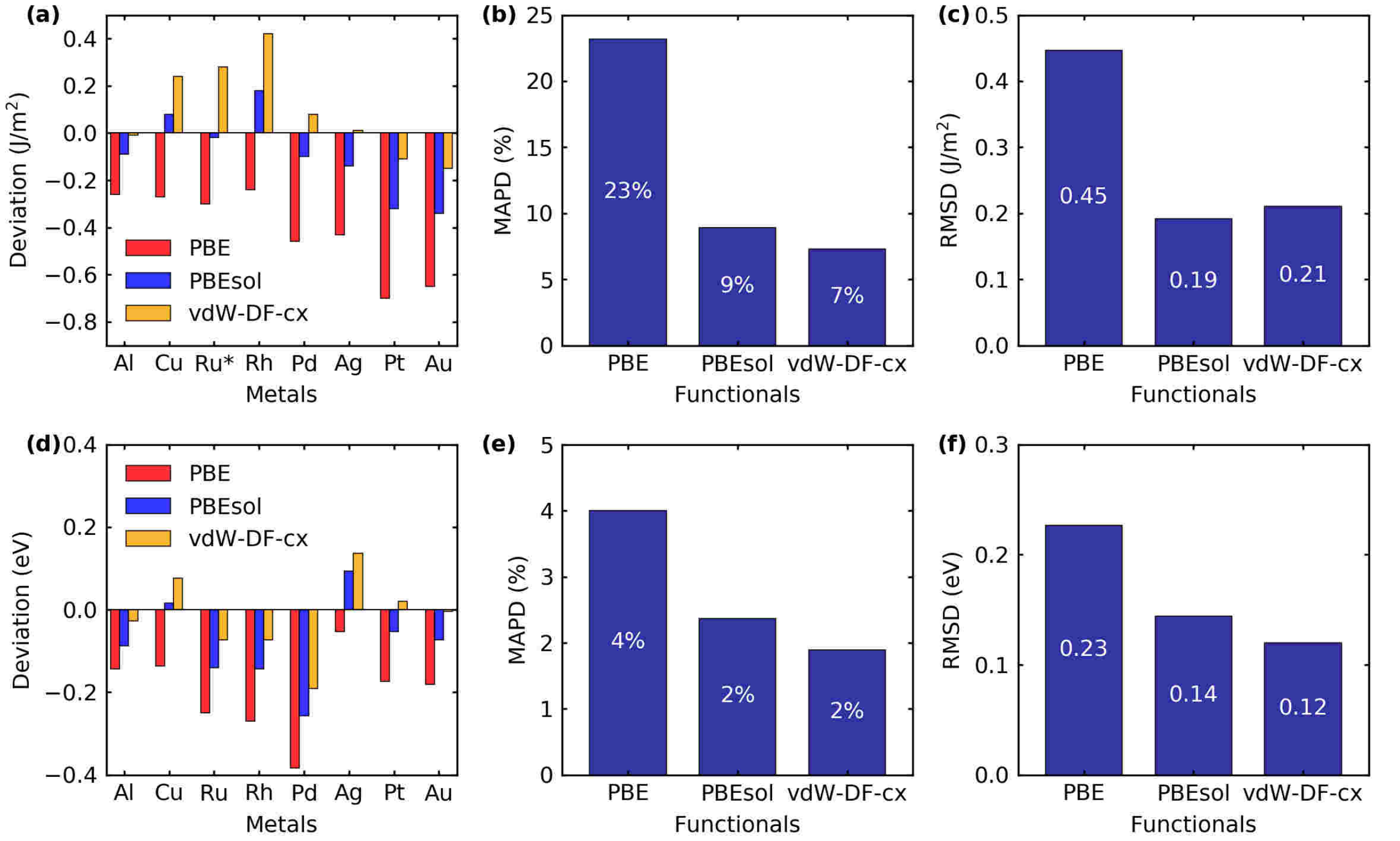

FIG. 12. Deviations of computed results from measurement values for mea surface energy ($\bar{\sigma}$) (top panels) and mean work functions ($\bar{\phi}$) (bottom panels) of the (111), (100), and (110) metal surfaces. The leftmost panels show the deviation from measured values (averaged over surface facets) while the middle and right panels show mean absolute percentage deviation (MAPD) and RMSD values for each of the here-characterized functionals. The error bars on the measured metal surface energies are listed in Table III. The average experimental uncertainty for surface energies is 0.22 J/m$^2$ or 12% for surface energies and 0.09 eV or 2.0 % for work functions [223].

DF-optb86r [42] are just ahead of vdW-DF-cx for layered solids, while vdW-DF2-b86r [42] is ahead of vdW-DF-cx for investigations of the set of molecular solids. The last point is consistent with the observation above that vdW-DF2-b86r has a small but finite performance advantage over vdW-DF-cx for molecules, and that it may be possible to further develop the class of consistent vdW-DF versions.

### B. Vertex corrections and surface properties

The real challenge for a nonlocal-correlation functional of the vdW-DF method, however, lies in the description of materials that have dense electron distributions. This follows because in these problems it is essential to describe subtle materials interactions as they occur in competition with one another [45]. For such systems we must balance both a) gradient-corrected exchange effects against truly nonlocal correlation effects and b) local as well as nonlocal vertex corrections in the response description.

Fortunately, the constraint-based PBE [28] and PBEsol [29] designs are believed (and proven) to be highly robust for materials with a dense electron distribution [30, 31]. As such, they open the door for another, computationally-efficient validation check on the robustness of vdW-DF-cx (and by extension, of the consistent vdW-DF formulations). These test is only available for dense-matter systems, like transitions-metal systems [95]. However, for such traditional systems, it is clear that vdW-DF-cx must keep up with the PBE/PBEsol performance to be trusted as a systematic extension. Most noticeable, we must check if vdW-DF-cx is as good for dense matter as PBE/PBEsol in avoiding the outliers in performance.

We mention that the meta-GGA SCAN functional [258] is also constraint-based and typically performs even better than PBE and PBEsol for dense matter, while the SCAN+rVV10 [69] fares well in comparison of broad extended-system performance [230]. A comparison of vdW-DF-cx with the SCAN/SCAN+rVV10 is certainly interesting. It is indirectly available in the following reporting but it is not highlighted in the discussion. This is because such a comparison is not a true assessment of soundness of present implementations of the consistent vdW-DF design logic: We repeat that with (spin) vdW-DF-cx we are starting with a GGA-level exchange description and have just enforced consequences of conservation laws and Lindhard screening in an electrodynamics formulation of the ACF. We seek comparisons with constraint-based GGA descriptions that have a GGA-level of exchange to test this idea of building from just one type of plasmons [9]. In summary, we do not here discuss a comparison of vdW-DF-cx and SCAN/SCAN+rVV10, but we are motivated to benchmark the vdW-DF-cx performance against experimental data while checking whether it is as good as PBE/PBEsol at avoiding truly bad characterizations.

In Ref. 95 one of us helped begin such validation work. The results are partly summarized in the lower panel of Fig. 5. This survey of thermo-physical properties has already been briefly discussed in Sec. IV.A and IV.D in the context of the accuracy of phonons, Fig. 5, and of the proper spin vdW-DF-cx extension. The thermo-physical properties study contrasts PBE, PBEsol, and vdW-DF-cx performance for the set of nonmagnetic transition metals. As mentioned above, the comparison stands out by treating vibrational ZPE and thermal effects through phonon calculations that are native to each of the semilocal and nonlocal functionals. The vdW-DF-cx is found to be highly accurate in comparison with experimental data. Moreover, it is found to have at least the same level of limited variation in the individual-system accuracy as do PBE and PBEsol [95].

We have already discussed spin vdW-DF-cx results for weak-chemisortion of graphene on Ni(111), Fig. 6. There, vdW-DF-cx provides results for the optimal adsorption distance that are closer to experimental observations [234] than what is obtained in RPA [235, 236]. Ref. 224 provides an additional comparison of vdW-DF-cx and of rVV10-corrected semilocal functionals, including PBEsol+rVV10 [72], with the corresponding RPA results for graphene adsorption on the (111) surfaces of Cu, Pt, Pd, and Ag [235, 236]. The vdW-DF-cx descriptions of the adsorption height are generally found in agreement with the RPA results for these surfaces [224]. The vdW-DF-cx results for the adsorption energy per carbon atom also follows the RPA trend, but the values for vdW-DF-cx (and for the set of investigated rVV10-corrected functionals) are found systematically smaller than the RPA results. Interesting, we have documented elsewhere that vdW-DF-cx, if anything, tends to overestimate the binding energy between two graphene sheets, as well as among other carbon nanostructures [12, 105].

The case of weak chemisorption on Ag(111) merits a separate discussion, since the RPA results [235, 236, 279] for the adsorption height do not line up with recent measurements [280]. Ref. 224 reports an adsorption height for vdW-DF-cx of 3.26 Å; Adapting our Ref. 11 study, we find here a vdW-DF-cx value of 3.24 Å for graphene on Ag(111). These vdW-DF-cx results are in fair agreement with the RPA value [235, 236], 3.31 Å, but not with the value 2.5 Å reported in Ref. 280.

Table II reports a comparison between the PBE, PBEsol, and vdW-DF-cx results for structure, cohesion, and elastic response in bulk in a set of ionic crystals, insulators, and alkali metals. The table also lists experimental results, back-corrected for ZPE and temperature effects [29, 272], to help us assert the accuracy and robustness of the functionals. This testing is included because we will provide a more detailed discussion of binding in some of these systems, later. The table supplements the performance comparison for semiconductors that is included in Ref. 15 and extends the transition-metal benchmarking [95].

Overall, the vdW-DF-cx seems to be extending the PBE [28] and PBEsol [29] strengths. It is useful for materials with a dense electron distribution [45, 95]. It brings this advantage to the much larger class of sparse problems [32], that is, systems that have important low-density regions.

The vdW-DF starting point, namely $E_{\mathrm{xc}}^{\mathrm{in}}$, retains all LDA-correlations but there are also relevant vertex corrections when the electron gas is weakly perturbed. An Occam's Razor argument led us to exclude these in $E_{\mathrm{xc}}^{\mathrm{in}}$ and let them emerge in the exponential re-summation that underpins the screening description. The approach is motivated, Appendix B, but the question remains: Does the cluster expansion retain enough nonlocal vertex corrections that vdW-DF-cx can work for traditional, dense matter (and not only for thermo-physical properties [95])?

We see the description of metal surface energies [273–275] and work functions, as summarized in Ref. 223, as an important test case. Also, a comparison of vdW-DF-cx and PBEsol performance is here a highly relevant assessment of the screening and vertex logic in the vdW-DF method. This follows because the gradient-corrected correlation in PBEsol itself is defined by a constraint-based fit to surface energies computed for a trusted model of metals [29].

Figure 11 compares our PBE, PBEsol, and vdW-DF-cx calculations of the surface energies for the (111), (100), and (110) surfaces of a range of metals. For comparison, we also include the computed variation of SCAN [258] and 'SCAN+rVV10' [69] (that is SCAN [258] combined with rVV10 [66, 67]) results, as obtained in Ref. 223 (with another plane-wave code but similar setup). For Ru surface energies, we focus on the fcc structure, denoted Ru*, as it constitutes the high-temperature face and therefore is relevant for the comparison with measurements [223]. All of the here-characterized functionals have the same general variation, although a consistent inclusion of vdW binding is expected to improve the description of the surface energies (and of surface-specific work functions) [223].

Table III reports a comparison of the performance relative to experimental observations, for the aforementioned metal set and for MgO. The experimental results are obtained by measuring the contact angle of molten droplets and have, as indicated, about a 10 percent uncertainty [273–275]. Except for MgO (where we focus on the nonpolar 110 surface), the computed surface energy is set by averaging over the facets identified in Fig. 11, as in Ref. 223. The performance of vdW-DF-cx is on par with that of SCAN+rVV10, both landing well within the error bars on the experimental results.

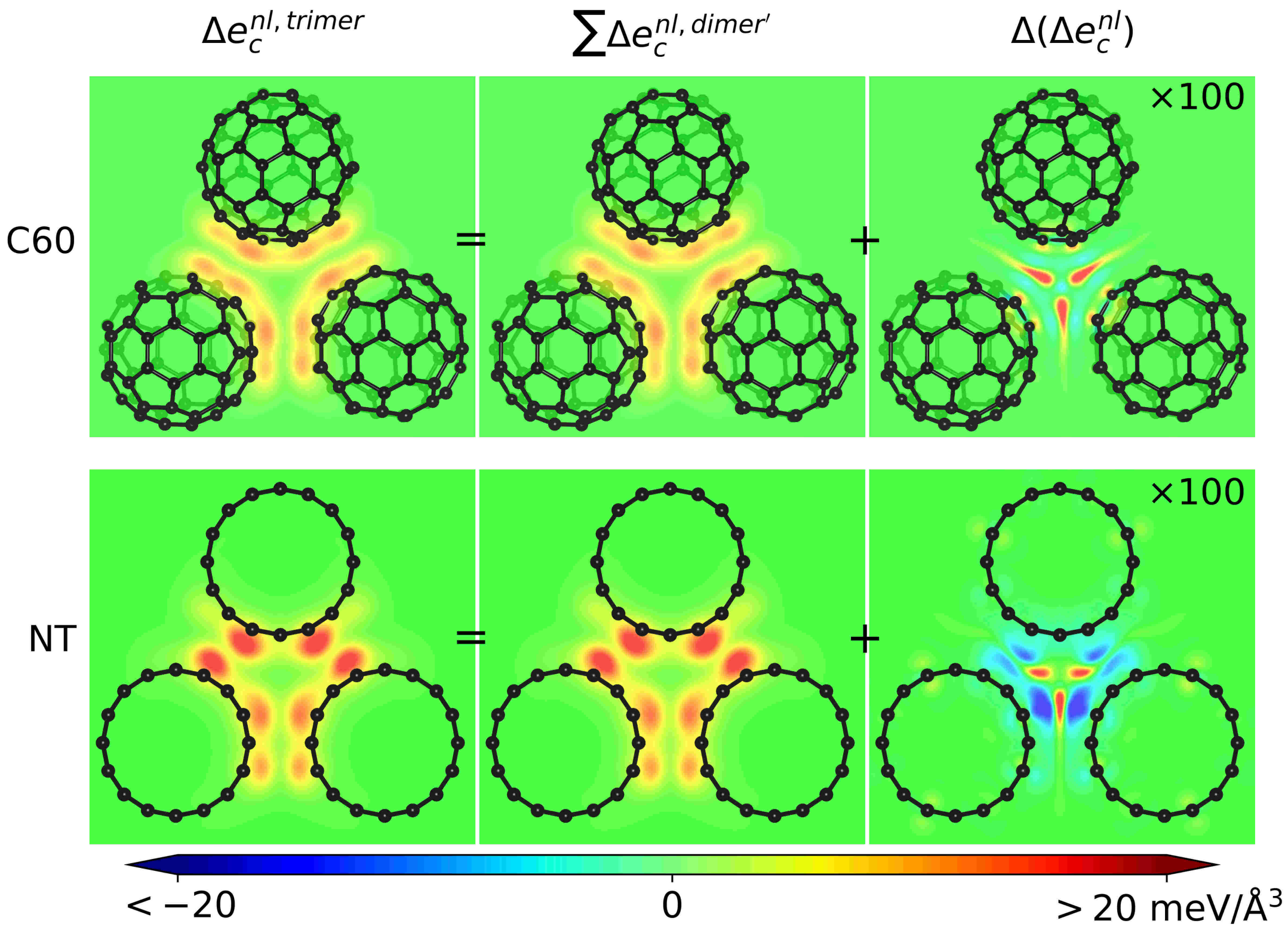

FIG. 13. Non-additivity of the nonlocal-correlation-energy binding contributions in the fullerene trimer (top panels) and a nanotube bundle (bottom panels). The left panels show the nonlocal-correlation contribution $\Delta e_{\rm c}^{\rm nl,trimer}$ to the binding energy in the trimers, directly. The middle panels show the results of simply making a superposition of three nonlocal-correlation binding contributions, $\Delta e_{\rm c}^{\rm nl,dimer'}$, each defined by a dimer. The right panels show the spatial variation in the difference between the actual trimer description and that of the dimer-based superposition, that is, $\Delta(\Delta e_{\rm c}^{\rm nl}) = \Delta e_{\rm c}^{\rm nl,trimer} - \sum \Delta e_{\rm c}^{\rm nl,dimer'}$.

Table IV reports our PBE, PBEsol, and vdW-DF-cx calculations of facet-specific metal surface work functions. The table also includes comparison with both experimental results and SCAN+rVV10 results, as summarized and obtained in Ref. 223. Here we include results both for the facets of the high-temperature Ru* phase and for the low-temperature hcp Ru phase. Overall, the inclusion of vdW interactions (in SCAN+rVV10 and in vdW-DF-cx) leads to improvements in the work-functional description relative to PBEsol, and certainly relative to PBE. We note that the error bars are small for the surface-specific work functions of such metals and that this is a strong test of the functional performance and robustness.

Fig. 12 reports a summary of the performance comparison that we have here provided for PBE, PBEsol and vdW-DF-cx. The top (bottom) panels concern surface energies (work function values), with the left column reporting signed errors; The middle and right column report corresponding MAPD and RMSD values, respectively. We note that this comparison of vdW-DF-cx, PBE, and PBEsol robustness is made from within the same DFT code and setup.

The central message of Fig. 12 and of Ref. 95 is that vdW-DF-cx delivers a robust and accurate account of traditional materials. The vdW-DF-cx performance matches or exceeds that of the best constraint-based GGAs, i.e., PBE and PBEsol. This is true for thermophysical properties of nonmagnetic transition metals [95] and for both surface energies and surface-specific work functions. The vdW-DF-cx errors are generally smaller and the standard deviation is comparable to those of PBEsol (and clearly better than those of PBE), Fig. 12. The vdW-DF-cx performance is better than that of PBEsol (and better than that of SCAN+rVV10) in the case of surface work functions where there is a smaller uncertainty on the experimental reference data, Table IV.

For a discussion of method promise, it is noteworthy that in vdW-DF-cx, we build all nonlocal vertex corrections (and vdW interactions) from an exponential re-summation on conservation laws. In contrast, the nonlocal-correlation part of PBEsol is extracted by a constraint-based fit to trusted surface energies [29]. Nevertheless, vdW-DF-cx still performs as good as PBEsol and it avoids outliers in both surface tests, Fig. 12, as well as for the thermo-physical properties of the nonmagnetic transition metals [95]. This vdW-DF-cx robustness is promising: It suggests a high degree of transferability for indirect (vdW-DF) functional designs that leverage the rules of screening in the electron gas.

### C. Non-additivity in vdW interactions

A recent carbyne-wire study, Ref. 246, finds that the TS-MBD method [80, 81] produces an additional enhancement of the interactions at shorter distances (out to about 1-2 nm,) even for cases where the $d^{-5}$ power law applies for the asymptotic vdW attraction. It is interesting to compare the TS-MBD enhancement near binding to that which we find among molecules, parallel wires, layered structures and at surfaces due to multipole and image-plane effects, as illustrated in Fig. 8 and as discussed in Refs. 6, 10, 14, 53, 56, 181, 210, 211, and 281.

The second-order expansion result Eq. (114) (used in the popular vdW-DF versions) is only the start of the formal vdW-DF method description for two disjunct fragments [10]. This cannot provide a non-additive description (nor in general, a complete description) of vdW interactions in the asymptotic limit [53, 105, 246, 247, 249, 250, 252–254].

At the same time, we trust even the expanded vdW-DF-cx version near binding separations. This is because of the foundation in an underlying Lindhard screening logic and because we are expanding in the plasmon propagator $S_{\rm xc}$, which (by definition) already reflects screening. This is true even if we typically approximate that by asserting the plasmon dispersion to a GGA-level internal functional [4, 8, 10]. Also, the vdW-DF method is electron-density based, not atom centered [53, 210, 211] and it has a natural inclusion of multipole enhancements at and around binding separations [3, 53, 181, 182, 210, 257]. The multipole enhancement effects are, for example, evident in Fig. 8: The vdW-attraction is dominated by regions that are between the graphene layers and not on the carbon atoms. This is exactly as expected when considering the importance of retaining image-plane effects on the vdW attraction [6, 10, 157, 179, 184–187].

Taken together, the observations of limitations and advantages of vdW-DF-cx merit a further discussion. We focus on the near-binding enhancements that are present both in TS-MBD [81, 246] and in vdW-DF-cx calculations, Fig. 9.

First, we illustrate the implicit vdW-DF screening effect that is discussed immediately after Eq. (95) and in Refs. 10, 56, 59, and 105. In Fig. 9, we report the vdW-DF-cx results for both the total nonlocal-correlation binding energy (given by $E_{\rm c}^{\rm nl}$) and for the pure-vdW binding component (given by $E_{\rm c,vdW}^{\rm nl}$) for the carbyne-wire interaction problem. This extraction permits us to analyze the interaction problem as if we had treated it in some would-be effective atom-centered pair-interaction form. That is, we adapt an analysis (in terms of standard vdW-corrected DFT [33, 80]) that was previously made for the TS-MBD method in Refs. 246 and 254.

Specifically, we note that all (carbon) atoms are equivalent by symmetry of the carbyne problem and define separation-dependent vdW-interaction coefficients $C_6^{\rm eff}$ and $C_6^{\rm damped}$ from our full vdW-DF-cx characterization of the mutual binding:

$$\Delta E_{\rm c}^{\rm nl}(d) = -C_6^{\rm scr}(d) \sum_{i,j} \frac{1}{|\mathbf{R}_i - \mathbf{R}_j|^6}\,, \tag{132}$$

$$\Delta E_{\rm c,vdW}^{\rm nl}(d) = -C_6^{\rm eff}(d) \sum_{i,j} \frac{1}{|\mathbf{R}_i - \mathbf{R}_j|^6} \tag{133}$$

$$\frac{\Delta E_{\rm c,vdW}^{\rm nl}(d)}{l_{C-C}} = -\frac{B_5^{\rm eff}(d)}{d^5}\,. \tag{134}$$

Here, $\mathbf{R}_i$ denotes a carbon-atom position and the summations over '$i$' and '$j$' are restricted to one or the other wire. In Eq. (134), we consider the interaction per unit-cell length $l_{C-C}$ of the wires [53, 246, 250].

The top right panel of Fig. 9 shows the wire-separation dependence of both $C_6^{\rm scr}(d)$ (black) and $C_6^{\rm eff}(d)$ (red) versus $\Delta d = d - d_0$. The insert gives the dependence for negative $\Delta d$ values. The $C_6^{\rm scr}(\Delta d)$ representation of the full $E_{\rm c}^{\rm nl}$ binding contribution can be compared to what emerges after the application of a damping function when using a atom-centered pair-wise summation description [33, 73, 74, 76–80] to a dispersion-corrected GGA; The line-up is not ideal, however, as there are also other screening effects in $E_{\rm c,\alpha}^{\rm nl}$.

The $C_6^{\rm eff}(d)$ representation of the pure vdW interaction, red curve in the upper right panel of Fig. 9, is interesting in light of recent discussions of MBD scaling effects in noncovalent-interaction problems [254]. This $C_6^{\rm eff}(d)$ representation has itself part of the multipole effects that are expected [53, 157, 184, 185, 257] and for which there are a scaling argument in a related C60 interaction problem [105]. Some dispersion-corrected DFTs seek to include such effects in higher-order atom-centered vdW coefficients [33, 78], but in the vdW-DF method we get them from looking directly at the electron-density variation itself [10, 14].

The bottom panel of Fig. 9 shows the separation dependence of the mutual attraction, as reflected directly by the effective wire-vdW-interaction coefficients $B_5$, defined in Eq. (134).

There is a clear suppression of $B_5$ (corresponding to the saturation of the vdW attraction) when distances become smaller than the equilibrium separation value, $d_0 = 3.81$ Å, indicated by the vertical line in the bottom panel. This suppression is expected from the fact that the vdW-DF method ensures a seamless integration with LDA [4, 7, 10]. There are no divergent vdW contributions originating from density points at small distances.

There is also a clear enhancement in the effective $B_5$ value (relative to the asymptotic value) near binding. As shown in the bottom panel of Fig. 9, it does not stretch as far out as found in the TS-MBD study [246] for a metallic wire, but it is there. In other words, calculations in vdW-DF and in vdW-DF-cx reflect more than a simple pair-wise summation of atom-centered contributations and they reflect more than a traditional $B_5$ vdW-coefficient for wire interactions. This finding is consistent with what one of us observed and documented for parallel, semi-conducting nanotubes in 2008, in Ref. 53.

The here-documented enhancement of binding from pure-vdW contributions in vdW-DF-cx is part of what can be captured in MBD formulations [2, 3, 75, 81]. These avoids the second-order truncation in the response description [4, 7, 17, 60]. We do not here correctly reflect the dependence on the wire dimerization and hence, nature of conduction [246, 247, 249, 250]. However, our vdW-DF-cx results, Fig. 9, show a qualitative agreement with MBD-type calculations presented in Refs. 10, 105, 246, 251, and 254, concerning the near-binding interaction enhancements. These enhancements are caused by screening and image-plane effects [53, 105] which vdW-DF-cx certainly mimics at a GGA level through the internal functional [10]. Our vdW-DF-cx calculations reflect additional screening and image-plane effects through the the cummulant-type nonlocal-correlation component.

This brings us back to the questions on the non-additivity in the vdW-DF-cx descriptions of the vdW interactions. We inquire if the vdW-DF-cx description

- has sufficient screening mechanisms that it yields a non-additive description at binding separations?
- can accurately handle binding in molecular crystals, where fragments will be sitting not only at the nearest-neighbor distances but also further apart?

Both questions are relevant since the design of the vdW-DF-cx is implicit: we focus on near-binding separations, assert the plasmon dispersion from an internal GGA-level functional, and can at most reflect a GGA-level of screening, when used in the typical truncated form [4, 7, 17]. The limitations and the ameliorating factors near the optimal binding separations were discussed in Ref. 10. Both questions are here discussed instead on the basis of computational results.

Figure 13 helps us address the first question. It shows that the attraction among three fullerenes (top panels) and among three carbon nanotubes is indeed non-additive at binding separations, that is, for fully relaxed geometries. The $E_c^{nl}$ binding for three objects exceeds that which results (middle panel) when considering a set of pair interactions. This demonstration of non-additivity for carbon structures supplements the affirmative answer we have previously obtained for a noble-gas atom [14].

We note that while there is symmetry in the binding of three fullerenes, the symmetry is broken for a triple-cluster of carbon nanotubes. The difference reflects a variation in the alignment of structural motifs in the carbon nanotube contact regions, as further discussed in Ref. 53 (which investigated a similar nanotube system, but with the original vdW-DF version).

Also, it is interesting that the non-additivity effects are, in fact, small for such large objects at binding separations. Here it is a 2% effect, whereas the non-additivity is a 10% effect in the case of a noble-gas trimer [14]. The smaller non-additivity is a consequence of a simple geometry effect: There is not enough room for the third structure to affect the binding of the other two when at or near binding separation.

Observe, however, that finding a small non-additivity effect for nanotubes at binding separations does not imply that screening effects are irrelevant in the present vdW-DF-cx characterization, Fig. (13). In fact, with the original vdW-DF version (which has the same $E_c^{nl}$ formulation as vdW-DF-cx), one of us has documented that the nanotube attraction is strongly affected by multipole (or image-plane effects,) as discussed in Ref. 53. The multipole effects are also directly evident in Fig. 9.

The observation that screening significantly affects the vdW attraction within nanotube bundles [53] and among carbyne wire, Fig. 9, is corroborated by the recent study of cohesion in fullerene crystals [105]. There we helped document and explain that multipole enhancements of the asymptotic intermolecular attraction are inescapable and essential, leading to modifications that have qualitative consequences [105]. For example, while the vdW-DF method underestimates the molecular $C_6$ interaction coefficient describing the asymptotic case, it still accurately reproduces known motifs in the structure of $C_{60}$ crystals [105].

For the second question, whether vdW-DF-cx is able to accurately describe molecular crystals, we make a number of observations. First is the fact that vdW-DF-cx produces the exact same asymptotic atom-$C_6$ coefficients as does the original vdW-DF version [4]. These atomic $C_6$ coefficients are not bad (unlike those of vdW-DF2) for small to medium size molecules [271]. Second, in a very recent paper, other members of the vdW-DF team have launched and investigated a vdW-DF version in which the atomic $C_6$ coefficients are further improved from the vdW-DF/vdW-DF-cx level [16]. It is observed that limited quality of the atomic $C_6$ coefficients is not an intrinsic feature of the vdW-DF method, and that the description of molecules does improve when optimizing the link between the plasmon description and the internal-functional energy-per-particle variation for better atomic $C_6$ coefficients. However, it is also observed that a different focus may be the key to improving the overall performance within the vdW-DF design strategy [16].

The vdW-DF-cx characterization can, under special conditions, fail to correctly describe the molecular asymptotic-vdW $C_6$ coefficients, for example, for C60 molecules [105, 252, 282] and for carbyne wires [246, 250]. However, there is also a set of arguments that the asymptotic performance is not a good discriminator for what happens at binding separations [10, 258]. The C60 case study makes these observation more clear: since there necessarily must exist large multipole enhancement effects it is not a priory certain that a poor asymptotic description must make vdW-DF-cx fail for the C60 crystal [105]. In fact, we observe that vdW-DF-cx performance is good for both the C60 and the polyethylene crystals [103, 105, 107]. The vdW-DF-cx performance is excellent for the oligoacene molecular crystals where there exists a full experimental mapping across temperatures and measurements at low temperatures [98, 99].

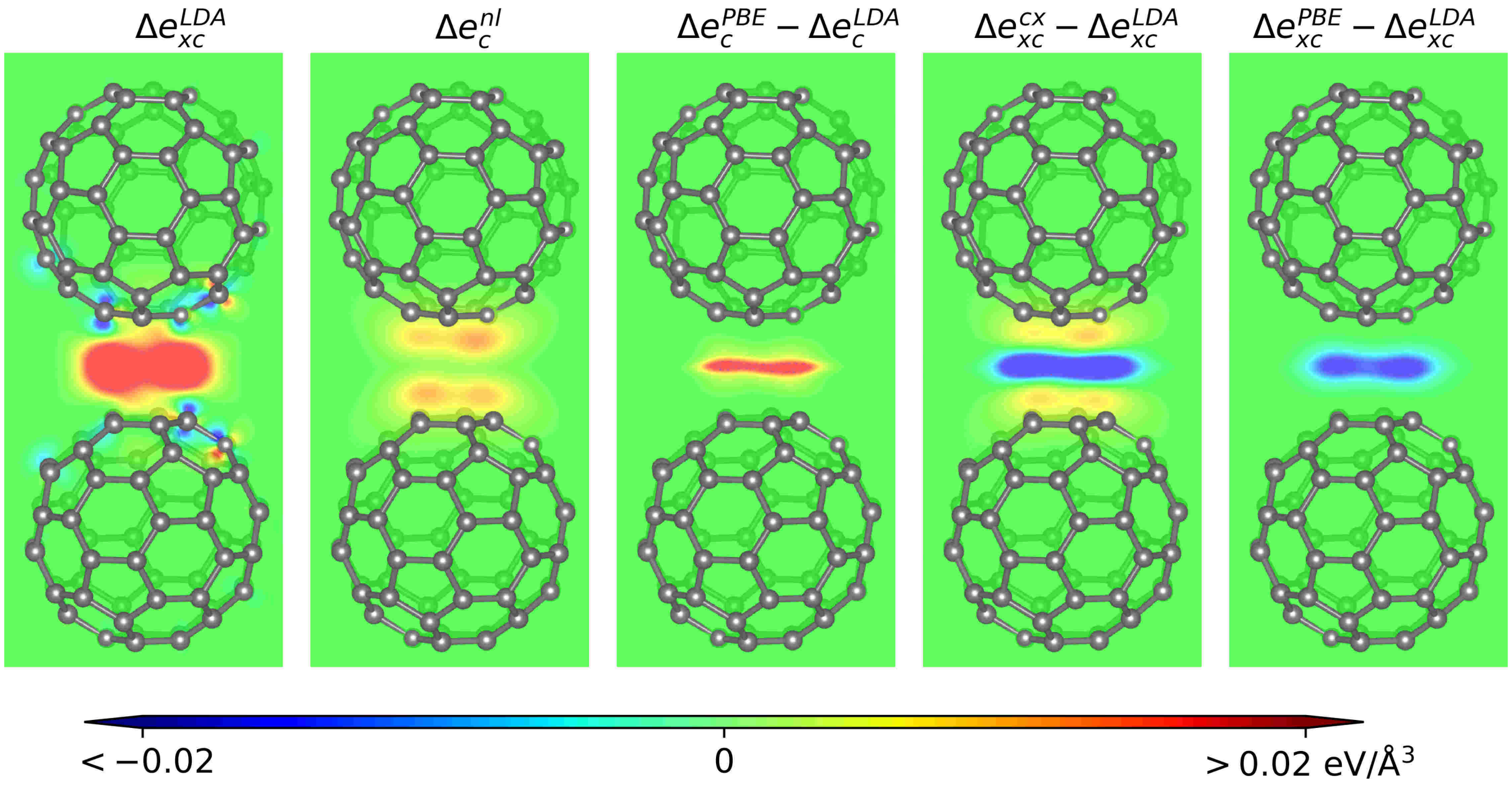

FIG. 14. Exchange and correlation contributions to the binding in $C_{60}$ dimers. Counting from the left, the first panel shows a mapping of binding contributions from the LDA XC energy. The last two panels contrast the corrections arising in the vdW-DF-cx and PBE XC functionals. Finally, the second and third panel contrast the variation in binding-energy contribution emerging from the vdW-DF(-cx) nonlocal-correlation term and from the gradient-corrected correlation part of PBE, respectively.

We have already observed in Sec. IV.A, that perhaps the strongest test of functional robustness is whether it makes accurate predictions for structure and vibrations (and surface work functions) that probe the quality of the density variation. We find that vdW-DF-cx correctly describes the known motifs of the C60 molecular crystals [105]. For the polyethylene crystals the experimental results vary to some extent. Nevertheless that vdW-DF-cx characterizations of structure and defects are accurate [103, 107].

For the oligoacenes there exists accurate low-temperature characterizations both of the low-temperature structure [98] and of the highly soft intermolecular rocking and libration modes [99], Fig. 5. With full stress and atomic-position relaxations, the vdW-DF-cx is able to predict lattice constants and angles in close agreement with measured structure data, and therefore useful for subsequent MBPT calculation of the optical response [98]. Moreover, as shown in Fig. 5, there is also (for the vdW-DF-cx predictions of the naphthalene structure) an almost complete agreement between neutron data and vdW-DF-cx phonons.

Finally, for the problem of molecular-crystal binding energies there are limitations of vdW-DF-cx being a truncated implementation of the vdW-DF method. The vdW-DF-cx predictions for the C60 molecular-crystal cohesive energies differ about 4 meV per carbon atoms [105]. The vdW-DF-cx predictions for oligoacene cohesive energies [98] differ 13% MAPD from low-temperature measurements, Table I. We are not aware of any MBD-based results for the oligoacenes but we mention, for comparison, that PBE-TS [80] was found to differ 28% MAPD. It is possible that the vdW-DF-cx description for molecular-crystals cohesive energies is indeed affected by the very weak (but also numerous) off-equilibrium binding contributions in molecular crystals. To capture those we must indeed go beyond the second-order truncation that secures its universal-kernel formulation [4, 7, 17].

A MBD extension of vdW-DF-cx, perhaps adapting the deas of the first (non-truncated) nonlocal-correlation functional for materials calculations [3], vdW-DF0, is desirable. It is a possible future direction for the vdW-DF development work.

Going foreward, however, there is a lesson to be drawn from the present exploration of consistent vdW-DF implementations. The existing MBD formulations rely on the Dyson equation to balance the lowest-order and higher-order vdW terms [3, 75, 81], but the lower-order contributions to XC designs are picked from other considerations. Meanwhile, the vdW-DF-cx stands out as the first consistent implementation of the vdW-DF method. It has clear accuracy advantages over vdW-DF1 [4] and we see these advantages as coming from using the Lindhard or Dyson logic to balance the exchange and correlation choices, as explained already in the introduction. To fully leverage the screening logic and the current-conservation criterion on all orders of relevant XC contributions seems important. There is thus a rationale for focusing on consistent implementations of the vdW-DF method, whatever other specific XC design choices we may make in the future.

### D. Nature of materials binding

We propose to use a spatial mapping of binding contributions [10, 14, 55–58] to explore physics implications and to detail differences between the direct and indirect XC functional design approaches.

It is natural to begin with an exploration of binding in sparse matter where some important interactions will be dominated by the vdW attraction. It is clear that the indirect-design vdW-DF-cx (compared with the GGAs) will here provide the more meaningful materials account. However, it is still interesting to track the differences to better understand the origin of the PBE shortcomings for this large class of materials.

Figure 14 presents such a comparison for the binding in a $C_{60}$ dimer. The leftmost or first panel shows the binding contribution that arises in LDA. A set of previous studies documents that this LDA binding in such weakly-interacting, sparse-matter systems is spurious and arises from an incorrect description of exchange [56, 59, 255, 283–285]. In fact, the PBE XC-functional correction, shown in the right-most fifth panel, compensates for and effectively offsets the spurious LDA binding.

The problem for the set of direct-design, gradient-corrected functionals is that they can only mimic sparse-matter binding at longer atom separation [10, 59, 91]. Moreover, by the nature of the approximation, they must do so in regions with a pronounced density overlap [3], that is, in the mid-region area between the fullerenes. PBE delivers a small enhancement of binding from PBE gradient-corrected correlation (central panel) in this mid-region. However, there is still a need to offset the spurious exchange binding, and that too must occur in the very same region. PBE likely strikes an optimal balance for the more strongly bonded systems, but the balance is delicate [59].

The vdW attraction is more widely distributed in space, Refs. 3, 10, 14, and 59 and Fig. 8, and the relative weights of exchange and correlation effects change as we vary the binding distances [12, 56, 59, 182, 210, 211]. A direct functional design (like PBE) is simply overloaded when put to the task of describing the general type of sparse-matter systems [6, 32].

On the other hand, the second panel shows that the vdW-DF-cx delivers something new, overcoming the limitations of such semilocal functionals. Having an indirect functional design logic, the vdW-DF-cx employs a truly nonlocal correlation XC functional component $E_c^{nl}$ and can thus reflect the nature of weak binding contributions. The vdW-DF functionals have a proper mechanism to represent image-plane effects [6, 12, 53, 181, 182, 184–186] and provide transferability over a range of binding separations. As such, they correctly describe the nonlocal-correlation binding enhancement as emerging in the density tails rather than just in the overlap region [3, 10, 14].

Contrasting the fourth and fifth panel we see that there are fundamental differences in the resulting binding description between the indirect-design vdW-DF-cx and the direct-design PBE functionals. Simply put, the vdW-DF logic allows us to put the nonlocal-correlation binding effect (second panel) in different spatial locations than the beyond-LDA-exchange correction that both functionals provide.

It is interesting that the $E_c^{nl}$ binding in the fullerene dimer is itself defined by the competition between pure vdW attraction effects and truly nonlocal vertex effects. The fullerene dimer and graphene bilayer systems are sufficiently similar that we can simply port the observations from Fig. 8 and thus conclude our analysis of binding in this sparse-matter problem. The nonlocal vertex corrections occur in the density overlap regions. The vdW attraction is widely distributed but the end result is that the total nonlocal-correlation binding is dominated by the near-fullerene regions. In short, the nonlocal vertex effects help concentrate the net attraction to regions that can be thought of as molecular image planes [10, 53, 56, 59].

Additional understanding of the nature of material interactions and of vdW-DF-cx/PBE differences can be gained by mapping the spatial variation in binding contributions in (more) strongly bound bulk materials. Figs. 15 and 16 contrast such mappings for Si, Na, and W. Since the vdW-DF-cx performance matches or exceeds the PBE performance here, it is relevant to ask if vdW-DF-cx has a similar or different balance between exchange and correlation than does PBE in those cases. It is also interesting to check how the balance between cumulant/vertex effects and pure-vdW-binding effects vary across regions with different density profiles.

As in the fullerene-dimer case, the set of panels in the first column of Fig. 15 shows the LDA binding contributions, while the fourth and fifth column detail the binding corrections as they emerge in vdW-DF-cx and PBE, respectively. Note that the figures present calculations of the spatial variation in binding contributions for a diagonal cut through the unit cells. The Si system has one extra atom in its bulk unit cell, located in this cut.

It is clear that the LDA description in these cases pro-vides the dominant features although there is some adjustments. Unlike in weakly bonded matter [283–285], this is the expected behavior. The indirect vdW-DF-cx and the direct PBE functionals both deliver accurate accounts of the basic properties, Table II and Refs. 15, 45, and 95. It is interesting that they still differ in important details of the binding description.

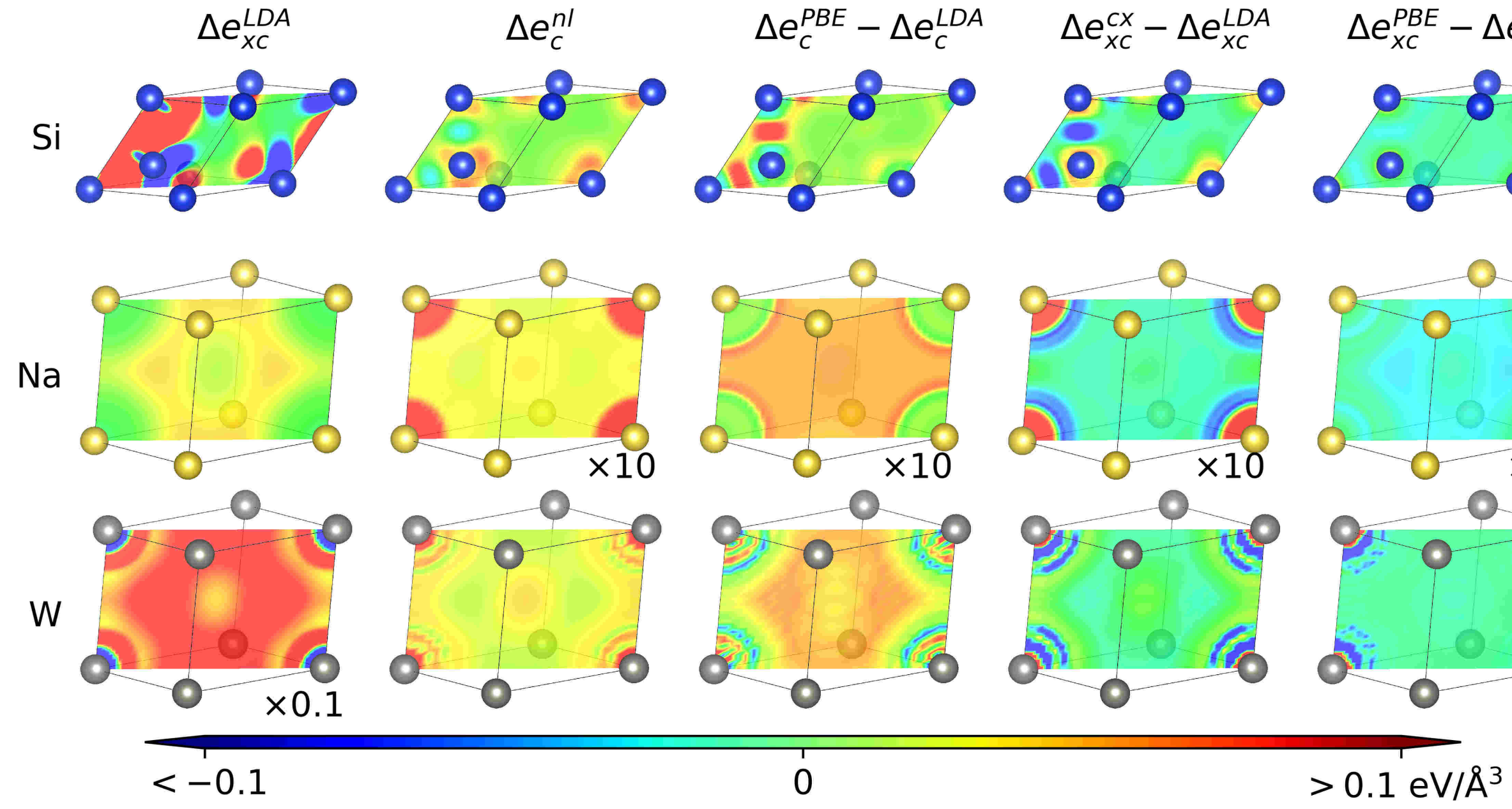

FIG. 15. Exchange and correlation contributions to the binding in bulk Si (top row), Na (middle row), and W (bottom row). The first column show mappings of the binding contribution from the LDA XC energy. The last two columns contrast the corrections arising in the vdW-DF-cx and PBE XC functionals. Finally, the second and third columns contrast the variation in binding-energy contribution emerging from the vdW-DF(-cx) nonlocal-correlation term and from the gradient-corrected correlation part of PBE, respectively.

Specifically, there is more spatial variation (or structure) in the vdW-DF-cx than in the PBE binding descriptions. The qualitative differences are most clear in the Na case, which has the lowest cohesive energy. Quantitatively the relative differences seem largest in the Si and W cases. Overall, the PBE corrections to LDA are essentially uniformly distributed and the vdW-DF-cx corrections have a similar moderating effect in the low-density regions. However, the vdW-DF-cx descriptions also include both repulsion and an interaction-enhancement region near the atoms.

Contrasting the nonlocal-correlation binding contributions of vdW-DF-cx and PBE (shown in the second and third column, respectively) details the origin of such differences. For example, the PBE nonlocal correlation does provide a binding contribution in the inter-atomic regions, where it is actually stronger than what arises in vdW-DF-cx. However, while the nonlocal correlation binding enhances near the atoms for vdW-DF-cx, it weakens in PBE. Meanwhile, the nonlocal-correlation binding in the middle of the cut (furthest away from the atoms) tends to be offset by the repulsion, set by gradient corrections. In the case of the vdW-DF-cx description, however, there is enough spatial variation in the nonlocal-correlation binding, second column, that the structure also survives in the nonlocal binding contribution, fourth column.

Figure 15 shows that the above-discussed usefulness of vdW-DF-cx (for traditional matter) does not arise simply because it mimics the PBE description in the balancing of exchange and correlation effects everywhere. Given the strong vdW-DF-cx performance, it is plausible that the vdW-DF-cx-based mapping of the nature and spatial variation of binding is trustworthy.

Diving in deeper, we analyze the nonlocal binding contributions also in terms of cumulant and pure vdW binding effects, Fig. 16. Again, we consider the binding contributions of Si, Na, and W and, for the reader's convenience, we repeat the PBE and vdW-DF-cx characterizations of nonlocal-correlation binding effects in the first and second columns.

We find that for these dense-matter systems there is a pronounced cancellation between vdW attraction and nonlocal-vertex corrections. We tend to ignore the relevance of dispersive interactions in traditional bulk. However, the Rapcewicz and Ashcroft observation [65] that vdW forces exist also in the itinerant electron gas, is clearly important [10, 12]. On the other hand, such pure vdW forces are also dramatically compensated by a repulsion defined by nonlocal vertex corrections (shown in the fourth panel). As in the case of the fullerene dimer, these vertex effects grow in the regions where there is, relatively speaking, a large density increase by the density overlap. For dense metals, the compensation ensures that the net $E_{\rm c}^{\rm nl}$ binding effect remains bounded at the PBE level, possibly with the exception of the description of noble metals [94, 95].

The last column of Fig. 16 shows our computed results for the spatial variations (for Si, Na, and W) of this local-field susceptibility term, $\Delta e_{{\rm c},\alpha}(\mathbf{r})$. Interestingly, the cumulant correction, or nonlocal vertex-correction, part (fourth column) does, to a large extend, offset the spatial variation in the LDA correlation binding contribution. For example, in Si we have pronounced signatures of LDA correlations but these are compensated by the vdW-DF-cx description of nonlocal-vertex corrections. The net effect is that there is only a moderate variation for $\Delta e_{{\rm c},\alpha}(\mathbf{r})$ in these dense-matter cases.

## VII. SUMMARY AND CONCLUSIONS

To set the stage, we mention again that the consistent vdW-DF versions, like (spin) vdW-DF-cx, presently have no input beyond QMC data entering in the LDA definition [23, 24] and a formal MBPT result on gradient-corrected exchange [4, 7, 12, 174]. As discussed in this paper, vdW-DF-cx must build all accounts of beyond-LDA correlation from an exponential re-summation. That is, it must rely on a connection with a cumulant-type expansion to get at more general vertex corrections (and other screening effects) that are already in a GGA. It is encouraging for the indirect-functional design strategy that vdW-DF-cx still manages to perform as good as PBE, even on the (dense-matter) home turf of this versatile gradient-corrected XC functional.

There are several key observations in this review paper concerning the screening nature of consistent vdW-DF functionals.

*First,* we argue that by asserting the shape of the model propagator, $S_{\rm xc} \equiv \ln(\epsilon)$ from the energy-per-particle variation of $E_{\rm xc}^{\rm in}$ the vdW-DF formulation, Eq. (3), we incorporate some of the vertex corrections and screening that are relevant to a GGA-type description. This observation supplements the previously published interpretation [10] that the vdW-DF-cx XC energy functional $E_{\rm xc}^{\rm DF}$ (also) represents a succinct formulation of pure vdW forces, in a ZPE-coupling picture of dispersion interactions [1, 6, 12, 64, 65, 158].

*Second,* we document the vdW-DF-cx usefulness for descriptions of bulk and molecular cases with covalent and noncovalent binding. We find that the indirect-XC-functional design of vdW-DF-cx provides a performance that is as good as or better than the trusted, constraint-based semilocal GGAs, even for surface properties.

*Third,* we demonstrate that there are similarities but also important differences between PBE [28] and vdW-DF-cx [9] characterizations of binding in a bulk semiconductor and metals. It is encouraging that the vdW-DF-cx goes beyond simply mimicking PBE for dense matter (as well as for molecular matter) but builds an equally accurate description from binding contributions with different spatial variations.

We think that the vdW-DF-cx characterization of materials binding can be trusted, and we therefore proceed to deliver tools for a deeper analysis. Specifically, we have developed code for universal-kernel evaluations that can separately map binding contributions arising from a) cumulant effects and b) pure-vdW interactions, given by the electron-XC-hole mechanism that is illustrated in Fig. 2.

We have shown that screening underpins not only the response description behind pure, long-range vdW forces but also the here-identified inclusion of a cumulant-type expansion. We have furthermore discussed how these terms provide consistent-exchange vdW-DF-cx with an account of beyond-LDA vertex corrections. As such, it provides vdW-DF-cx with a seamless integration [2, 12, 166] to a GGA-type description. This is true not only in the HEG limit (as discussed in Refs. 4 and 7) but also in the case of the weakly perturbed electron gas, i.e., conditions that are important for the description of bonding inside molecules and in bulk.

As for the vdW-DF-cx robustness, we expect that having better compliance with the Dyson equation provides vdW-DF-cx [9, 10, 45] with an advantage over the original vdW-DF version [3, 4]. This advantage is relevant when we seek to describe traditional materials [95] and surface problems [134, 257], i.e., cases where the nonlocal correlations can be expected to play a greater relative role than in intra-molecular binding. We note that both versions have identical nonlocal correlation energy $E_{\rm c}^{\rm nl}$ (for any given density). However, vdW-DF-cx stands out in also effectively achieving compliance with the Dyson equation. This implies in turn that vdW-DF-cx is closer in nature both to the formal ACF recast, Eq. (18) and to the cumulant-expansion logic [86].

In practical terms, this means for vdW-DF-cx (for vdW-DF1) that $E_{\rm c}^{\rm nl}$ reflects an exponential re-summation and XC-hole coupling that are more relevant (less relevant). In the consistent vdW-DF versions, like vdW-DF-cx, the total correlation provides a better balance to the exchange component that is contained in the semilocal functional component $E_{\rm xc}^{0} \approx E_{\rm xc}^{\rm in}$.

Finally, we mention an implication regarding handling of semilocal correlation effects in indirect functional designs (like vdW-DF-cx). In picking the plasmon propagator model $S_{\rm xc}$, there is a formal choice concerning retaining or discarding gradient-corrected correlations. In the vdW-DF-cx design we avoid this explicit inclusion to avoid a double counting because we are also trying to tap into the vertex-correction handling of the cumulant-expansion logic. It is encouraging that this simple approach works, for it simplifies construction of new versions in the class of consistent vdW-DFs.

*Overall,* this paper documents the central role that screening plays in specifying the various terms that enter in consistent formulations of the vdW-DF design. It seems that the Occam's-razor approach of the vdW-DF method and of vdW-DF-cx is a good starting point.

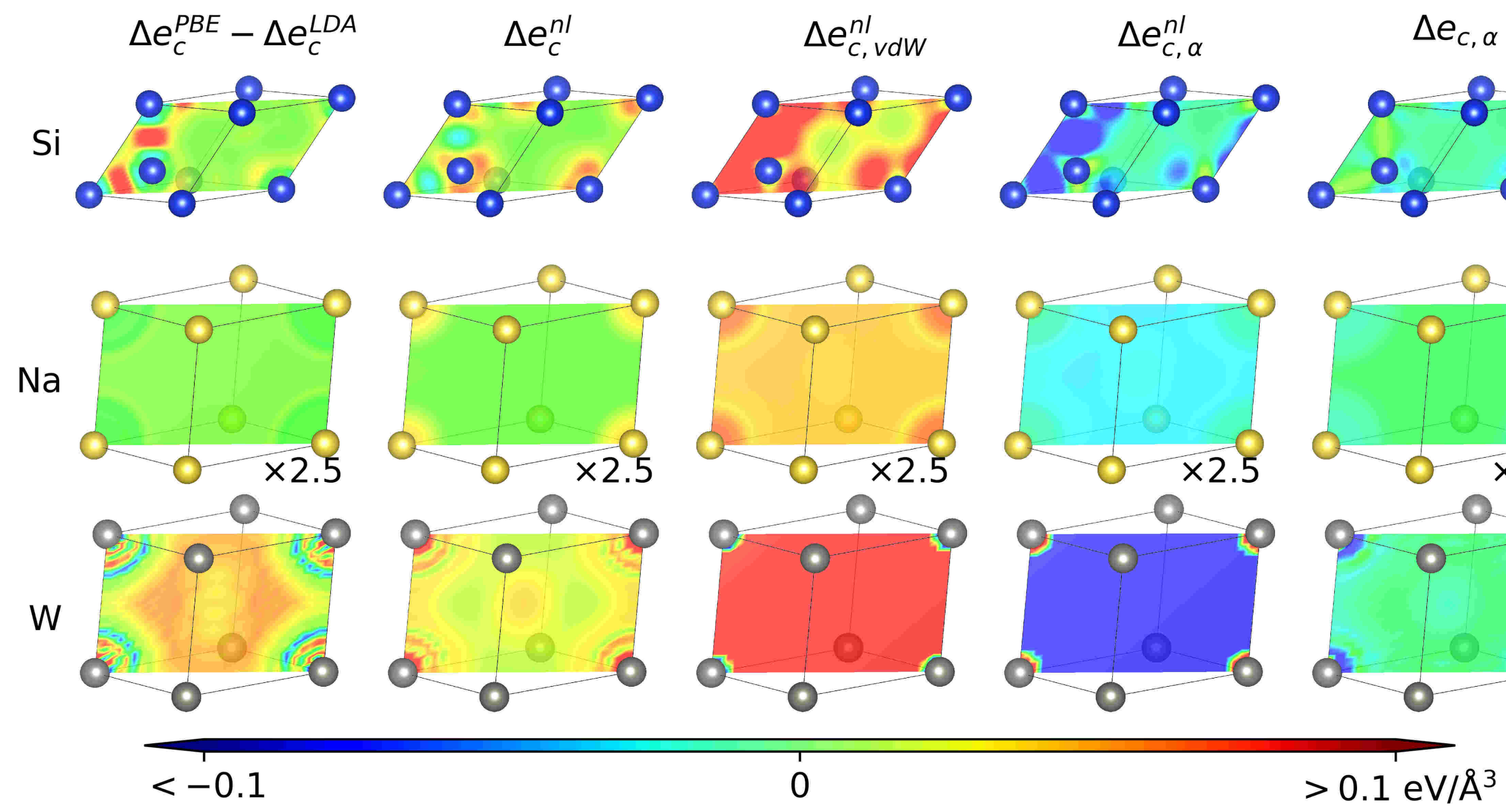

FIG. 16. Correlation contributions to the binding in solid Si, Na and W. The first and second columns contrast the contribution arising from the gradient-corrected correlation part of PBE and from the total nonlocal correlation in vdW-DF-cx. The third and fourth columns depict binding energy contributions from pure vdW interactions and from cumulant effects. The last column shows the binding contribution originating from the vdW-DF-cx description of the model susceptibility $\alpha(\omega)$.

Finally, it is worth discussing our calculation of the binding contribution that arises from the term, Eq. (109), reflecting the local-field susceptibility behavior. This term is defined by the LDA-correlation input but corrected for additional screening that we represent in an exponential re-summation. A corresponding local energy density, termed $e_{c,\alpha}(\mathbf{r})$ is given by the sum of Eq. (119) and the corresponding energy density term for LDA correlation.

## ACKNOWLEDGEMENT

This review of consistent implementations of the vdW-DF method is dedicated to John W. Wilkins.[2] The analysis of the screening nature of the vdW-DF method was seeded many years ago by discussions by the first author with David C. Langreth, shortly before his passing in 2011. We thank Florian Brown-Altvater for adapting the upper panel of Fig. 5 from Fig. 3 of his naphthalene phonon paper, Ref. 99. We also thank Dr. Andreas Hansen for computing a new CCSD(T) reference-energy value for the $C_{22}H_{46}$ alkane unfolding energy (now obtained for the structure computed with vdW-DF-cx relaxations, rather than at the reference geometry that is listed in IDISP benchmark set of GMTKN55 [35]). We are grateful for discussions over many years with Kristian Berland, Tonatiuh Rangel, Paul Erhart, Henrik Grönbeck, Jung-Hoon Lee, Jeffrey B. Neaton, Gerald D. Mahan, Elsebeth Schröder, and Bengt I. Lundqvist. Work supported by the Swedish Research Council (VR), the Chalmers Area-of-Advance-Materials theory activity, and the Swedish Foundation for Strategic Research (SSF) under contracts SE13-0016 and ITM17-0324. The authors also acknowledge computer allocations from the Swedish National Infrastructure for Computing (SNIC), under contracts 2016-10-12, 2018-5-138, and 2019-1-36, and from the Chalmers Centre for Computing, Science and Engineering (C3SE).

### Appendix A: Screening nature of electron dynamics

Consider the Dyson equation for the electron Green function

$$g(1,1') - g_0(1,1') = \int d2 \int d3\, g_0(1,2)\sigma(2,3)g(3,1')\,, \tag{A1}$$

[2] John Wilkins passed away December 6, 2019. John helped make the electron-gas tradition thrive in DFT. He impacted the lifes of the vdW-DF seniors.

and the effective Dyson equation that exists for the screened Coulomb interaction

$$W(1,1')-V(1-1') = \int d2 \int d3\, V(1-2)P(2,3)W(3,1')\,. \tag{A2}$$

We use Hedin notation [188] where interaction points 1, 2, 3, and 4 are normally thought of as time and space (plus spin) coordinates, but we can work with any complete representation of space.

Figure 4 explains an equivalence between the exact descriptions of the screened interaction and that of the quasi-particle electron dynamics, i.e., the Green function. The descriptions are related because they both fully retain the electron-electron interacting vertex (triangle) and are, as such, both defined by the Feynman diagram for the linked-cluster-expansion evaluation of the exact total energy (shown in panel a).

As used in the Hedin equations [188], we can express both the exact electron self energy $\sigma$ and the local-field response function $\tilde{\chi}$ in terms of one diagram structure but using an electron-Coulomb vertex function $\Gamma(1'2';3')$. Here, $1'$ and $2'$ denote the coordinates of connected electron Green functions while $3'$ denotes the coordinate of a connected screened-interaction line $W$. This leads to formal results [188]:

$$\sigma(1,2) = i\int_{3,4} g(1,4)\,\Gamma(42;3)\,W(3,1)\,, \tag{A3}$$

$$P(2,1) = -i\int_{3,4} g(1,3)\,\Gamma(34;2)\,g(4,1)\,. \tag{A4}$$

A diagram expansion of the second term in the square bracket of Eq. (A1) can thus be expressed

$$\int_2 \sigma(1,2)g(2,1) = i\int_{2,3,4} g(1,4)\,\Gamma(42;3)\,g(2,1)\,W(3,1)\,. \tag{A5}$$

Meanwhile, for the description of the screened interaction, we have

$$\int_2 W(1,2)P(2,1) = -i\int_{2,3,4} g(1,3)\,\Gamma(34;2)\,g(4,1)\,W(2,1)\,, \tag{A6}$$

where we have used $W(2,1)=W(1,2)$.

Taking Eqs. (A5) and (A6) together, and applying a cyclic permutation of internal integration variables 2, 3, and 4, we conclude

$$\int_2 W(1,2)P(2,1) = -\int_2 \sigma(1,2)\,g(2,1)\,. \tag{A7}$$

The result, Eq. (A7), can also be formulated

$$\int \frac{d\omega}{2\pi}\,\langle \mathbf{r}_1|\sigma(\omega)g(\omega)|\mathbf{r}_1\rangle = -\sum_\mu \int \frac{d\omega}{2\pi}\,\langle \mathbf{r}_1|W(\omega)P^{\mu,\nu}(\omega)|\mathbf{r}_1\rangle, \tag{A8}$$

where the quasi-particle dynamics is described for spin $\nu$. This follows by Fourier transformation because we work with a time-translationally invariant problem. The result Eq. (A8) is used, for example, in discussion of the nature of the electron-phonon interaction problem in Ref. 167. It is also consistent with the two ways Ref. 169 provides for the evaluation of the ground-state expectation value of the electron-electron interaction energy.

Moreover, since Eq. (A8) holds for any complete (coordinate) representation, we have the formal operator relation

$$\int \frac{d\omega}{2\pi}\,\sigma(\omega)g(\omega) = -\int \frac{d\omega}{2\pi}\,W(\omega)P(\omega) \rightarrow -\int \frac{d\omega}{2\pi}\,V\chi(\omega)/2\,, \tag{A9}$$

where $P(\omega) \equiv \sum_\mu P^{\mu,\nu}(\omega)$. The last line holds when $P(\omega) = \tilde{\chi}(\omega)/2$, which we shall assume to simplify the discussion.

Finally, Eq. (72) relates the frequency variations of the exponential-re-summation factors of the Green function and of the dielectric-function descriptions. The exact result Eq. (A9) suggests the connection Eq. (72) as the simplest assumption on the per-frequency variations in these re-summation factors.

## Appendix B: Plasmon-propagator approximation for the indirect vdW-DF XC functional design

In vdW-DF we work with a double-pole model for the plasmon propagator, termed $S_{\rm xc}(\omega)$. We assume a local dispersion of the plasmon dispersion $\omega_q(\mathbf{r})$ in a weak-perturbed electron gas, as detailed below. We note that $S_{\rm xc}(\omega)$ must reflect a semilocal XC hole and we rely on the ideas of the formal MBPT gradient expansion [87]. In effect, in the vdW-DF method, we employ a folding of plasmon-pole contributions from two momenta [4, 47]:

$$S_{\rm xc}(\omega) = \bar{S}^{\rm RLL}_{\mathbf{q},\mathbf{q}'}(\omega) \equiv \frac{1}{2}\left[S_{\mathbf{q},\mathbf{q}'}(\omega) + S_{\mathbf{q},\mathbf{q}'}(-\omega)\right], \tag{B1}$$

$$S_{\mathbf{q},\mathbf{q}'}(\omega) = \int_{\mathbf{r}} e^{i(\mathbf{q}-\mathbf{q}')\cdot\mathbf{r}} \frac{\omega_p^2(\mathbf{r})}{[\omega+\omega_{\mathbf{q}}(\mathbf{r})][-\omega+\omega_{\mathbf{q}'}(\mathbf{r})]} \tag{B2}$$

This plasmon propagator is dominated by the diagonal components [83, 85, 86] used in the early specification of LDA [21, 22], but the off-diagonal terms reflect the effects of density gradients on the screening, i.e., the dielectric function, and hence on the resulting XC hole, Eq. (20).

In the popular general-geometry versions and variants [4, 8, 9, 40–42], this Rydberg, Lundqvist, Langreth (RLL) plasmon-progator approximation is used directly with one of two specifications of $\omega_q(\mathbf{r})$, Refs. 4 and 8. Earlier nonlocal functional generations [2, 3, 6] (for example the layered-geometry version) used a slightly modified response form that is more true to the Lundqvist single-pole plasmon-propagator form [21, 22].

The RLL form, Eqs. (B1) and (B2), explicitly complies with time-reversal symmetry [4]. By properly specifying the dispersion form $\omega_q(\mathbf{r})$ this form complies with all known constraints on the plasmon-response behavior [4, 56, 82]. The idea is to set the assumed dispersion

$$\omega_q = q^2/[2h(q/q_0(\mathbf{r}))]\,, \tag{B3}$$

$$h(y) = 1 - \exp[-\gamma y^2]\,, \tag{B4}$$

in terms of an inverse length scale $q_0(\mathbf{r})$ that reflects the energy-per-particle density of the internal functional [4, 7]:

$$\varepsilon_{\rm xc}^{\rm in}(\mathbf{r}) = \pi \int \frac{d^3\mathbf{q}}{(2\pi)^3}\left[\frac{1}{\omega_q(\mathbf{r})} - \frac{1}{\omega_{\rm self}(q)}\right] = -\frac{3}{4\pi}q_0\,. \tag{B5}$$

Here $\omega_{\rm self}(q) = q^2/2$ is just the free-electron energy. It defines a contribution which corresponds to the $E_{\rm self}$ term in Eq. (1). The choice of $\gamma$ is truly arbitrary, set to $\gamma = (4\pi/9)$, but any other choice would simply adjust the appearence – not the physics content – of the internal-functional description. In making the connection from plasmon-dispersion to the energy-per-particle variation, Eq. (B5), the vdW-DF method adapts the logic that Langreth and Perdew used to improve computations of surface energies beyond an underlying LDA account [25, 46, 89].

Finally, it is interesting to contrast this RLL model for a GGA-based double-pole plasmon propagator with the generalized plasmon pole (GPP) model [88] that is used, for example, in discussing GW calculations. To that end it is relevant to first consider how the RLL form should be described in a periodic system, using a gradient expansion in the arguments. Thus we express $\mathbf{q}$ in terms of a major reciprocal-lattice vector $\mathbf{G}$ component and the residual wavevector component $\mathbf{k}$ (reflecting the component inside the first Brillouin zone), in a uniquely defined separation $\mathbf{q} = \mathbf{G} + \mathbf{k}$. We note that $\mathbf{q} - \mathbf{q}' = \mathbf{G} - \mathbf{G}'$ and recast the RLL double-pole plasmon propagator

$$\bar{S}^{\rm RLL}_{\mathbf{G},\mathbf{G}'}(\mathbf{k};\omega) = \int_{\mathbf{r}} e^{i(\mathbf{G}-\mathbf{G}')\cdot\mathbf{r}}\,\frac{\omega_p^2(\mathbf{r})(\omega_{\mathbf{G}+\mathbf{k}}(\mathbf{r})\omega_{\mathbf{G}'+\mathbf{k}}(\mathbf{r}) - \omega^2)}{[\omega^2_{\mathbf{G}+\mathbf{k}}(\mathbf{r}) - \omega^2][\omega^2_{\mathbf{G}'+\mathbf{k}}(\mathbf{r}) - \omega^2]}\,. \tag{B6}$$

This RLL form, Eq. (B6), should be compared to the GPP formulation:

$$S^{\rm GPP}_{\mathbf{G},\mathbf{G}'}(\mathbf{k}) = \frac{\Omega^2_{\mathbf{G},\mathbf{G}'}(\mathbf{k})}{\omega^2_{\mathbf{G},\mathbf{G}'}(\mathbf{k}) - \omega^2} \tag{B7}$$

$$\Omega^2_{\mathbf{G},\mathbf{G}'}(\mathbf{k}) = \omega^2_{pl}(\mathbf{G}-\mathbf{G}')\frac{(\mathbf{G}+\mathbf{k})\cdot(\mathbf{G}'+\mathbf{k})}{|\mathbf{G}'+\mathbf{k}|^2}\,, \tag{B8}$$

where $\omega^2_{pl}(\mathbf{q})$ is the Fourier transform of $\omega^2_{\rm pl}(\mathbf{r}) \propto n(\mathbf{r})$. This GPP form is set by asserting the ratio of bare and screened plasmon frequencies

$$S^{\rm GPP}_{\mathbf{G},\mathbf{G}'}(\mathbf{k},\omega=0)\,\omega^2_{\mathbf{G},\mathbf{G}'}(\mathbf{k}) = \Omega^2_{\mathbf{G},\mathbf{G}'}(\mathbf{k})\,. \tag{B9}$$

For semi-conductors and insulators, this is often done in an initial RPA calculation for $S^{\rm GPP}_{\mathbf{G},\mathbf{G}'}(\mathbf{k},\omega=0)$. The GPP is believed to be broadly applicable for a characterization of response. The RLL plasmon propagator is similar to this GPP form with one important difference: the absence of an explicit longitudinal projection $(\mathbf{G}+\mathbf{k})\cdot(\mathbf{G}'+\mathbf{k})$ in the RLL form, Eq. (B6).

The vdW-DF difference from the GPP is deliberate. The longitudinal projection is an absolute must for any proper description for the electron gas response. In the GPP it is hardwired from the start by the need to comply with the $f$-sum rule. In the vdW-DF, we delay the enforcement of the longitudinal projection in $S_{\rm xc} \equiv \bar{S}^{\rm RLL}$, and put it instead in the $\kappa_{\rm long}$ criterion. The projection is no less important for the vdW-DF-cx version. However, by lifting it to an explicit constraint on the electrodynamics description, Eq. (34), we position the vdW-DF framework and the vdW-DF-cx version for an automatic inclusion of vdW forces. This is done in the presence of screening by the electron gas, that is, in the setting first analyzed by the Ashcroft group [1, 10, 12, 64, 65].

## Appendix C: Nature of a full vdW-DF method implementation

The analysis of the optical-response description, Eqs. (102) and (105), suggests a path to go beyond the second-order expansion that is reflected in the popular general-geometry vdW-DF version [4]. While the existing functionals are given by Eqs. (2) and (3), a full implementation of the vdW-DF method is formally given

$$E^{\rm DF}_{\rm xc} = \sum_{n=1}^{\infty}\sum_{m\geq n}^{\infty}\sum_i E^{{\rm DF},(i)}_{{\rm xc},m,n}\,, \tag{C1}$$

$$E^{{\rm DF},(i)}_{{\rm xc},m,n} = -\int_0^\infty \frac{du}{2\pi}\,\frac{1}{m}\,{\rm Tr}\{\nabla\cdot\boldsymbol{\alpha}^{(i)}_{m,n}(iu)\cdot\nabla V\}\,. \tag{C2}$$

We may, for example, add some or all triples (identified as reflecting $n = 3$ factors of $S_{\rm xc}$), since each of the terms Eq. (C2) is, in itself, charge conserving.

In seeking such generalizations, however, we recommend always keeping all terms '$(i)$' containing $n$ powers of $S_{\rm xc}$ (up to a given maximum exponent value $n \leq l$) if truncating an expansion like Eq. (C2). This is done in vdW-DF1, vdW-DF2, and (s)vdW-DF-cx [4, 8, 9, 11]. We see the plasmon propagator $S_{\rm xc}$ as the natural physics property to use as a building block in a vdW-DF-based XC density functional [2–4, 10, 64, 65, 85, 86, 163, 164]. It is designed to reflect the simpler physics of the weakly perturbed electron gas. That is, the choice of $S_{\rm xc}$ as our expansion variable facilitates seamless integration with a near-HEG description [4, 47].

In any case, it is straightforward to sort XC functional contributions subject to the Dyson logic (even when components are expanded in $S_{\rm xc}$ factors). Each and every susceptibility term $\boldsymbol{\alpha}^{(i)}_{m,n}$ in Eq. (102) corresponds to a given order $(m)$ in Eq. (30) – but that is also the order it takes in the expansion of the logarithm in the full vdW-DF method specification.

### Appendix D: Nonlocal-correlation contributions

Finally, we explore the nature of the partial nonlocal-correlation energy contributions, Eqs. (110) and (113), when restricted to exactly a second-order expansion in $S_{\rm xc}$. Together they define the standard $E_{\rm c}^{\rm nl}$ approximation, Eq. (2). Using a Fourier-transformed representation [4] of the plasmon-propagator model $S_{\rm xc}(\omega,\mathbf{q},\mathbf{q}')$ these components can be expressed

$$E_{\rm c,\alpha}^{\rm nl} \approx \frac{1}{2}\int \frac{d\mathbf{q}}{(2\pi)^3}\int \frac{d\mathbf{q}'}{(2\pi)^3} \times \tag{D1}$$

$$\int_0^\infty \frac{du}{2\pi}\, S_{\rm xc}(iu,\mathbf{q},\mathbf{q}')\, S_{\rm xc}(iu,\mathbf{q}',\mathbf{q})\,, \tag{D2}$$

$$E_{\rm c,vdW}^{\rm nl} \approx -\frac{1}{2}\int \frac{d\mathbf{q}}{(2\pi)^3}\int \frac{d\mathbf{q}'}{(2\pi)^3}\,(\hat{\mathbf{q}}\cdot\hat{\mathbf{q}}')^2 \times$$

$$\int_0^\infty \frac{du}{2\pi}\, S_{\rm xc}(iu,\mathbf{q},\mathbf{q}')\, S_{\rm xc}(iu,\mathbf{q}',\mathbf{q})\,, \tag{D3}$$

where $\hat{\mathbf{q}} = \mathbf{q}/|\mathbf{q}|$. The factor $(\hat{\mathbf{q}}\cdot\hat{\mathbf{q}}')^2$ reflects the longitudinal projection in the vdW term.

Each of the terms in the second order expansion for $E_{\rm c}^{\rm nl}$ can be expressed in terms of corresponding partial kernels,

$$E_{\rm c,\alpha}^{\rm nl}[n] = \frac{1}{2}\int d\mathbf{r}\, d\mathbf{r}' n(\mathbf{r})\, \phi_\alpha(\mathbf{r},\mathbf{r}')\, n(\mathbf{r}'), \tag{D4}$$

$$E_{\rm c,vdW}^{\rm nl}[n] = \frac{1}{2}\int d\mathbf{r}\, d\mathbf{r}' n(\mathbf{r})\phi_{\rm vdW}(\mathbf{r},\mathbf{r}')n(\mathbf{r}'), \tag{D5}$$

as in the original vdW-DF determination [4, 47, 56, 171]. As discussed in the main text and elsewhere [4, 5, 10], the evaluation of $E_{\rm c}^{\rm nl}$, and of the here-sought components, is naturally cast by tracking the variation in a set of inverse length scales $q_0(\mathbf{r})$ and $q_0(\mathbf{r}')$. Meanwhile, the partial and full kernels are universal and can be tabulated in advance in terms of an effective separation $D = [(q_0 + q_0')/2]|\mathbf{r}-\mathbf{r}'|$ and in terms of an asymmetry parameter $\delta = (q_0 - q_0')/(q_0 + q_0')$, Ref. 4.

To specify the form of the universal (partial) kernels, we adapt the derivation presented in Refs. 56 and 60. We first extract a density component from the classical-plasmon mumerator $\omega_{\rm pl}^2(\mathbf{r}) = 4\pi n(\mathbf{r})/m$ from the plasmon-propagator model [4]

$$S_{\rm xc}(\omega,\mathbf{q},\mathbf{q}') = \int_{\mathbf{r}} e^{-i(\mathbf{q}-\mathbf{q}')\cdot\mathbf{r}} n(\mathbf{r}) \times$$

$$\frac{2\pi}{m}\,[f(\omega,\mathbf{q},\mathbf{q}',\mathbf{r}) + f(\omega,\mathbf{q}',\mathbf{q},\mathbf{r})]\,, \tag{D6}$$

$$f(\omega,\mathbf{q},\mathbf{q}',\mathbf{r}) = \frac{1}{[\omega+\omega_{\mathbf{q}}][-\omega+\omega_{\mathbf{q}'}]}\,, \tag{D7}$$

where $\omega_{\mathbf{q}}$ denotes the assumed plasmon dispersion [4, 47, 171]. This allows us to split the special kernels into a spatial and a frequency-dependent part [4, 56, 171]

$$\begin{aligned}&\phi_\alpha(\mathbf{r},\mathbf{r}') = \\ &\int_0^\infty \frac{dq}{(2\pi)^3} q^2 \int_0^\infty \frac{dq'}{(2\pi)^3} (q')^2 \times \\ &(4\pi)^2 W_\alpha(a = q|\mathbf{r}-\mathbf{r}'|, b = q|\mathbf{r}-\mathbf{r}'|) \times \\ &I_{\rm freq}(q,q',\mathbf{r},\mathbf{r}')\,,\end{aligned} \tag{D8}$$

in the same way as can be done for the total $E_{\rm c}^{\rm nl}$ energy [47, 56, 171], Eq. (2). The frequency-integration term is by construction unchanged from the original kernel evaluation:

$$I_{\rm freq}(q,q';\mathbf{r},\mathbf{r}') = \frac{(2\pi)^2}{m^2} T[\omega_q(\mathbf{r}),\omega_{q'}(\mathbf{r}),\omega_q(\mathbf{r}'),\omega_{q'}(\mathbf{r}')] \tag{D9}$$

with $T[w,x,y,z] > 0$ as defined in Ref. 4, Eq. (15).

The kernels from the two partial contributions differ in how we handle the spatial integrations in Eq. (D8). For the $(S_{\rm xc})^2$ contribution, we simply obtain [56]

$$\begin{aligned}&(4\pi)^2 W_\alpha(a = q|\mathbf{r}-\mathbf{r}'|, b = q|\mathbf{r}-\mathbf{r}'|) \\ &\quad \equiv \int d\Omega_1 \int d\Omega_2 e^{-i\mathbf{q}(\mathbf{r}-\mathbf{r}')} e^{i\mathbf{q}'(\mathbf{r}-\mathbf{r}')}\,. \\ &\quad = \frac{\sin(a)\sin(b)}{ab}\,.\end{aligned} \tag{D10}$$

For the pure vdW-attraction term we need

$$\begin{aligned}&(4\pi)^2 W_{\rm vdW}(a = q|\mathbf{r}-\mathbf{r}'|, b = q|\mathbf{r}-\mathbf{r}'|) \\ &\quad \equiv -\int d\Omega_1 \int d\Omega_2 (\hat{\mathbf{q}}\cdot\hat{\mathbf{q}}')^2 e^{-i\mathbf{q}(\mathbf{r}-\mathbf{r}')} e^{i\mathbf{q}'(\mathbf{r}-\mathbf{r}')}\end{aligned} \tag{D11}$$

This evaluation is also indirectly detailed in Ref. 56. The resulting specification is

$$\begin{aligned}&(4\pi)^2 W_{\rm vdW}(a = q|\mathbf{r}-\mathbf{r}'|, b = q|\mathbf{r}-\mathbf{r}'|) \\ &= -\frac{1}{(ab)^3}\big(2a\cos(a)\left[3b\cos(b) + (b^2-3)\sin(b)\right] \\ &\quad +2\sin(a)\cos(b)(a^2-3)b \\ &\quad +\sin(a)\sin(b)\left[a^2(b^2-2) - 2(b^2-3)\right]\big)\,.\end{aligned} \tag{D12}$$

This determination follows simply from performing the wavevector integrations in polar coordinates [47, 56, 60, 171].

Figure 7 presents our numerical evaluation of both the full vdW-DF kernel $\phi$ and of the kernel components, $\phi_\alpha$ and $\phi_{\rm vdW}$. The results are given in terms of universal kernel forms, $\Phi_\alpha(D,\delta)$, $\Phi_{\rm vdW}(D,\delta)$ and $\Phi(D,\delta) = \Phi_\alpha + \Phi_{\rm vdW}$, that permit evaluations under various density conditions [4, 7, 17]. For example, $\phi_{\rm vdW}(\mathbf{r},\mathbf{r}') = \Phi_{\rm vdW}(D,\delta)$ where the functional forms of $D(\mathbf{r},\mathbf{r}')$ and $\delta(\mathbf{r},\mathbf{r}')$ are given above.

The set of blue and red curves compares the variation of $\Phi_\alpha$ and of $\Phi_{\rm vdW}$ as a function of $D$ for a selection of indicated $\delta$ values, respectively. Finally, the set of the black curves tracks the corresponding variation in the full vdW-DF kernel, $\Phi$. It will, in the typical case, be relevant to use such kernels also when $q_0(\mathbf{r})$ differs from $q_0(\mathbf{r}')$, and $\delta$ will seldom be zero.

The kernel $\Phi_{\rm vdW}$ is always negative, and $E_{\rm c,vdW}^{\rm nl}[n]$ reflects a pure-vdW attraction. This is expected as it is formed from the term that has an explicit longitudinal projection (that cannot be removed by a partial integration as in Eq. (92)). The full vdW-DF behavior is offset by the other kernel $\Phi_\alpha$ which is always positive. The $E_{\rm c,\alpha}^{\rm nl}[n]$ term does not comply with the logic of Eq. (55) and does not secure a seamless integration, by itself.

One of the advantages of the vdW-DF method, and of vdW-DF-cx in particular, is that seamless integration is provided when combining the terms, as $E_{\rm c}^{\rm nl}$ does. The solid black curve in Fig. 7 shows the total kernel at $\delta = 0$, i.e., the condition that $q_0(\mathbf{r}) = q_0(\mathbf{r}')$, which is relevant for discussing the homogeneous-gas limit. For this curve, the short-range repulsion and longer-ranged attraction components fully compensate each other, the curve for $E_{\rm c}^{\rm nl}$ (given by Eq. (2)) integrates to zero in HEG, as it was designed to do [4, 5, 171] .

---